\documentclass[preprint,11pt]{article}
\usepackage{blindtext}
\usepackage[a4paper, total={6in, 8in}]{geometry}
\usepackage{array,multirow,graphicx}
\usepackage[backend=bibtex, sorting=none, maxbibnames=99]{biblatex}
\usepackage{float}
\usepackage{amsfonts, amsmath, amssymb, amsthm}
\usepackage{comment}
\usepackage{caption}
\usepackage{subcaption}
\usepackage{siunitx}
\usepackage{arydshln}
\usepackage{bbm}
\usepackage{lineno}
\usepackage{setspace}
\usepackage{gensymb}
\usepackage{xargs}                      
\usepackage[colorinlistoftodos,prependcaption,textsize=tiny]{todonotes}
\newcommandx{\unsure}[2][1=]{\todo[linecolor=red,backgroundcolor=red!25,bordercolor=red,#1]{#2}}
\newcommandx{\info}[2][1=]{\todo[linecolor=OliveGreen,backgroundcolor=OliveGreen!25,bordercolor=OliveGreen,#1]{#2}}
\newcommandx{\improvement}[2][1=]{\todo[linecolor=Plum,backgroundcolor=Plum!25,bordercolor=Plum,#1]{#2}}
\usepackage{array}
\usepackage{authblk} 
\newtheorem{problem}{Problem}

\newtheorem*{routine}{Routine}

\DeclareMathOperator*{\argmin}{argmin}
\DeclareMathOperator*{\x}{\boldsymbol{x}}
\DeclareMathOperator*{\y}{\textbf{y}}
\DeclareMathOperator*{\fnn}{\mit f_{nn}}
\DeclareMathOperator*{\diff}{\mathrm{d}}

\newcommand{\avsum}{\mathop{\mathpalette\avsuminner\relax}\displaylimits}

\makeatletter
\newcommand\avsuminner[2]{%
  {\sbox0{$\m@th#1\sum$}%
   \vphantom{\usebox0}%
   \ooalign{%
     \hidewidth
     \smash{\vrule height\dimexpr\ht0+1pt\relax depth\dimexpr\dp0+1pt\relax}%
     \hidewidth\cr
     $\m@th#1\sum$\cr
   }%
  }%
}
\makeatother

\title{A model learning framework for inferring the dynamics of transmission rate depending on exogenous variables for epidemic forecasts}

\author[a]{Giovanni Ziarelli\thanks{Corresponding author: giovanni.ziarelli@polimi.it}}
\author[a]{Stefano Pagani}
\author[a]{Nicola Parolini}
\author[a]{Francesco Regazzoni}
\author[a]{Marco Verani}

\affil[a]{MOX, Department of Mathematics, Politecnico di Milano}

\date{}

\begin{document}
\maketitle

\begin{abstract}
    % intro
Recent advancements in scientific machine learning offer a promising framework to integrate data within epidemiological models, offering new opportunities for the implementation of tailored preventive measures and the mitigation of the risks associated with epidemic outbreaks.
% problem
Among the many parameters to be calibrated and extrapolated in an epidemiological model, a special role is played by the transmission rate, whose inaccurate extrapolation can significantly impair the quality of the resulting forecasts.
% solution
In this work, we aim to formalize a novel scientific machine learning framework to reconstruct the hidden dynamics of the transmission rate, by incorporating the influence of exogenous variables (such as environmental conditions and strain-specific characteristics).
We propose an hybrid model that blends a data-driven layer with a physics-based one.
The data-driven layer is based on a neural ordinary differential equation that learns the dynamics of the transmission rate, conditioned on the meteorological data and wave-specific latent parameters.
The physics-based layer, instead, consists of a standard \textit{SEIR} compartmental model, wherein the transmission rate represents an input.
The learning strategy follows an end-to-end approach: the loss function quantifies the mismatch between the actual numbers of infections and its numerical prediction obtained from the \textit{SEIR} model incorporating as an input the transmission rate predicted by the neural ordinary differential equation.
We validate this novel approach using both a synthetic test case and a realistic test case based on meteorological data (temperature and humidity) and influenza data from Italy between 2010 and 2020.
In both scenarios, we achieve low generalization error on the test set and observe strong alignment between the reconstructed model and established findings on the influence of meteorological factors on epidemic spread.
Finally, we implement a data assimilation strategy to adapt the neural equation to the specific characteristics of an epidemic wave under investigation, and we conduct sensitivity tests on the network’s hyperparameters.
\end{abstract}

\section{Introduction}
Mathematical models have been extensively employed in epidemiology in order to answer to the most common questions arising from both policy-makers and the scientific community.
As the recent SARS-CoV-2 pandemic has highlighted, key-problems in this sense are the allocation of pharmaceutical \cite{lemaitre2022optimal, souto2022assessing, ziarelli2023optimized} and non-pharmaceutical control interventions \cite{giordano2021modeling, richard2021age, hinch2021openabm} and, especially, the forecasting of infection trends \cite{pasetto2018near, parolini2021suihter}.
Long-term scenario analyses are valuable for making strategic decisions regarding, \textit{e.g.}, treatment facilities placement, non-pharmaceutical interventions allocation, and the social and economical burden.
On the other hand, short-term forecasts, ranging from days to weeks, help in predicting the immediate need for resources such as protective gear, ventilators, hospital beds and vaccinations.
However, these forecasts pose significant challenges due to the many uncertainties that arise over extended time horizons.

Over the years, different modelling strategies have been proposed to tackle the problem of epidemic forecasting.
For instance, compartmental models are a popular choice and can be tailored for taking into account complex dynamics, \textit{e.g.} transmission mechanisms \cite{marziano2021effect}, concurring variants and illness-specific peculiarities \cite{parolini2022modelling}, or geographic dependencies of infections' diffusion \cite{bertuzzo2020geography}.
An alternative paradigm which does not involve adopting equation-based models is represented by Agent-Based Models \cite{hoertel2020stochastic, lima2021impact}.
ABMs follow a bottom-up approach for evaluating various interventions \cite{kerr2021covasim}, like home-schooling \cite{lasser2022assessing}, mobility-restrictions and so forth.
Recently, machine-learning-based methods have become widespread and reliable tools for making predictions and scenario analyses based on available data \cite{olumoyin2021data, shaier2022data}.
In some cases, the demand for accurate epidemic surrogate models has led to the development of innovative architectures, such as Asymptotic Preserving Neural Networks (APINN) \cite{bertaglia2022asymptotic}.
These networks are designed to address the challenges posed by multiscale hyperbolic PDE problems by incorporating heterogeneities, such as geographic features, into the model.
Additionally, drawing inspiration from fields like meteorology, forecasts generated by various techniques have recently been aggregated into forecasting hubs \cite{sherratt2023predictive, fiandrino2024collaborative}.
By combining different models using standardized formats, these ensemble predictions have shown greater accuracy than most individual models.

In general, the reliability of epidemic forecasts is intertwined with the choice of the model, the calibration strategy, and the method used to extrapolate the estimated parameters beyond the calibration interval.
A common approach for extrapolation is to keep parameters constant over time.
This strategy turns out to be effective when applied to parameters that can be determined through specific clinical trials and cohort studies, as they are related to the virological characteristics of the disease, such as the average recovery rate or incubation period.
Alternatively, linear, polynomial or exponential extrapolations are commonly used techniques, but suffer when the parameters exhibit a complex time-dependent behaviour.
Among the many parameters to be calibrated and extrapolated in an epidemiological model, a special role is played by the transmission rate, which is a time dependent parameter crucially influencing the overall transmission mechanism.
A poor extrapolation strategy of the transmission rate can seriously harm the quality of the overall forecasting process \cite{sherratt2023predictive}.

In recent years, different approaches have been proposed to extract from available data the time evolution behavior of the transmission rate.
For instance, in \cite{millevoi2023physics} the authors propose a new approach for training PINNs tracking temporal changes in epidemiological data and reconstructing the transmission rate.
Instead, a new architecture called Transmission-Dynamics-Informed Neural Network (TDINN) has been formalized for simulating the COVID19 epidemic in five Chinese cities, combining scattered data and $SIR$-like models in a physics-informed neural network (PINN) fashion \cite{he2023combining}.
A more straightforward approach, adopted in \cite{jing2021covid}, consists in prescribing a time dependent law for the transmission rate based on exponential decay models \cite{chowell2016characterizing}, which need to be fitted by available data on infections.

Actually, the time-dependent law derived from the aforementioned approaches does not fully account for the fact that the transmission rate depends on varying external factors.
By definition, the transmission rate represents the number of infected individuals per unit of time per susceptible individual necessary to spread the disease.
Thus, its value is influenced by both the intrinsic transmissibility, which remains constant over short time horizons without the emergence of new variants, and by external, or exogenous, factors such as environmental conditions (\textit{e.g.}, climate) \cite{mecenas2020effects}, social dynamics shaped by mobility patterns \cite{chinazzi2020effect}, the enforcement of non-pharmaceutical interventions and the distribution of vaccine doses across different age groups.
Consequently, incorporating these factors into the dynamic evolution of the transmission rate is crucial to improve the reliability of the forecasts.

It is worth mentioning that calibrating time-dependent coefficients influenced by exogenous variables presents a significant technical challenge across many scientific and engineering fields, ranging from epidemiology \cite{hazelbag2020calibration, gleeson2022calibrating}, to cardiac applications \cite{sainte2006modeling, regazzoni2020biophysically}, climate modeling \cite{sanso2009statistical}, and control systems \cite{saiteja2022critical}.

\subsection*{Original contributions}
\begin{figure}[t]
    \centering
    \includegraphics[width=\textwidth]{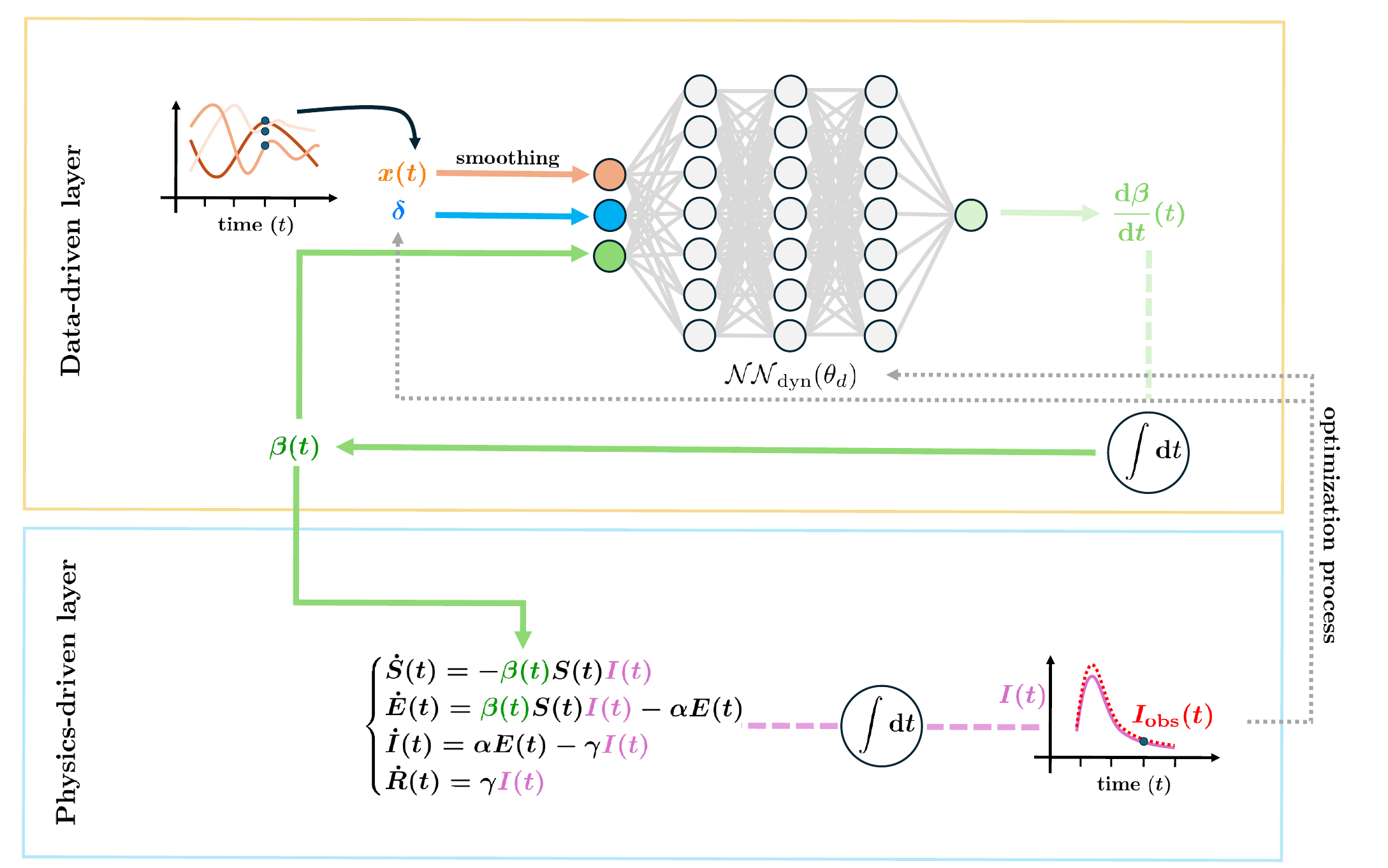}
    \caption{Schematic representation of the method.}
    \label{fig:scheme1}
\end{figure}

In this work, we propose an innovative neural network architecture, in a model learning fashion (see, \textit{e.g.}, \cite{regazzoni2019machine}), designed to learn the evolution of the transmission rate associated to a prescribed compartmental model.
A schematic overview of the architecture is provided in Figure \ref{fig:scheme1}.
The proposed architecture is hybrid in nature, as it consists of a data-driven layer (learning the transmission rate dynamics) coupled with a physics-based layer (predicting the epidemic scenario).
The data-driven differential model for the transmission rate is governed by a neural network that takes also exogenous variables as input.
The output is then integrated using a standard numerical integration scheme to generate the full time series of the transmission rate.
This time series is subsequently inserted into a physics-based epidemiological ODE model to predict disease incidence.
For the ease of presentation, we employ the classical \textit{SEIR} compartmental model, commonly used for epidemic diseases with an incubation period, but the framework is general enough to be employed with other different compartmental (or not compartmental) models.
All training variables are calibrated, in an end-to-end training, by minimizing an error metric based on infection data, specifically by reducing the mean squared error between the observed number of new cases (weekly incidence) and the corresponding values reconstructed by the hybrid model.
Additionally, we introduce a data assimilation strategy to recover, during both training and testing phases, a latent parameter that affects the transmission rate.
This parameter is interpreted as a measure of the intrinsic transmissibility of the illness, influenced by various factors such as disease strains, the immunisation profile and individuals' behavior in response to the outbreak, and it characterizes each wave of the same illness independently.

The proposed approach enables learning and exploration of the hidden dynamics of the transmission rate in relation to external factors, while providing reliable forecasts through the \textit{SEIR} model without depending on common, often inaccurate, extrapolation techniques. 

\subsection*{Outline}
This paper is structured as follows. In Section \ref{sec:methodLD}, we introduce our strategy, focusing specifically on how we addressed the training and testing stages, as well as the notation that will be used throughout the work.
In Section \ref{sec:resLD}, we carry out an extensive numerical testing campaign to evaluate the sensitivity of the architecture to the neural network’s topology hyperparameters in a synthetic test case (see Section \ref{subsec:tc1_das}), where the differential law of the transmission rate is prescribed \textit{a priori}, and temperature is treated as an exogenous variable.
We also discuss the validity of this approach in a real-world scenario using data from influenza epidemics in Italy between 2010 and 2020 (see Section \ref{subsec:realInflu}), investigating the role of meteorological factors such as temperature and relative humidity on the differential law governing the transmission rate (the impact of both factors on the transmission mechanism has been largely studied in epidemiology).
Finally, in Section \ref{sec:concLD}, we summarize the main contributions of this work and highlight potential future directions and extensions.

\section{Methods}
\label{sec:methodLD}
In this section, we introduce the notation used in this work and illustrate our proposed method for making reliable forecasts by analyzing the evolving dynamics of the transmission rate.
We define the vector of exogenous variables as,
\begin{equation}
    \x(t) = \begin{bmatrix}
        x_1(t)\\
        x_2(t)\\
        \vdots\\
        x_n(t)\\
    \end{bmatrix}  : [t_0, T] \subset \mathbb{R}^+ \rightarrow \mathbb{R}^{n}.
\end{equation}
Each $x_i(t), \; \forall i = 1,2,\hdots, n$ represents a different exogenous quantity on which we assume the transmission rate depends.
The choice of $x_i$ variables depends on the conditions that modellers want to embody in the analysis and on the disease at stake.
In the following, we will consider the case of infectious respiratory diseases as influenza, that shows a marked wintertime seasonality \cite{lowen2014roles}.
In this respect, some of the factors driving seasonality dependence are: variations in the host's ability to withstand immune system stress caused by extremes in temperature, as measured by melatonin and vitamin D levels; environmental factors, such as temperature itself, humidity, UV irradiation, and the direction of ambient air movement, which influence spread of the disease and seasonal variations in the host's behavior.
Another key quantity which impacts on the transmission rate is the relative transmissibility of the influenza virus itself, depending on the major lineages and on the virological composition of the spreading disease (cf. \textit{e.g.}, \cite{martcheva2015introduction}).
More precisely, in \textit{Test case 1} (cf. Section \ref{subsec:tc1_das}) we choose $x_1(t)$ as $T_p: [t_0, T] \rightarrow \mathbb{R}$ \textit{i.e.} the temperature measured in Celsius degree, taken as a suitable index for seasonality, whilst a scalar $\delta$ can be thought as an unknown parameter taking into account for unmodelled quantities, including lineages variability.
Furthermore, in \textit{Test case 2} (cf. Section \ref{subsec:realInflu}) we consider an additional exogenous variable $x_2(t)$ as $U: [t_0, T] \rightarrow \mathbb{R}$, which is the average relative humidity value.
We postpone to Section \ref{subsec:realInflu} the discussion on the motivations behind our choices.
Additionally, in both test cases we assume that transmission rate depends on a \textit{latent parameter} $\delta \in \mathbb{R}$, which is a possible unknown exogenous variable (to be learned).

Hence, our goal is to learn the function $f$ governing the unknown dynamics of the transmission rate:
\begin{equation}
\begin{cases}
    \dot{\beta}(t) = f(\beta(t); \x(t), \delta) \quad \forall \, t \in (t_0, T]\\
    \beta(t_0) = \beta_0.
    \end{cases}
    \label{eq:trmodel}
\end{equation}
The identification of $f$ is further complicated by the lack of direct measures of $\beta$, with the only available data being the effect on the epidemic (like, \textit{e.g.}, the number of weekly new infected patients).
This problem can be reformulated as a statistical learning problem based on the accessible and observable data by properly choosing the functional space $\mathcal{F}$ containing $f$.

\begin{problem}
Let $\{\Tilde{\boldsymbol{x}}^{(s)}, \Tilde{y}^{(s)}\}_{s\leq N} $ the sequence of pairs of observable data, where $\Tilde{\boldsymbol{x}}^{(s)} \in \mathcal{X}_d = \{ \boldsymbol{x} : [t_0, T]\rightarrow \mathbb{R}^n\}$ represents an input exogenous quantity and $ \Tilde{y}^{(s)} \in \mathcal{Y}_d = \{ y :[t_0, T] \rightarrow \mathbb{R}\}$ is the associated output functions.
Given $\Tilde{\boldsymbol{x}}^{(s)}$, we define $\beta^{(s)}$ as the corresponding solution of \eqref{eq:trmodel} for a certain $f \in \mathcal{F}$.
Each $\beta^{(s)}$ is uniquely determined once the couple of associated initial value and latent parameter $\{\beta_0^{(s)}, \delta^{(s)} \}$ is picked.
Furthermore, we assume that an epidemic model, mapping the transmission rate to the output quantity $y^{(s)}(t) = y(\beta^{(s)})(t)$, is prescribed.
Our learning problem reads as follows: 
\begin{equation}
    f^*, \, \{ \beta_0^{(s),*} \}_{s=1}^N, \{\delta^{(s),*}\}_{s=1}^N = \argmin_{f \in \mathcal{F}, \, \{ \beta_0^{(s)} \}_{s=1}^N \in \mathbb{R}, \, \{ \delta^{(s)} \}_{s=1}^N \in \mathbb{R}} \mathcal{L}(y^{(s)}, \Tilde{y}^{(s)}),
\end{equation}
where $\mathcal{L}: \mathcal{Y}_d \times \mathcal{Y}_d \rightarrow \mathbb{R}$ is a properly defined discrepancy measure or error metric to be minimized (see equations \eqref{eq:lossE1} and \eqref{eq:lossE2} for precise definitions in our context). 
\label{prob:Lprob}
\end{problem}

Inspired by the recently introduced Latent Dynamics Neural Networks (LDNets) \cite{regazzoni2024learning}, we choose $\mathcal{F}$ equal to the space of neural network functions and we assume that the map from the transmission rate to the output is given by a classical compartmental model.
A schematic representation of our methodology is depicted in Figure \ref{fig:scheme1}.
Additionally, we propose a computational strategy to estimate online the $\delta$ variable for both training and testing stages.
Specifically, we set
\begin{equation}
\begin{split}
    & \mathcal{F} = \biggl \{ \fnn:\mathbb{R} \times \mathbb{R}^{n+1} \rightarrow \mathbb{R} \; \mathrm{s.t.} \; \fnn(\beta, \x, \delta; \theta_d) = W_L z^{L} + b_L, \\
    & \mathrm{where} \; z^{l} = \sigma(W_{l-1} z^{l-1} + b_{l-1}) \; \mathrm{for} \; l=1,2,\hdots,L, \, z^0 = \begin{bmatrix}
        \beta\\
        \x\\
        \delta
    \end{bmatrix} \biggl \},
\end{split}
\end{equation}
\textit{i.e.} the set of neural network functions with $L$ layers and $N_l$ neurons per layer.
We denote by the generic $\theta_d$ the set of all trainable network parameters, namely the matrices $W_l$ and vectors $b_l$ for $l = 0,\hdots,L$.

We frame Problem \ref{prob:Lprob} in the specific context of this work. 
We define a map from the input $ [x_1, \, x_2, \, \hdots , \, x_n]^T \in \mathcal{X} \subset [L^{\infty}([t_0,T])]^n$ to the vector $\y = [S, \, E, \, I, \, R]^T \in \mathcal{Y} \subset [\mathcal{C}^0([t_0,T])]^4$ by solving the following system of ODEs:
    \begin{equation}
        \begin{cases}
            \dot{\beta}(t) = \fnn(\beta(t), \x(t), \delta; \theta_d) & \forall t \in (t_0, T]\\
            \beta(t_0)  = \beta_0\\
            \dot{\y}(t) = \mathrm{SEIR}(\y(t), \beta(t)) & \forall t \in (t_0,T]\\
            \y(t_0) = \y_0,
        \end{cases}
    \label{eq:sys1}
    \end{equation}
    where the \textit{SEIR} epidemic model \cite{dukic2012tracking, lekone2006statistical, buonomo2011simple, kumar2021wavelet} reads as: \begin{equation}
    \begin{cases}
        \dot{S}(t) = - \beta(t) S(t) I(t), & \forall t \in (t_0, T], \\
        \dot{E}(t) = \beta S(t) I(t) - \alpha E(t), & \forall t \in (t_0, T],\\
        \dot{I}(t) = \alpha E(t) - \gamma I(t), & \forall t \in (t_0, T],\\
        \dot{R}(t) = \gamma I(t), & \forall t \in (t_0, T],\\
        [S(t_0), E(t_0), I(t_0), R(t_0)]^T = [S_0, E_0, I_0, R_0]^T.
    \end{cases}
    \label{eq:SEIRsys}
\end{equation}
The states $S, E, I, R : [t_0,T] \rightarrow \mathbb{R}^{+}$ represent the relative amount of susceptible, exposed, infected and recovered individuals with respect to the illness under investigation.
In this work we assume that a suitable calibration stage for initial conditions $[S_0, E_0, I_0, R_0]^T$ has been already undertaken before the simulated scenarios.
The parameter $\alpha = \frac{1}{T_I} > 0$ is the inverse of the average incubation time. %, which is physiological of the disease of concern.
Instead, $\gamma = \frac{1}{T_R}>0 $ is the recovery rate, where $T_R$ represents the recovery time, \textit{i.e.} the mean time infected individuals remain infected and infectious.
Both parameters can be estimated starting from contact tracing studies during the epidemic waves, through cohort studies following groups of people exposed to the disease, or through statistical and Bayesian inference, \textit{e.g.} exploiting Kaplan-Meier estimation as in \cite{virlogeux2015estimating}.

Before addressing the solution of Problem \ref{prob:Lprob}, let us mention that for integrating \eqref{eq:sys1} we employ the Forward Euler method with a constant time step, $\Delta t$, selecting the step size to ensure both stability and accurate representation of the system's dynamics.
For this purpose, it may be necessary to integrate the differential equations on a finer grid than the observation grid.
In our case we perform a suitable resampling of each $\boldsymbol{x}^{(s)}(t)$ exploiting piecewise linear interpolation.

We are now ready to describe the computational strategy for addressing Problem \ref{prob:Lprob}.
In doing this we provide details about the training and test phases.
During the training stage we look for the optimal values of the network trainable variables (\textit{i.e.} weights and biases), the initial conditions for the transmission rate and the corresponding latent parameter for each training trajectory.
To achieve this goal, we employ a two-step optimization routine.
The two steps differ in the choice of the cost functional to be minimized: the output of the first step will employed as initial guess for the second step.
The testing stage on unseen inputs is, in turn, subdivided in two steps, as it is schematically synthesized in Figure \ref{fig:scheme2}: (a) estimation stage, combined with data-assimilation techniques, for retrieving initial transmission rate and latent parameter of each testing trajectory; (b) prediction stage to guess the trend of the epidemic model for each input by numerically solving the transmission and epidemic ODEs using the corresponding estimated parameters.
In the following, we detail the numerical procedure of each stage:

\begin{enumerate}
    \item \textbf{Training}: 
The optimization training routine is constituted by two different optimization problems: we first optimize $\mathcal{L}_{\mathrm{train}, \mathcal{E}_1}$ and then $\mathcal{L}_{\mathrm{train}, \mathcal{E}_2}$, where
       \begin{equation}
    \begin{split}
        \mathcal{L}_{\mathrm{train}, \mathcal{E}_1}(\theta_d; \{\beta_0^{(i)}\}_{i=1}^{N}, \{\delta^{(i)}\}_{i=1}^{N}) &= \avsum_{i =1, \hdots, N} \avsum_{\omega \in \mathcal{W}} \mathcal{E}_1(\{ \textbf{y}^{(i)}(\tau)\}_{\tau \in (\omega, \omega+1]},\overline{\Delta I}^{(i)}(\omega)) + \alpha_{\mathrm{reg}} \mathcal{R}(\theta_d)\\
        &+ \mathcal{B}(\theta_d; \{\beta_0^{(i)}\}_{i=1}^{N}, \{\delta^{(i)}\}_{i=1}^{N}) + \mathcal{D}(\theta_d; \{\beta_0^{(i)}\}_{i=1}^{N}, \{\delta^{(i)}\}_{i=1}^{N})\\
        &+ \mathcal{I}(\theta_d; \{\beta_0^{(i)}\}_{i=1}^{N}, \{\delta^{(i)}\}_{i=1}^{N}) + \mathcal{A}(\{\delta^{(i)}\}_{i=1}^{N}),
        \end{split}
        \label{eq:lossE1}
    \end{equation}
    and 
    \begin{equation}
    \begin{split}
        \mathcal{L}_{\mathrm{train}, \mathcal{E}_2}(\theta_d; \{\beta_0^{(i)}\}_{i=1}^{N}, \{\delta^{(i)}\}_{i=1}^{N}) &= \avsum_{i =1, \hdots, N} \avsum_{\omega \in \mathcal{W}} \mathcal{E}_2(\{ \textbf{y}^{(i)}(\tau)\}_{\tau \in (\omega, \omega+1]},\overline{\Delta I}^{(i)}(\omega)) + \alpha_{\mathrm{reg}} \mathcal{R}(\theta_d)\\
        &+ \mathcal{B}(\theta_d; \{\beta_0^{(i)}\}_{i=1}^{N}, \{\delta^{(i)}\}_{i=1}^{N}) + \mathcal{D}(\theta_d; \{\beta_0^{(i)}\}_{i=1}^{N}, \{\delta^{(i)}\}_{i=1}^{N})\\
        &+ \mathcal{I}(\theta_d; \{\beta_0^{(i)}\}_{i=1}^{N}, \{\delta^{(i)}\}_{i=1}^{N}) + \mathcal{A}(\{\delta^{(i)}\}_{i=1}^{N}),
        \end{split}
        \label{eq:lossE2}
    \end{equation}
    with $\mathcal{W}$ representing the set of weeks considered in each scenario.
    Each term has its own specific weight factor to be tuned ($\alpha_*$).
    The numerical values corresponding to each test case will be provided in their respective sections.
    All the optimization problems are solved numerically through two iterative schemes applied in sequence: the first order Adam and the second order BFGS (for additional information see, \textit{e.g.}, \cite{regazzoni2024learning}).
    To ensure BFGS convergence, we perform a sufficient number of Adam iterations, following manual selection when the cost functional reaches a plateau.
    In contrast, BFGS terminates based on an early stopping criterion.
    
    The two loss functions differ for the discrepancy metrics $(\mathcal{E}_*)$ employed.
    Indeed, for the first $N_{ep,1}$ iterations, we minimize $\mathcal{L}_{\mathrm{train}, \mathcal{E}_1}$ with $\mathcal{E}_1$:
    \begin{equation}
        \mathcal{E}_1(\{\textbf{y}^{(i)}(\tau)\}_{\tau \in (\omega, \omega+1]},\overline{\Delta I}^{(i)}(\omega), \delta^{(i)}) = \alpha_{\mathcal{E},1} \dfrac{(\Delta I^{(i)}(\omega) - \overline{\Delta I}^{(i)}(\omega))^2}{ \overline{\Delta I}^{(i)}(\omega)^2}.
        \label{eq:disc1}
    \end{equation}
    This metric measures the relative discrepancies among weekly new cases, the computed ($\Delta I^{(i)}(\omega)$) and the target ($\overline{\Delta I}^{(i)}(\omega))$) ones.
    The choice of considering the weekly datum is guided by the way available data about infected individuals are commonly delivered for epidemic outbreaks (\textit{e.g.} influenza's new infections are collected weekly in a public repository as it is described in Section \ref{subsec:realInflu}).
    
    Afterwards, we focus on the second optimization process, which is crucial for accurately determining peaks, a primary objective in epidemiology.
    Indeed, in the first optimization process, we prioritized learning the start and end times of epidemic wave phases by minimizing relative discrepancies ($\mathcal{E}_1$).
    As a result, at the end of the first stage, we often obtain inaccurate trajectories in terms of peak values.
    Therefore, after completing $N_{ep,1}$ iterations, we initiate the second optimization process, using the current values of the trainable variables, and aiming to minimize $\mathcal{L}_{\mathrm{train}, \mathcal{E}_2}$ over the remaining $N_{ep,2}$ iterations.
    The metric $\mathcal{E}_2$ is defined as:
    \begin{equation}
        \mathcal{E}_2(\{\textbf{y}^{(i)}(\tau)\}_{\tau \in (\omega, \omega+1]},\overline{\Delta I}^{(i)}(\omega), \delta^{(i)}) = \alpha_{\mathcal{E},2} \left ( \Delta I^{(i)}(\omega) - \overline{\Delta I}^{(i)}(\omega)\right )^2.
        \label{eq:disc2}
    \end{equation}
    Thus, the training optimization routine reads as follows:
    \begin{routine}
    Initialize $(\theta_d; \{\beta_0^{(i)}\}_{i=1}^{N}, \{\delta^{(i)}\}_{i=1}^{N})$.
    Then, solve
    \begin{equation}
        (\theta_{d,1}^*, \{ \beta_{0,1}^{i,*}\}_{i=1}^{N}, \{ \delta_1^{i,*}\}_{i=1}^{N}) = \argmin_{\theta_d, \{ \beta_0^{(i)}\}_{i=1}^{N}, \{ \delta^{(i)}\}_{i=1}^{N}} \mathcal{L}_{\mathrm{train}, \mathcal{E}_1}(\theta_d; \{\beta_0^{(i)}\}_{i=1}^{N}, \{\delta^{(i)}\}_{i=1}^{N}).
        \label{eq:optTrain1}
    \end{equation}
    Starting from $(\theta_{d,1}^*, \{ \beta_{0,1}^{i,*}\}_{i=1}^{N}, \{ \delta_1^{i,*}\}_{i=1}^{N})$, we solve
    \begin{equation}
        (\theta_d^*, \{ \beta_0^{i,*}\}_{i=1}^{N}, \{ \delta^{i,*}\}_{i=1}^{N}) = \argmin_{\theta_d, \{ \beta_0^{(i)}\}_{i=1}^{N}, \{ \delta^{(i)}\}_{i=1}^{N}} \mathcal{L}_{\mathrm{train}, \mathcal{E}_2}(\theta_d; \{\beta_0^{(i)}\}_{i=1}^{N}, \{\delta^{(i)}\}_{i=1}^{N}).
        \label{eq:optTrain1}
    \end{equation}
    \end{routine}
    In each loss function, we also consider other terms in order to help the training process. 
    \begin{itemize}
        \item The term $\alpha_{\mathrm{reg}} \mathcal{R}(\theta_d)$ represent the Tikhonov regularization term on the trainable variables (both weights and biases);
        \item
        A (Tikhonov-like) regularization enforcing $\{\delta^{(i)}\}$ centered in 1:
        \begin{equation}
            \mathcal{A}(\{\delta^{(i)}\}_{i=1}^{N}) = \alpha_{\mathcal{A}} \avsum_{i =1, \hdots, N} (\delta^{(i)} - 1)^2.
        \end{equation}
        This approach can help us in interpreting the unknown parameters as positive factors among different epidemic waves.
        This choice is arbitrary: indeed, a different type of prior could be used without affecting the algorithm's performance;
        \item A regularization term penalizing whether transmission rates exceed a given threshold:
        \begin{equation}
        \mathcal{B}(\theta_d; \{ \beta_0^{(i)}\}, \{\delta^{(i)}\}_{i=1}^{N}) = \alpha_{\mathcal{B}} \avsum_{i =1, \hdots, N} \mathrm{max}((\beta^{(i)}(t) - \beta_{th})^q, 0).
        \end{equation}
        Indeed, $\beta_{th}$ is an informed threshold, and $q \in \mathbb{N}$ is an odd number to be tuned (in our case $q = 5$).
        This term is fundamental for obtaining physical results when dealing with realistic data, whilst it is unnecessary in our considered synthetic scenario;
        \item A term that looks for parsimonious reconstructions of the dynamics, penalizing unexpected unphysical oscillations in the transmission rate:
        \begin{equation}
        \mathcal{D}(\theta_d; \{\beta_0^{(i)}\}_{i=1}^{N}, \{\delta^{(i)}\}_{i=1}^{N}) = \alpha_{\mathcal{D}} \, | \beta |^2_{H^1(t_0,T)},
        \end{equation}
        where 
        \begin{equation}
            | f |^2_{H^1(t_0,T)} = \int_{t_0}^T f'(t) \diff t.
            \label{eq:h1seminorm}
        \end{equation}
        We opt to approximate $f'$ in \eqref{eq:h1seminorm} using forward finite differences;
        \item 
        A regularization loss addendum in order to maintain a final level of retrieved susceptible ($\{S_{\infty,i}\}_i$) over a given threshold:
        \begin{equation}
        \mathcal{I}(\theta_d; \{\beta_0^{(i)}\}_{i=1}^{N}, \{\delta^{(i)}\}_{i=1}^{N}) = \alpha_{\mathcal{I}} \avsum_{i =1, \hdots, N} \mathrm{max}((S_{th} - S_{\infty,i})^q, 0).
        \end{equation}
        This helps in avoiding excessive starting spikes in the transmission rate which could lead to the unfeasible emptying of the Susceptible compartment.
    \end{itemize}
    We remark that 
    \begin{equation}
            \Delta I^{(i)}(\omega) = S^{(i)}(\omega) - S^{(i)}(\omega +1)+ E^{(i)}(\omega) - E^{(i)}(\omega + 1),
            \label{eq:reform}
    \end{equation}
    exploiting the fundamental theorem of calculus in the continuous \textit{SEIR} model.
    Hence, the amount of new infections can be determined as the difference in the extrema of a given week of those individuals which are still susceptible to the illness, \textit{i.e} individuals in $S$ and in $E$. 
    Therefore, the discrepancy metrics can be rewritten in a much computationally efficient form from the point of view of gradients to be computed during optimization.
    
    \item 
        \begin{figure}[t]
        \centering
        \includegraphics[width=\textwidth]{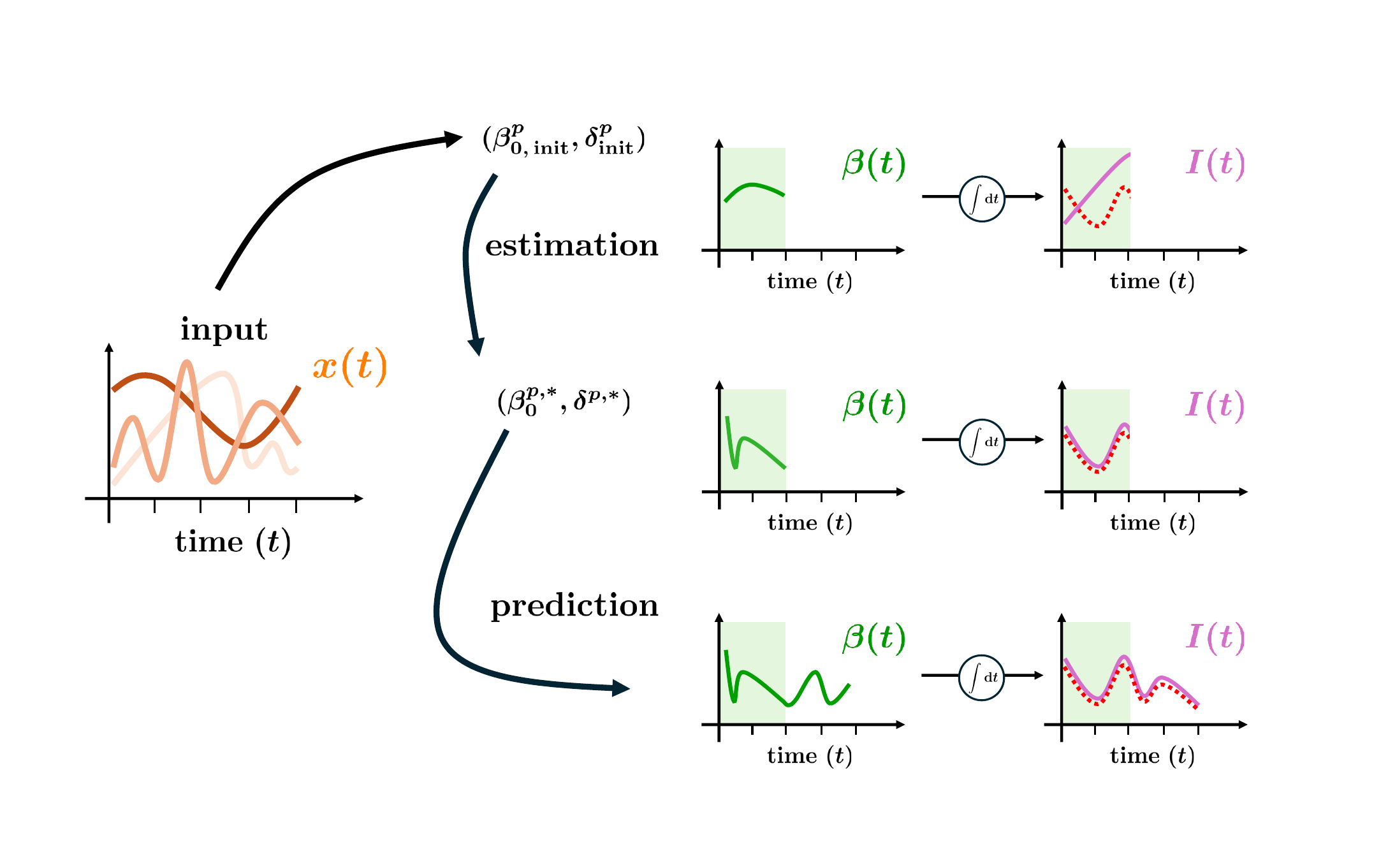}
        \caption{Schematic representation of the estimation phase. The dashed red line represents the target trajectories of the transmission rate (left pictures) and infected individuals (right pictures). The green line indicate the reconstructed transmission rate, while the pink lines depict the reconstructed infected individuals. The green shaded area corresponds the observation window $[t_0, T_{\mathrm{obs}}]$.}
        \label{fig:scheme2}
    \end{figure}
    \textbf{Estimation}: A schematic representation of both estimation and prediction steps can be found in Figure \ref{fig:scheme2}.
    Once a testing sample is drawn, its associated $\delta^{(p)}$ needs to be estimated, together with the initial value for the transmission rate $\beta_0^{(p)}$.
    For this purpose, we rely on the approach proposed in \cite{regazzoni2021combining} which is based on data-assimilation techniques.

    Starting from given initial guesses $(\beta_{0, \mathrm{init}}^{p,*}, \delta_{\mathrm{init}}^{p,*})$, we solve another optimization problem for determining the unknown couple of coefficients $(\beta_0^{(p)}, \delta_p)$, while keeping frozen the parameters of the trained model.
    In the interval $[0,T_{\mathrm{obs}}]$, where $T_{\mathrm{obs}} \leq T$ is a prescribed observation time, we minimize the following estimation loss:
    \begin{equation}
    \begin{split}
        \mathcal{L}_{\mathrm{estim}}(\beta_0^{(p)},\delta^{(p)}) & = \avsum_{\omega \in \mathcal{W}, \; \omega \leq T_{\mathrm{obs}}} \mathcal{E}_2(\{\textbf{y}^{(p)}(\tau)\}_{\tau \in (\omega, \omega+1]},\overline{\Delta I}^{(p)}(\omega))\\
        &+ \alpha_{\mathcal{A}} \dfrac{(\delta^{(p)} - \Bar{\delta})^2}{C_{\delta}} + \alpha_{\beta_0} \dfrac{(\beta^{(p)} - \Bar{\beta_0})^2}{C_{\beta}},
        \end{split}
        \label{eq:lossest}
    \end{equation}
    where $(\alpha_{\mathcal{A}}, \; \alpha_{\beta_0} )$ are the weights of the \textit{a priori} regularization.
    In $\mathcal{L}_{\mathrm{estim}}$ we have prescribed an empirical gaussian-like prior distribution for the (disjoint) pair $( \delta, \beta_0)$ based on the observed values  $\{ \delta^{(i)}, \beta_0^{(i)}\}_{i=1}^{N}$ after the training stage.
    In this case $(\bar{\delta}, \; \bar{\beta_0})$ are the sample mean values of the reconstructed parameters after training $\{ \delta^{(i)}, \beta^{(i)}\}_{i=1}^{N}$ and $(C_{\delta}, \; C_{\beta})$  the sample covariances.
    
    We remark that the model for the transmission rate has already been determined, \textit{i.e.} weights and biases of the network do not belong to the set of trainable variables.
    Hence, the estimation problem reads as
    \begin{equation}
        (\beta_0^{p,*}, \delta^{p,*}) = \argmin_{\beta_0^{(p)}, \delta^{(p)}} \mathcal{L}_{\mathrm{estim}}(\beta_0^{(p)},\delta^{(p)}).
        \label{eq:optEstim}
    \end{equation}
    \item \textbf{Prediction}: Once the trainable variables have been optimized and the sample-dependent couple of parameters has been estimated, we solve the forward system \eqref{eq:sys1} to predict the new sample evolution in the whole interval $[t_0,T]$.
\end{enumerate}
Without loss of generality, in the following section we keep $t_0 = 0$.

\section{Results}
\label{sec:resLD}
In this section we present some numerical results in order to validate our novel approach.
In particular, in Section \ref{subsec:tc1_das} we test our approach in an artificial scenario in which we aim at recovering the prescribed differential dynamics of the transmission rate which depends on an exogenous variable (temperature).
In this context we also show the ability of the architecture to reconstruct hidden dynamics of the transmission rate in presence of data affected by noise.
Additionally, we perform an extensive campaign of numerical simulations in order to assess the sensitivity and robustness of our learning procedure with respect to some network hyperparameters, to the width of the observation window during the estimation stage and to the increase in the amount of training samples.
We assess the performance of our approach by comparing different cases based on the test error, measured on a fixed batch of previously unseen samples.

In Section \ref{subsec:realInflu} we consider a realistic scenario investigating on the differential model relating meteorological quantities, \textit{i.e.} temperature and relative humidity, to the recovered transmission rate for the case of influenza spreading in Italy during 2010-2020, with waves of average length of 28 weeks.
Moreover, we propose to study the equilibrium points of the reconstructed dynamical system associated with the transmission rate, considering different values of the input exogenous variables.

\subsection{Test case 1: Artificial scenario}
\label{subsec:tc1_das}
\subsubsection{Data generation}
\label{subsec:dataGenSynt}
\begin{figure}[t]
    \centering
    \includegraphics[width=0.6\textwidth]{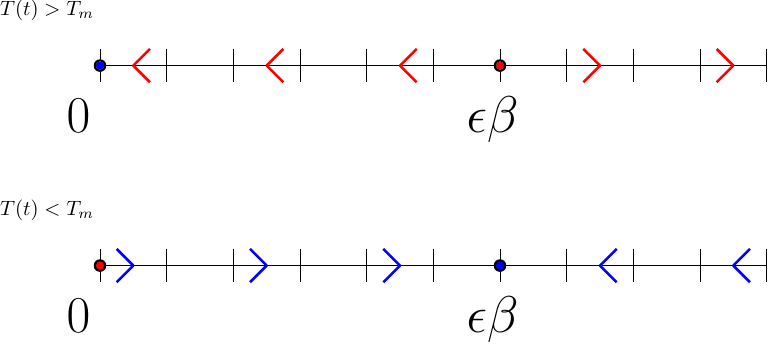}
    \caption{Phase line of the imposed model for synthetic transmission rate.}
    \label{fig:equilibria}
\end{figure}
We generate a training dataset of temperatures taking seasonal variations into account.
In particular, we consider each input time history belonging to the following family:
\begin{equation}
    T^{(i)}(t) = T_{m} + A^{(i)} \sin(2 \pi \,f \,t + \phi^{(i)}), 
    \label{eq:tempGen}
\end{equation}
where $T_{m}$ represents a medium value of temperature, $A^{(i)}$ is the amplitude of the seasonal effect, $f$ is the seasonal wave frequency and $\phi^{(i)}$ its phase.
The $i$-th superscript indicates that the quantities are associated with the $i$-th sample.
Moreover, we assume that the transmission rate satisfies a logistic equation with fluctuations driven by temperature:
\begin{equation}
    \dot{\beta^{(i)}}(t) = \left( \dfrac{T^{(i)}(t)}{T_m}-1\right)\left(\dfrac{\beta^{(i)}(t)^2}{ \epsilon^{(i)} \beta_r} - \beta^{(i)}(t)\right),
    \label{eq:modelTRsynth}
\end{equation} 
where $\beta_r$ is a scaling factor for the transmission rate, and $\epsilon^{(i)} > 0$ represents a suitable scaling factor embodying the specific infectivity of different epidemic waves.
The differential law \eqref{eq:modelTRsynth} has been retrieved starting from the modelling choices of \cite{johnsen2022seasonal} as explained in \ref{app:der_dl}.
The equilibrium points of \eqref{eq:modelTRsynth} are
\begin{equation}
    \begin{split}
        & T^{(i)}(t) \neq T_m, \, \beta^{i,\mathrm{eq}} = \{0, \epsilon^{(i)} \beta^{(i)} \};\\
        & T^{(i)}(t) = T_m, \, \beta^{i,\mathrm{eq}} = \mathbb{R}^+.
    \end{split}
\end{equation}
In case $T^{(i)}(t) \neq T_m$, the stability of the two equilibrium points depends on the sign of $T^{(i)}(t) - T_m$ quantity.
Indeed,
\begin{equation}
\begin{split}
    & T^{(i)}(t) > T_m \Rightarrow \beta^{i, \mathrm{eq}} = 0 \; \mathrm{asymptotically \; stable}, \, \beta^{i, \mathrm{eq}} = \epsilon^{(i)} \beta^{(i)} \; \mathrm{unstable},\\
     & T^{(i)}(t) < T_m \Rightarrow \beta^{i, \mathrm{eq}} = 0 \; \mathrm{unstable}, \, \beta^{i, \mathrm{eq}} = \epsilon^{(i)} \beta^{(i)} \; \mathrm{asymptotically \; stable},\\   
\end{split}
\label{eq:anal_synt_db}
\end{equation}
as it is represented in Figure \ref{fig:equilibria}.
We remark that the regime where $T(t) > T_m$ with $\beta^{(i)}(t) > \epsilon^{(i)} \beta_r$  would lead the transmission rate to diverge, which is clearly an unphysical behavior.
In the following, we only consider transmission rates with initial condition in the physical regime. 
We remark that during both training and testing phases we will estimate the latent parameter $\delta$ for each sample, that we expect to be linked to the  lineage parameter $\epsilon$ -- the only other factor driving our prescribed synthetic dynamics.

In order to retrieve the target values corresponding to each couple $\{(T^{(i)}(t), \epsilon^{(i)})\}_i$ we solve the \textit{SEIR} model \eqref{eq:SEIRsys}, by keeping fixed the parameters: $$[S_0, E_0, I_0, R_0]^T = [0.97, 0.01, 0.02, 0]^T, \; \alpha = 1/10, \; \; \mathrm{and} \; \; \gamma = 1/20. $$
We simulate each phase until the final time $T = 364$, \textit{i.e.} considering a time window of length one year.
Unless otherwise specified, we deal with training datasets of size 50 and with a fixed test dataset of size 50.
Both datasets are generated by randomly sampling the input parameters following the probability distributions in Table \ref{tab:tabParams}.
Instead, the reference value for the transmission rate has been computed through the following algebraic expression:
\begin{equation}
    \beta_r = \gamma \dfrac{\log \left ( \dfrac{S_0}{S_{\infty}}\right )}{1 - S_{\infty}},
\end{equation}
which is derived from a standard \textit{SEIR} model with constant coefficients by imposing a final size of Susceptible $S_{\infty}$.
In our case we keep $S_{\infty} = 0.80$, therefore $\beta_r \approx 0.061$.

\renewcommand{\arraystretch}{2}
\begin{table}[t]
\tiny
\centering
\begin{tabular}{|l|l|l|l|}
\hline
Parameter & Distribution & Interval & Unit of measurement \\ \hline \hline
$T_m$    & Uniform          &          $[10,15]$   & \degree$C$\\  \hline
$A$      & Uniform          &          $[5,10]$ & \degree$C$\\  \hline
$f$      & Constant         & $\frac{1}{365}$ &   $[\frac{1}{\mathrm{days}}]$     \\  \hline          
$\phi$   & Uniform          &             $[\frac{1}{6}\frac{\pi}{365}, \frac{3\pi}{365}]$ & $[-]$\\ \hline               
$\delta$ & Uniform          &         $[0.1, 3]$   & $[-]$\\ \hline
$\beta_0$ & Uniform          & $[0, 0.1]$& $[\frac{1}{\mathrm{days}}]$             \\   \hline
\end{tabular}
\caption{Table of the parameters for generating synthetic data (Test case 1).}
\label{tab:tabParams}
\end{table}
\subsubsection{Hyperparameters setup}
\label{subsec:hyp_set_synt}
The network's architecture in terms of width and layers has been selected following the sensitivity analysis that will be presented in Section \ref{sec:sensAn}, considering the hyperbolic tangent as activation function.
For each Adam optimization stage we fix the learning rates as $\eta_{\mathrm{train}, \mathcal{E}_1} = 5 \cdot 10^{-4}$, $\eta_{\mathrm{train}, \mathcal{E}_2} = 10^{-5}$ and $\eta_{\mathrm{testg}} = 10^{-3}$.
Those values have been set up following a trial-and-error approach.
The number of maximum epochs of both Adam and BFGS optimizations and weights for each cost functional addendum for each stage are set as in Table \ref{tab:tabHyperParam}.
Initial values of neural network's weights and biases are randomly initialized (standard deviation equal to 0.001).
The observation window for the test set has been fixed to 11 weeks, according to the sensitivity analysis.
\begin{table}[t]
\tiny
\centering
\begin{tabular}{|l|l|l|l|l|l|l|l|l|l|}
\hline
& Epochs & $\alpha_{\mathcal{A}}$ & $\alpha_{\mathcal{B}}$ & $\alpha_{\mathcal{D}}$ & $\alpha_{\mathcal{E},1}$ & $\alpha_{\mathcal{E},2}$ & $\alpha_{\mathcal{I}}$ & $\alpha_{\mathrm{reg}}$ & $\alpha_{\beta}$\\ \hline \hline
Training ($\mathcal{E}_1/\mathcal{E}_2$) Adam   &  200/200 & $5.0 \cdot 10^{-3}$ & 0 & $5.0 \cdot 10^{-2}$ & $1.1$& $1.1 \cdot 10^{2}$ & $10^{-6}$ & $10^{-6}$ & 0\\  \hline
Training BFGS  ($\mathcal{E}_1/\mathcal{E}_2$) &  200/500 & / & / & / & / & / & / & / & /\\  \hline
Estimation Adam &  500 &  6.2 & $5 \cdot 10^{-1}$& 0 & $7.0$ & 0& 0& 0& $5.0 \cdot 10^{-2}$ \\  \hline          
Estimation BFGS &  500 & /&/ &/ & /& /&/  & / &  / \\  \hline         
\end{tabular}
\caption{Table of the hyperparameters of each minimization stage (Test case 1).}
\label{tab:tabHyperParam}
\end{table}

\subsubsection{Results: Impact of uncertain data}

\begin{figure}[t]
    \centering
    \begin{subfigure}[b]{0.45\textwidth}
        \includegraphics[width=0.8\textwidth]{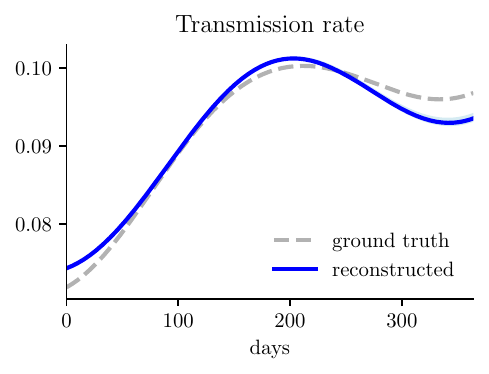}
        \caption{}
        \label{fig:beta_train}
    \end{subfigure}
    \hfill
    \begin{subfigure}[b]{0.45\textwidth}
        \includegraphics[width=0.8\textwidth]{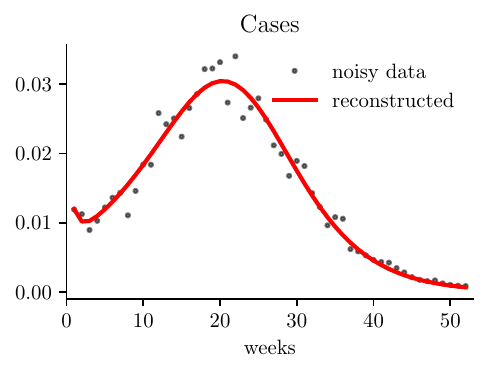}
        \caption{}
        \label{fig:cas_train}
    \end{subfigure}
    \vspace{0.5cm}
    \begin{subfigure}[b]{0.45\textwidth}
        \includegraphics[width=0.8\textwidth]{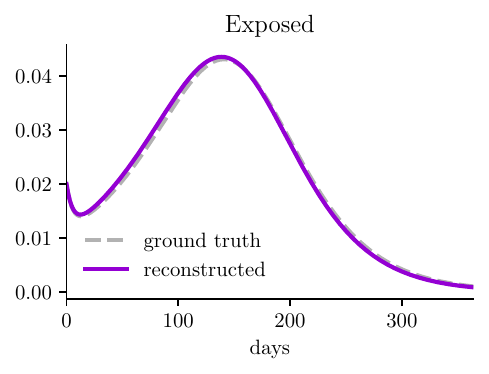}
        \caption{}
        \label{fig:exp_train}
    \end{subfigure}
    \hfill
    \begin{subfigure}[b]{0.45\textwidth}
        \includegraphics[width=0.8\textwidth]{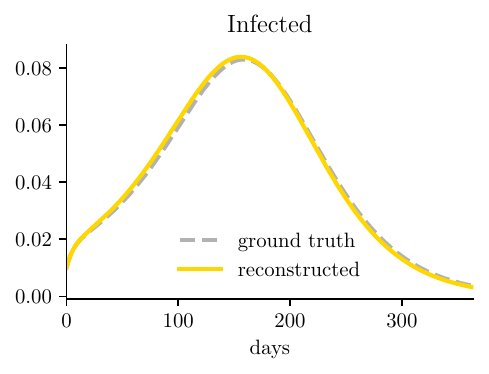}
        \caption{}
        \label{fig:inf_train}
    \end{subfigure}
    \caption{Uncertain data: Transmission Rate (a), Weekly cases (b), Exposed (c) and Infected (d) of a given training sample.}
    \label{fig:trainPlot}
\end{figure}
In this section we provide numerical results dealing with noisy data, to test robustness of the proposed approach.
This analysis has a crucial impact in scenarios embodying real infectious data, whose quality is typically affected by under-reporting or reporting delays, potentially causing misinformation to public health authorities \cite{white2010reporting}.
Moreover, infections are difficult to be isolated during epidemic waves, and the proxies that we use, such as cases and deaths, often provide noisy approximations of the unknown real amount of infections.

To this aim, we train the model adding artificial noise in a multiplicative way to the target amount of new cases:
\begin{equation}
    \overline{\Delta I}^{(i)}(\omega) = \widetilde{\overline{\Delta I}^{(i)}}(\omega) (1 + u \kappa), 
    \label{eq:data_mod_unc}
\end{equation}
where $u > 0$ is the absolute entity of uncertainty,  $\widetilde{\overline{\Delta I}^{(i)}}(\omega)$ is the detected amount of new cases and $\kappa \sim \mathcal{N}(0,1)$ is a standard Gaussian error.
In this synthetic case, we retrieve each $\widetilde{\overline{\Delta I}^{(i)}}(\omega)$ by numerically solving the \textit{SEIR} model coupled with \eqref{eq:modelTRsynth}.
According to \eqref{eq:data_mod_unc}, we assume higher probability of detection errors in the neighbourhood of peak values, since, due to limited detection resources, it is often more difficult to detect the correct amount of infected when the illness has already largely spread.
In Figure \ref{fig:trainPlot}, the evolution of the reconstructed compartments of infected and exposed and the amount of cases is presented for one simulation obtained with a random initializations of weights and biases (cf. Subsection \ref{subsec:hyp_set_synt}).
The reconstructed transmission rate, as a function of temperature as in Equation \eqref{eq:tempGen}, has been recovered with a mean relative error in the whole reconstruction $ e_{r} \approx 1.3\%$, even if the amount of cases are affected by uncertainty of module $u = 0.1$.
With the reconstructed transmission rate, infected and exposed compartments overlap with the denoised original data.

We then consider various levels of uncertainty 
\begin{equation}
    u \in \{0, 0.001, 0.005, 0.01, 0.05, 0.1\},
\end{equation}
and, after the training process, we evaluate the prediction error \eqref{eq:disc2} in the test dataset specified in Section \ref{subsec:dataGenSynt}.
The original denoised test set is altered in each case with the same level of uncertainty error of the training set: for obtaining each test dataset for a given level of uncertainty $u$, we start from the denoised 50 trajectories of the neat test dataset described in Subsection \ref{subsec:dataGenSynt} and we apply \eqref{eq:data_mod_unc} corresponding to the respective $u$.

\begin{figure}[H]
    \centering
    \begin{subfigure}[b]{0.45\textwidth}
    \includegraphics[width=\textwidth]{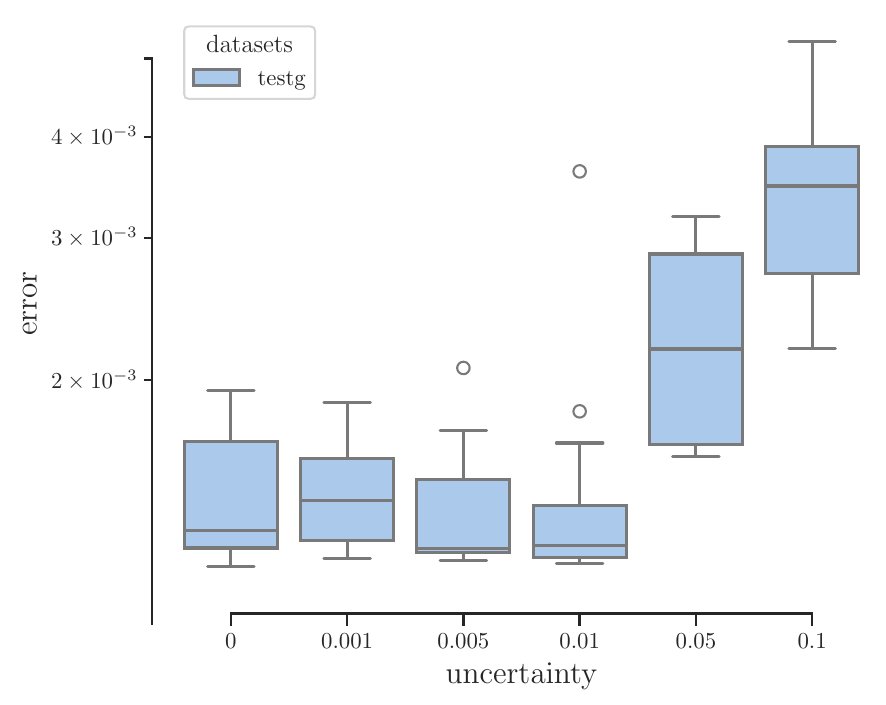}
    \caption{Test errors with different uncertainties.}
\label{fig:boxUnc}
    \end{subfigure}
    \hfill
    \begin{subfigure}[b]{0.45\textwidth}
    \includegraphics[width=\textwidth]{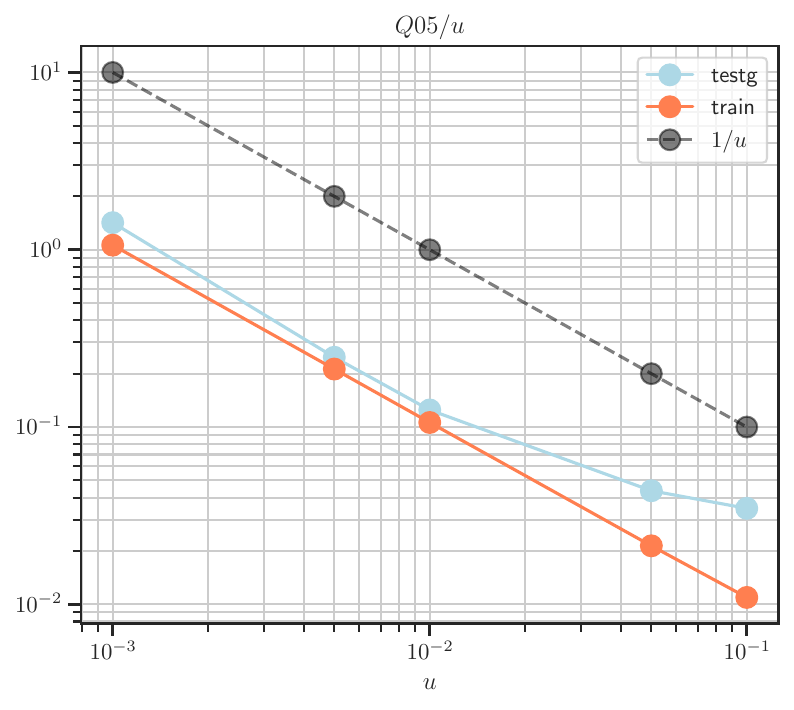}
    \caption{Median over amplitude of uncertainty.}
    \label{fig:q05}
    \end{subfigure}
    
    \vspace{0.2cm}

        \centering
    \begin{subfigure}[b]{0.8\textwidth}
    \includegraphics[width=\textwidth]{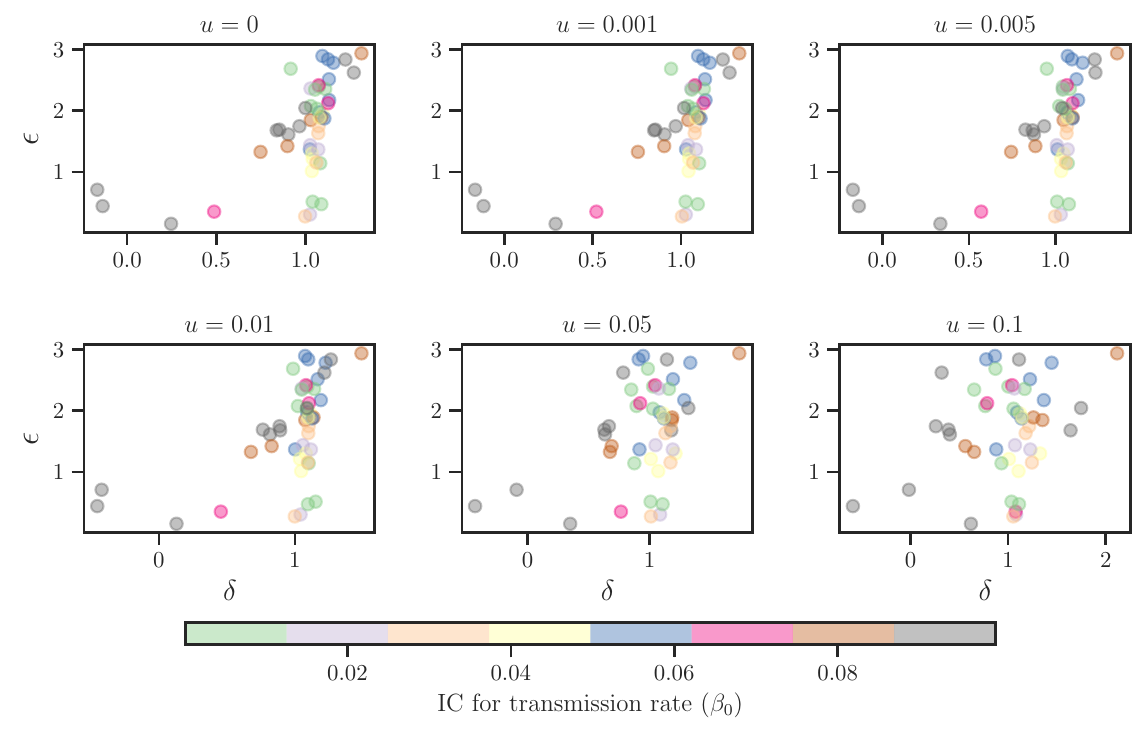}
    \caption{Lineage parameters generating the test set versus the mean value of the reconstructed latent parameters with different uncertainties. Different colors correspond to different value of the the initial condition (IC) for the transmission rate dynamics  according with the colorbar.}
    \label{fig:deltas_unc}
    \end{subfigure}

    \caption{Analysis of uncertainty.}
    \label{fig:unc_all}
\end{figure}

In Figure \ref{fig:boxUnc} we considered 20 different runs with varying initializations of the trainable variables, presenting test errors in boxplot form for the difference between the reconstructed new infection cases and the denoised ground truth values.
The IQR of the boxplots widens across the five cases as uncertainty increases, and the median error value gets higher.
However, in Figure \ref{fig:q05} we observe that the median test error ($Q05$), when scaled by the average magnitude of uncertainty ($u$), decreases as uncertainty grows.
Additionally, the median training error remains constant despite uncertainty rises.

In Figure \ref{fig:deltas_unc}, we examine the latent parameter distribution on the test set ($\delta^{(p)}$).
The x-axis represents the values of the lineage parameter $\epsilon$  used to generate the dataset through \eqref{eq:modelTRsynth}, while the y-axis shows the mean value of the reconstructed latent parameter across simulations with different initializations of the trainable variables.
As noted in \cite{regazzoni2021combining}, the reconstruction value of the latent parameter through the data-assimilation technique introduced above is not unique, but for different target values the reconstructions should lie on a parametric curve.
Thus, for similar values of $\beta_0^{(p)}$, the initial value for the transmission rate dynamics of each testing sample, a clear trend is expected.
The network should increasingly recognize these trends as more training data becomes available and noise decreases, as it happens, \textit{e.g.} for the black and brown latent parameters in the first row of Figure \ref{fig:deltas_unc}.
However, in some cases the parameter reconstructions for similar $\beta_0^{(p)}$ values present a cloudy shape, \textit{e.g.} the green latent parameters in Figure \ref{fig:deltas_unc}.
Actually, these values correspond to declining epidemics with low initial transmission rates: our model performs weakest in approximating these epidemic  trajectories, since, starting from a low transmission initial value, the epidemic wave does not outbreak, and it evolves independently on the transmission rate dynamics, making the parameter $\delta$ hardly identifiable.
We also observe that increasing uncertainty in the target data impacts the latent parameters, causing their reconstructions to become more diffuse with respect to the lineage parameter as uncertainty increases within the same range of initial transmission rates.

\begin{figure}[H]
    \begin{subfigure}[b]{0.25\textwidth}
    \centering
\includegraphics[width=0.48\textwidth]{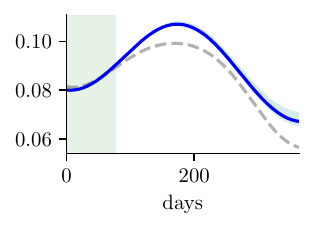}
\includegraphics[width=0.48\textwidth]{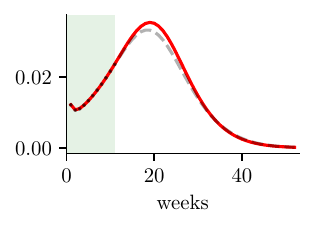}
        \captionsetup{labelformat=empty}
        \caption{(a1) $u = 0$.}
        \label{fig:figbeta0_1}
    \end{subfigure}%
    %\hfill
    \begin{subfigure}[b]{0.25\textwidth}
    \centering
\includegraphics[width=0.48\textwidth]{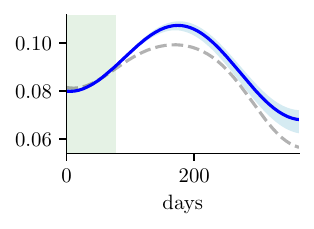}
\includegraphics[width=0.48\textwidth]{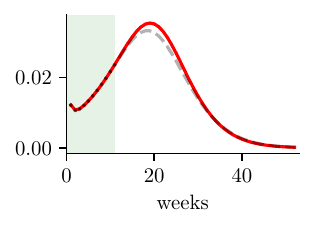}
        \captionsetup{labelformat=empty}
        \caption{(a2) $u = 0.001$.}
        \label{fig:figbeta1_1}
    \end{subfigure}%
    %\hfill
    \begin{subfigure}[b]{0.25\textwidth}
    \centering
\includegraphics[width=0.48\textwidth]{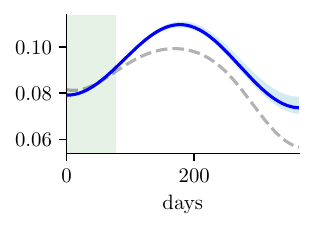}
\includegraphics[width=0.48\textwidth]{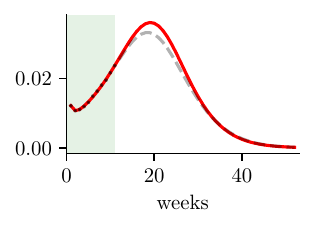}
        \captionsetup{labelformat=empty}
        \caption{(a3) $u = 0.01$.}
        \label{fig:figbeta2_1}
    \end{subfigure}%
    %\hfill
    \begin{subfigure}[b]{0.25\textwidth}
    \centering
\includegraphics[width=0.48\textwidth]{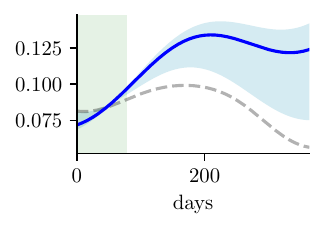}
\includegraphics[width=0.48\textwidth]{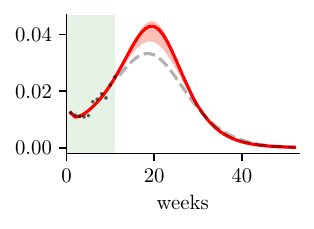}
        \captionsetup{labelformat=empty}
        \caption{(a4) $u = 0.1$.}
        \label{fig:figbeta3_1}
    \end{subfigure}%
    %\hfill
    
    \vspace{0.5cm}
    
        \begin{subfigure}[b]{0.25\textwidth}
    \centering
\includegraphics[width=0.48\textwidth]{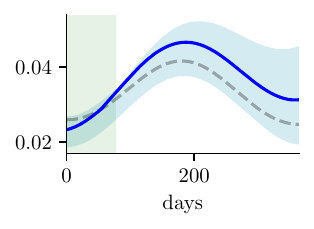}
\includegraphics[width=0.48\textwidth]{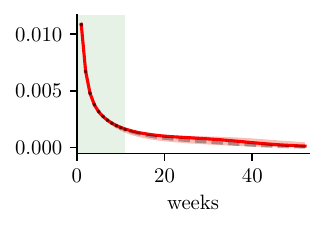}
        \captionsetup{labelformat=empty}
        \caption{(b1) $u = 0$.}        \label{fig:figbeta0_1}
    \end{subfigure}%
    %\hfill
    \begin{subfigure}[b]{0.25\textwidth}
    \centering
\includegraphics[width=0.48\textwidth]{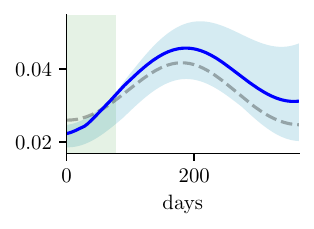}
\includegraphics[width=0.48\textwidth]{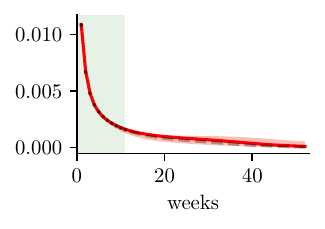}
        \captionsetup{labelformat=empty}
        \caption{(b2) $u = 0.001$.}  
        \label{fig:figbeta1_1}
    \end{subfigure}%
    %\hfill
    \begin{subfigure}[b]{0.25\textwidth}
    \centering
\includegraphics[width=0.48\textwidth]{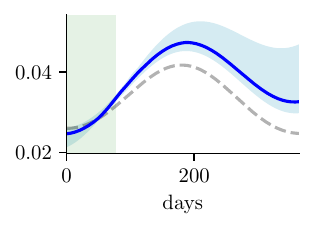}
\includegraphics[width=0.48\textwidth]{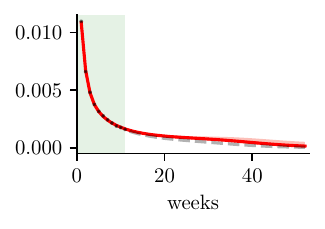}
        \captionsetup{labelformat=empty}
        \caption{(b3) $u = 0.01$.}  
        \label{fig:figbeta2_1}
    \end{subfigure}%
    %\hfill
    \begin{subfigure}[b]{0.25\textwidth}
    \centering
\includegraphics[width=0.48\textwidth]{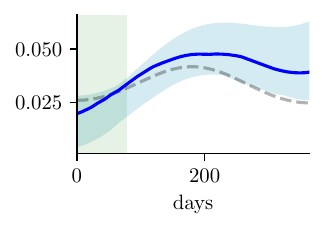}
\includegraphics[width=0.48\textwidth]{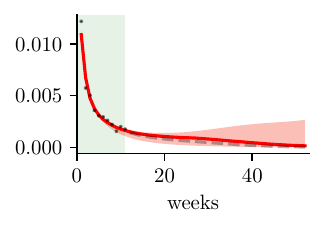}
        \captionsetup{labelformat=empty}
        \caption{(b4) $u = 0.1$.}  
        \label{fig:figbeta3_1}
    \end{subfigure}%

        \vspace{0.5cm}
    
        \begin{subfigure}[b]{0.25\textwidth}
    \centering
\includegraphics[width=0.48\textwidth]{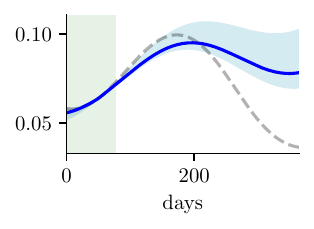}
\includegraphics[width=0.48\textwidth]{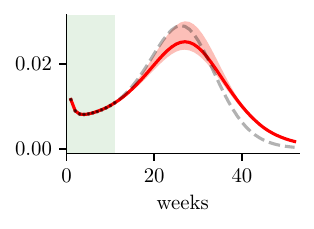}
        \captionsetup{labelformat=empty}
        \caption{(c1) $u = 0$.}          \label{fig:figbeta0_1}
    \end{subfigure}%
    %\hfill
    \begin{subfigure}[b]{0.25\textwidth}
    \centering
\includegraphics[width=0.48\textwidth]{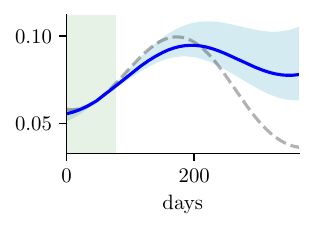}
\includegraphics[width=0.48\textwidth]{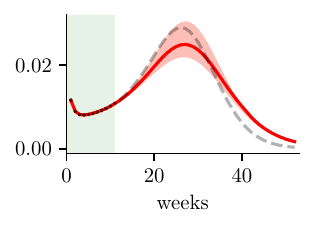}
        \captionsetup{labelformat=empty}
        \caption{(c2) $u = 0.001$.}  
        \label{fig:figbeta1_1}
    \end{subfigure}%
    %\hfill
    \begin{subfigure}[b]{0.25\textwidth}
    \centering
\includegraphics[width=0.48\textwidth]{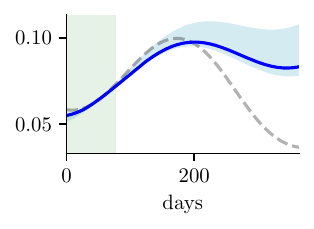}
\includegraphics[width=0.48\textwidth]{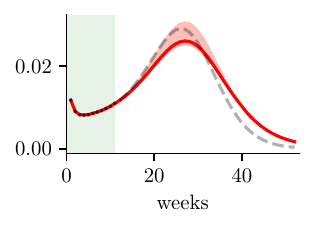}
        \captionsetup{labelformat=empty}
        \caption{(c3) $u = 0.01$.}  
        \label{fig:figbeta2_1}
    \end{subfigure}%
    %\hfill
    \begin{subfigure}[b]{0.25\textwidth}
    \centering
\includegraphics[width=0.48\textwidth]{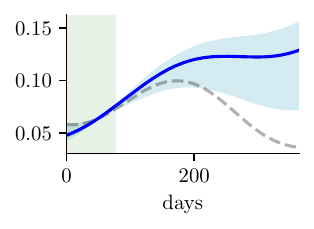}
\includegraphics[width=0.48\textwidth]{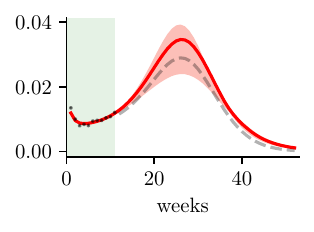}
        \captionsetup{labelformat=empty}
        \caption{(c4) $u = 0.1$.}  
        \label{fig:figbeta3_1}
    \end{subfigure}%
    
        \vspace{0.5cm}
    
        \begin{subfigure}[b]{0.25\textwidth}
    \centering
\includegraphics[width=0.48\textwidth]{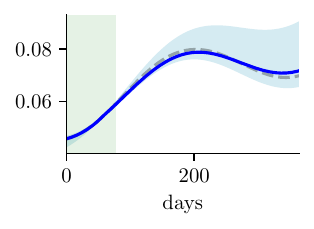}
\includegraphics[width=0.48\textwidth]{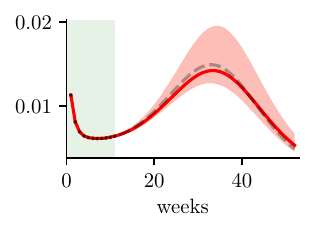}
        \captionsetup{labelformat=empty}
        \caption{(d1) $u = 0$.}  
        \label{fig:figbeta0_1}
    \end{subfigure}%
    %\hfill
    \begin{subfigure}[b]{0.25\textwidth}
    \centering
\includegraphics[width=0.48\textwidth]{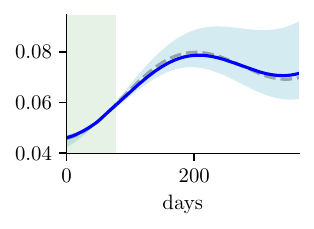}
\includegraphics[width=0.48\textwidth]{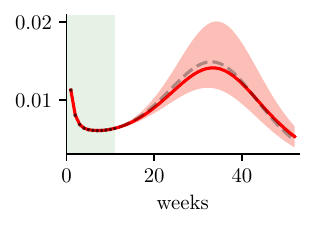}
        \captionsetup{labelformat=empty}
        \caption{(d2) $u = 0.001$.}  
        \label{fig:figbeta1_1}
    \end{subfigure}%
    %\hfill
    \begin{subfigure}[b]{0.25\textwidth}
    \centering
\includegraphics[width=0.48\textwidth]{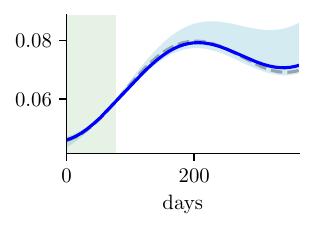}
\includegraphics[width=0.48\textwidth]{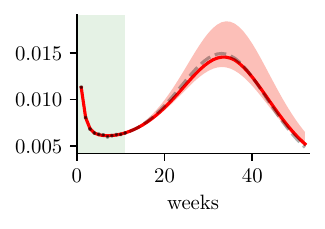}
        \captionsetup{labelformat=empty}
        \caption{(d3) $u = 0.01$.}          \label{fig:figbeta2_1}
    \end{subfigure}%
    %\hfill
    \begin{subfigure}[b]{0.25\textwidth}
    \centering
\includegraphics[width=0.48\textwidth]{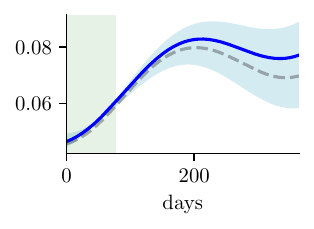}
\includegraphics[width=0.48\textwidth]{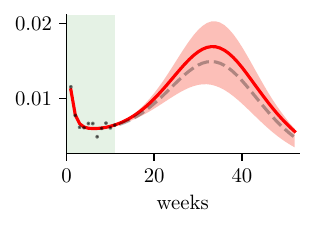}
        \captionsetup{labelformat=empty}
        \caption{(d4) $u = 0.1$.}          \label{fig:figbeta3_1}
    \end{subfigure}%
    
        \vspace{0.5cm}
    
        \begin{subfigure}[b]{0.25\textwidth}
    \centering
\includegraphics[width=0.48\textwidth]{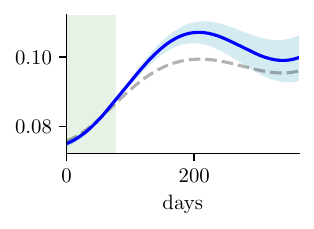}
\includegraphics[width=0.48\textwidth]{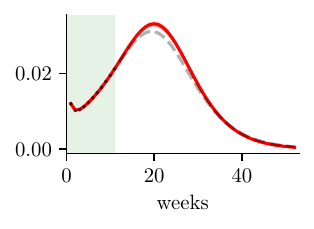}
        \captionsetup{labelformat=empty} % Rimuove la didascalia predefinita (a)
        \caption{(e1) $u = 0$.}          \label{fig:figbeta0_1}
    \end{subfigure}%
    %\hfill
    \begin{subfigure}[b]{0.25\textwidth}
    \centering
\includegraphics[width=0.48\textwidth]{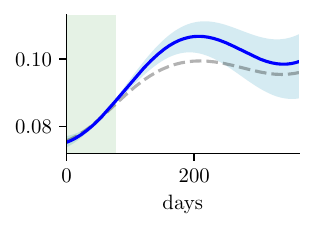}
\includegraphics[width=0.48\textwidth]{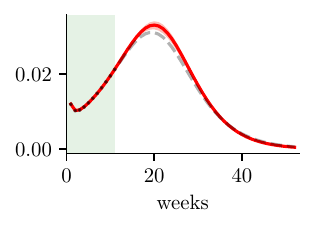}
        \captionsetup{labelformat=empty}
        \caption{(e2) $u = 0.001$.}             \label{fig:figbeta1_1}
    \end{subfigure}%
    %\hfill
    \begin{subfigure}[b]{0.25\textwidth}
    \centering
\includegraphics[width=0.48\textwidth]{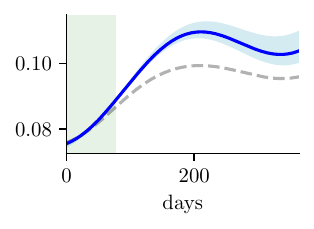}
\includegraphics[width=0.48\textwidth]{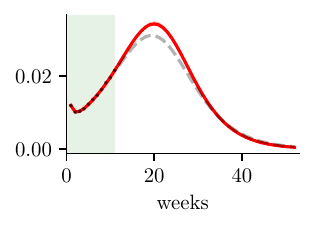}
        \captionsetup{labelformat=empty}
        \caption{(e3) $u = 0.01$.}             \label{fig:figbeta2_1}
    \end{subfigure}%
    %\hfill
    \begin{subfigure}[b]{0.25\textwidth}
    \centering
\includegraphics[width=0.48\textwidth]{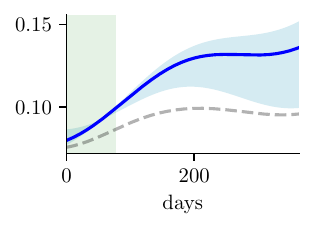}
\includegraphics[width=0.48\textwidth]{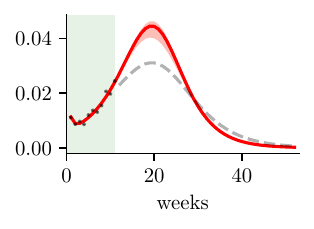}
        \captionsetup{labelformat=empty}
        \caption{(e4) $u = 0.1$.}             \label{fig:figbeta3_1}
    \end{subfigure}%
    
        \vspace{0.5cm} % Adjust the space between rows
    
        \begin{subfigure}[b]{0.25\textwidth}
    \centering
\includegraphics[width=0.48\textwidth]{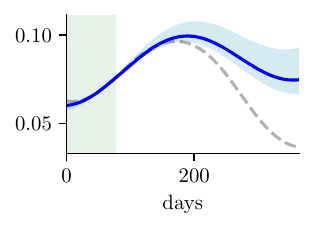}
\includegraphics[width=0.48\textwidth]{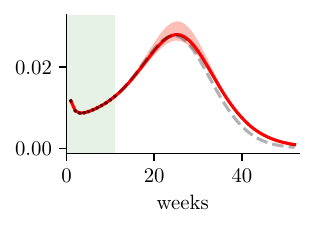}
        \captionsetup{labelformat=empty}
        \caption{(f1) $u = 0$.}             \label{fig:figbeta0_1}
    \end{subfigure}%
    %\hfill
    \begin{subfigure}[b]{0.25\textwidth}
    \centering
\includegraphics[width=0.48\textwidth]{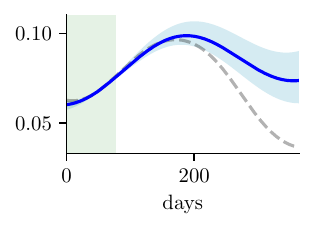}
\includegraphics[width=0.48\textwidth]{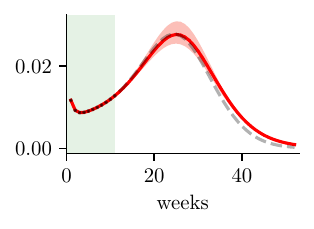}
        \captionsetup{labelformat=empty}
        \caption{(f2) $u = 0.001$.}
        \label{fig:figbeta1_1}
    \end{subfigure}%
    %\hfill
    \begin{subfigure}[b]{0.25\textwidth}
    \centering
\includegraphics[width=0.48\textwidth]{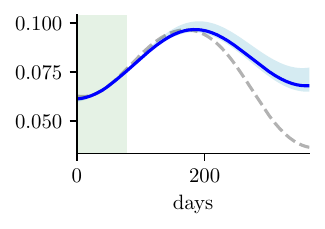}
\includegraphics[width=0.48\textwidth]{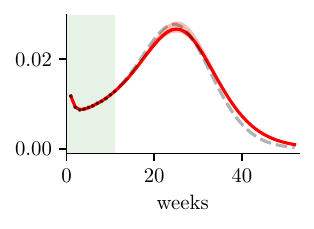}
        \captionsetup{labelformat=empty}
        \caption{(f3) $u = 0.01$.}
        \label{fig:figbeta2_1}
    \end{subfigure}%
    %\hfill
    \begin{subfigure}[b]{0.25\textwidth}
    \centering
\includegraphics[width=0.48\textwidth]{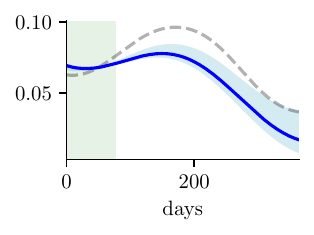}
\includegraphics[width=0.48\textwidth]{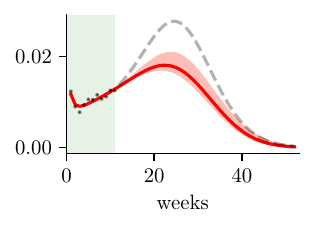}
        \captionsetup{labelformat=empty}
        \caption{(f4) $u = 0.1$.}          \label{fig:figbeta3_1}
    \end{subfigure}%
    \caption{Testing transmission rates (blue) and cases (red) with data with different uncertainties.
    Each row shows the same sample with growing uncertainty: $u \in \{0, \, 0.001,  \, 0.01, \,  0.1\}$. }
    \label{fig:cases_unc_s}
\end{figure}

In Figure \ref{fig:cases_unc_s} we reported six samples, one per row, of transmission rates and the respective weekly amount of cases belonging to the test dataset for three different levels of uncertainty.
The dot points correspond to noised observed data, whilst the dashed line is the unknown original (denoised) target trajectory.
The uncertainty bands correspond to the 0.1-0.9 quantiles of the trajectories considering 20 different models trained starting with different random initializations of weights and biases.
We note that the learned evolutive model for the transmission rate is able to capture the frequency and phase of the harmonic signal inferred by temperature.
We observe that the transmission rate is coherent with the ground truth, even if, as one can expect, noisy measurements influence the reconstructed dynamics.
After the epidemic peak is reached, the reconstructed transmission rates tend to be less accurate and to be more stationary than the target curve.
This behavior is explained by the fact that, with the compartment of infected individuals nearly depleted, constant non-zero transmission rates have little effect on the epidemic, which is already in decline.
As a result, the trained model tends to be more accurate during the early stages of the epidemic wave, less in the long-time horizons.

\subsubsection{Results: Sensitivity analysis}
\label{sec:sensAn}
\begin{figure}[H]
    \begin{subfigure}[b]{0.43\textwidth}
    \centering
    \includegraphics[width=\textwidth]{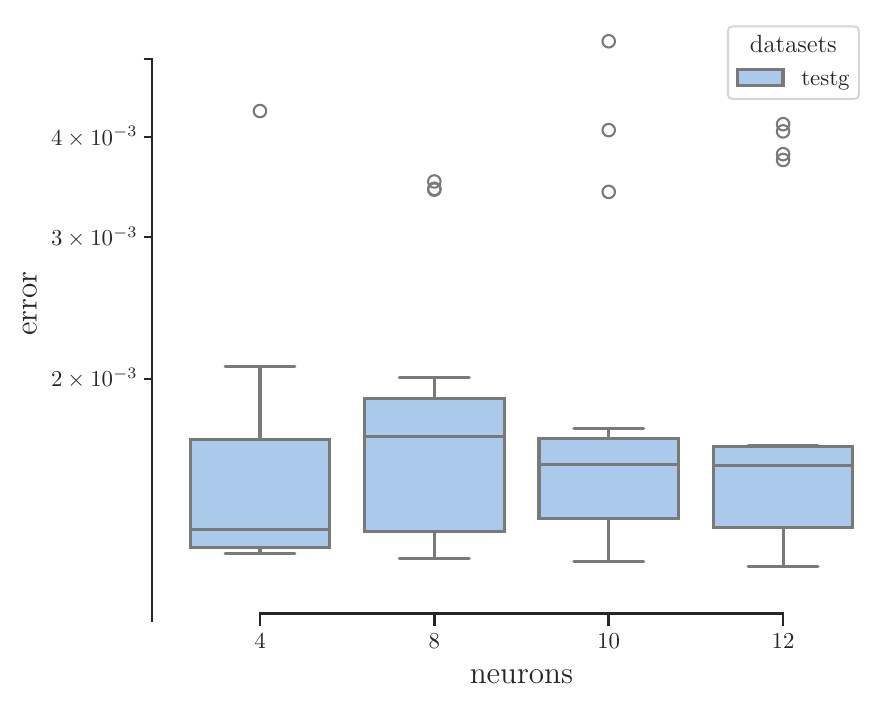}
    \caption{Number of neurons.}
    \label{fig:boxNeu}
    \end{subfigure}
    \hfill
    \begin{subfigure}[b]{0.43\textwidth}
    \centering
    \includegraphics[width=\textwidth]{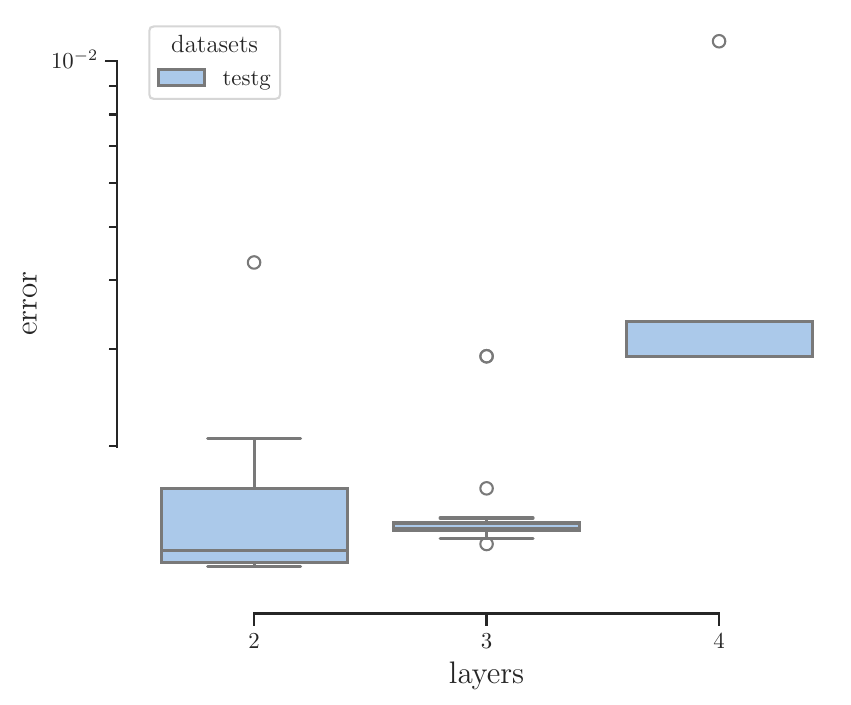}
    \caption{Number of layers.}
    \label{fig:boxLay}
    \end{subfigure}
    
    \vspace{0.5cm} 

    \begin{subfigure}[b]{0.43\textwidth}
    \centering
    \includegraphics[width=\textwidth]{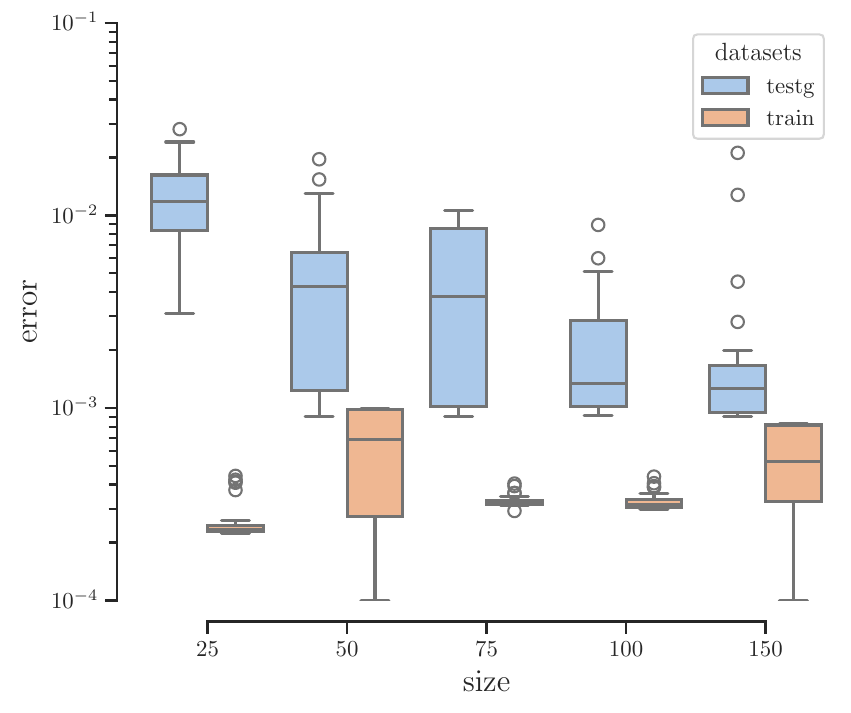}
    \caption{Number of training samples.}
    \label{fig:boxSize}
    \end{subfigure}
    \hfill
    \begin{subfigure}[b]{0.43\textwidth}
    \centering
    \includegraphics[width=\textwidth]{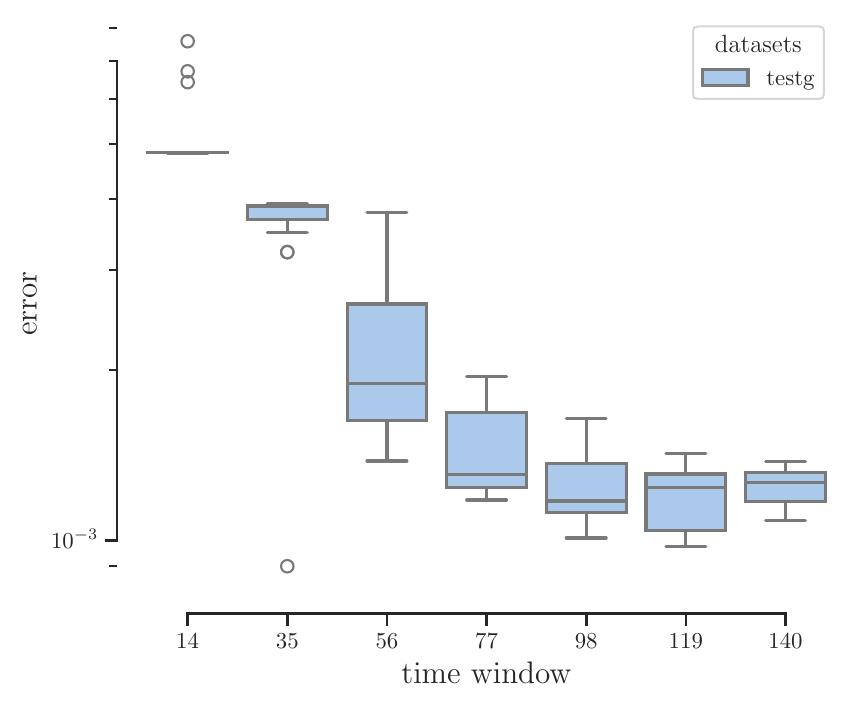}
    \caption{Observation window on test error. Time windows are measured in days.}
    \label{fig:boxTobs}
    \end{subfigure}
    \caption{Sensitivity analysis on hyperparameters. Blue boxplots correspond to the test error, orange ones to the training error.}
    \label{fig:sensAnal_all}
\end{figure}

We perform a sensitivity analysis with synthetic data in order to motivate our practical choices for the involved hyperparameters.
For this purpose we consider the following hyperparameters and evaluate their impact in terms of testing and training error over 20 different runs with random initializations of weights: (i) number of neurons of the neural network; (ii) number of layers of the neural network; (iii) number of training samples; (iv) observation window width (only on test error).

In Figure \ref{fig:boxNeu} we gathered the results in terms of amount of neurons with a two-layers neural network.
Following the \textit{Occam's razor principle}, reducing the number of neurons, for example to 4, leads to higher median accuracy by preventing overfitting of the training trajectories.
Similarly, increasing the model's nonlinearity by adding more layers does not contribute to a reduction in test error.
Indeed, in Figure \ref{fig:boxLay} the median error of the two-layers neural network with 4 neurons is lower with respect to the other considered cases.

Furthermore, increasing the training size enhances the possibilities of achieving higher test accuracy, since the neural network has more samples to learn from.
This behavior is evident from the median trends of test errors in Figure \ref{fig:boxSize}, although the training process becomes more time consuming (see Table \ref{tab:comp_tmes}).
\begin{figure}[t]
    \centering
    \includegraphics[width=0.8\linewidth]{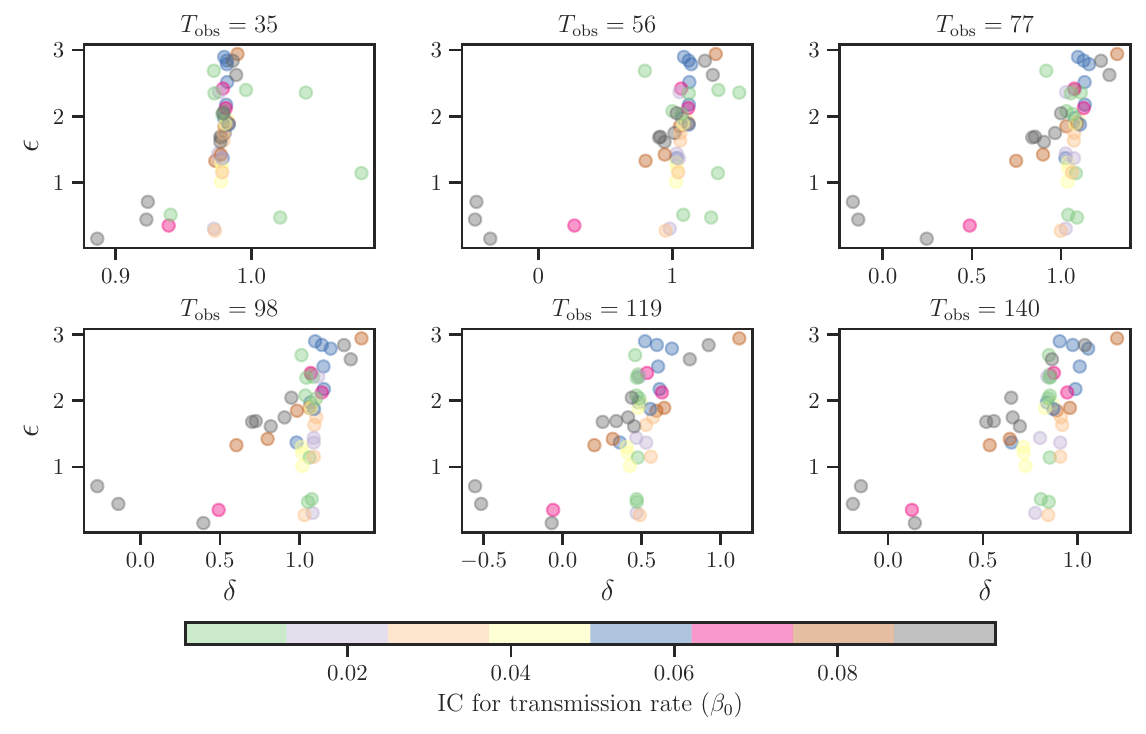}
    \caption{Lineage parameter versus the reconstructed latent parameters with observation windows. Different colors correspond to different value of the the initial condition (IC) for the transmission rate dynamics according with the colorbar.}
    \label{fig:delta_tobs}
\end{figure}
\begin{table}[t]
\tiny
    \centering
    \begin{tabular}{|c|c|c|c|}
    \hline
         Size &  Median training time [s] & Q01 training time [s] & Q09 training time [s] \\
         \hline 
         \hline
         25 & 629.7 & 579.8 & 724.3\\
         \hline
         50 & 780.7 & 536.4 & 802.8\\
         \hline
         75 & 887.5 & 820.6 & 900.6\\
         \hline
         100 & 957.3 & 946.2 & 973.3\\
         \hline
         150 & 1229 & 1179 & 1235\\
         \hline
    \end{tabular}
    \caption{Median over 20 runs of the computational times for training the architecture with different training sizes on a 8-core machine Intel i7.}
    \label{tab:comp_tmes}
\end{table}

At last, we consider the impact of the width of the observation window for estimating the couples of parameters when a new trajectory has to be simulated.
From Figure \ref{fig:boxTobs} we deduce that prediction accuracy suffers when windows which are not large enough.
On the other hand, after almost 11 weeks, corresponding to  of one-fourth of the total simulation time, the prediction error tends to saturate, thus motivating our choice for the synthetic scenario.
Similar conclusions can be drawn from Figure \ref{fig:delta_tobs}, where we can note that the learnt latent parameters, corresponding to similar values of initial transmission rates, identify one-dimensional curves starting from an observation time $T_{\mathrm{obs}} = 77$ days.
This behavior is not observed with shorter time windows.

\subsection{Test case 2: Realistic scenario dealing with influenza waves}
\label{subsec:realInflu}

\subsubsection{Data generation}
\label{subsec:real_dg}
\begin{figure}[t]
    \centering
    \includegraphics[width=0.4\textwidth]{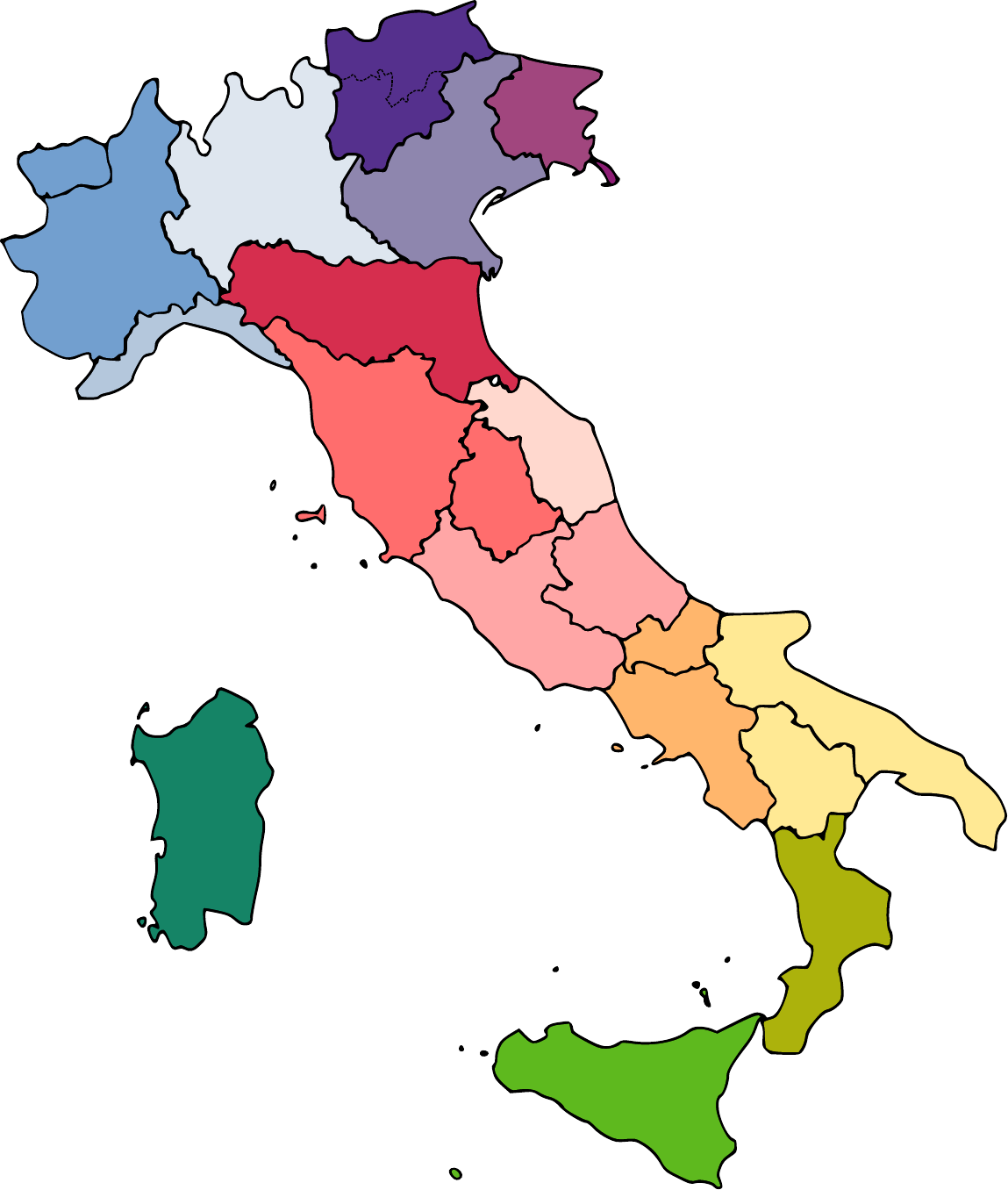}
    \caption{Repartition of Italy in 15 areas in order to retrieve weighted average national temperature and relative humidity.
    Indeed, only meteorological data from the largest cities were available.
    Different colors are associated with different areas in the averaged data by population.}
    \label{fig:italianRegions}
\end{figure}
Seasonal influenza is a contagious, infectious disease of the respiratory tract, which is estimated to cause tens of millions of infections and linked respiratory illnesses every year, as well as nearly 250.000-500.000 deaths worldwide \cite{world2003influenza}.
Early detection  and reliable predictions of disease evolution, when followed by a rapid response, can drastically reduce consequences of both seasonal and pandemic influenza, and encourage prevention measures such as vaccinations. 
We exploit the data reported by the Italian epidemiological and virological influenza surveillance system from 2010-2011 to 2019-2020, available in the influnet open access Github repository\footnote{\texttt{https://github.com/fbranda/influnet}}.
These data have been collected in the following way: using standardized forms, general practitioners and pediatricians are asked to report weekly influenza like illness cases occurring from week 42 to week 17 of successive years.
These data are delivered divided in four age groups (0-4, 5-14, 15-64, $>$64 years), together with age-specific data about influenza vaccine status.
Moreover, in order to surveil the circulating influenza virus strains, random swabs of the first influenza-like-illness (ILI) individuals have been analyzed by the regional Reference Laboratories in 15 different Italian regions. 
Every season almost 2000 samples have been collected, with a proportion of positive specimens of around 34\%.
In \cite{trentini2022characterizing} it was estimated the surveillance system average detects from 18.4\% to 29.3\% of actually infected by influenza.
Therefore, we divided total new cases by a mean factor $u_{\mathrm{inf}} = 0.23$, in order to be coherent with a \textit{SEIR} mathematical model which does not take into account for undetected infected.
Additionally, it was estimated that the average duration of the incubation period for influenza is 1.5 days ($\frac{1}{\alpha}$) \cite{lessler2009incubation} and the infectious period 1.2 days ($\frac{1}{\gamma}$) , in order to have a total generation time of 2.7 days, in accordance with existing literature \cite{wallinga2007generation}.
Concerning the initial conditions for the \textit{SEIR} model, we assume as in \cite{trentini2022characterizing} to prescribe a small seed corresponding to $10^{-6}$ initial percentage of infected individuals, and 
\begin{equation}
    E_0 = \dfrac{\overline{\Delta I}(1)}{7 \alpha},
\end{equation}
\textit{i.e.} considering a plausible value of initial exposed individuals in order to approximately generate in 7 days the initial amount of new cases. 
\begin{figure}[t]
    \centering
    \begin{subfigure}[b]{0.19\textwidth}
    \includegraphics[width=\textwidth]{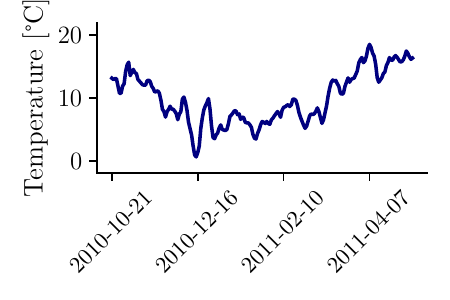}
    \end{subfigure}%
    \hfill
    \begin{subfigure}[b]{0.19\textwidth}
    \includegraphics[width=\textwidth]{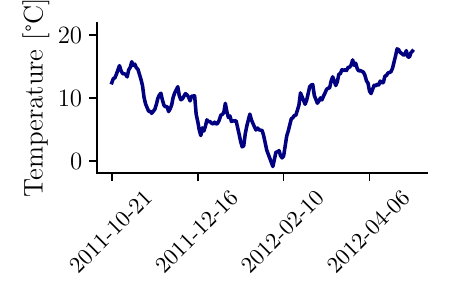}
    \end{subfigure}%
    \hfill
        \begin{subfigure}[b]{0.19\textwidth}
    \includegraphics[width=\textwidth]{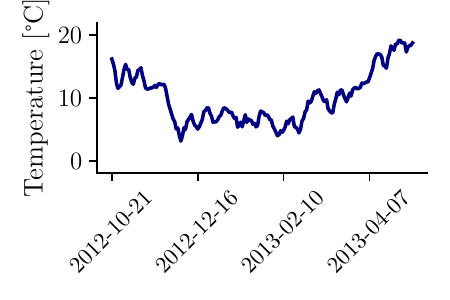}
    \end{subfigure}%
    \hfill
        \begin{subfigure}[b]{0.19\textwidth}
    \includegraphics[width=\textwidth]{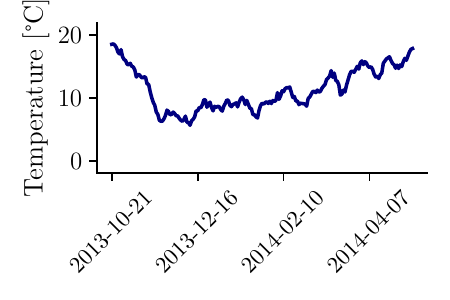}
    \end{subfigure}%
    \hfill
        \begin{subfigure}[b]{0.19\textwidth}
    \includegraphics[width=\textwidth]{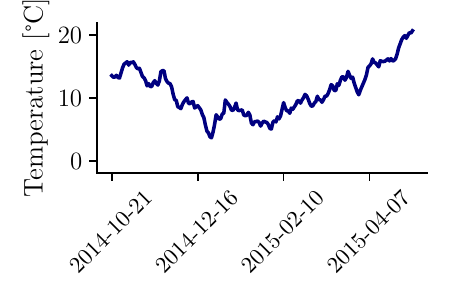}
    \end{subfigure}%
    \vspace{0.5cm} 
        \begin{subfigure}[b]{0.19\textwidth}
    \includegraphics[width=\textwidth]{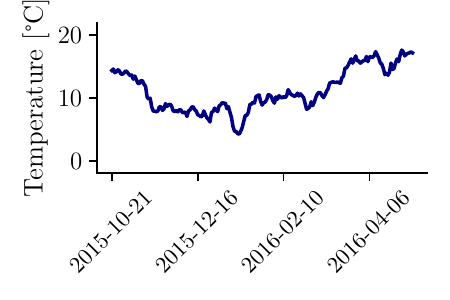}
    \end{subfigure}%
    \hfill
    \begin{subfigure}[b]{0.19\textwidth}
    \includegraphics[width=\textwidth]{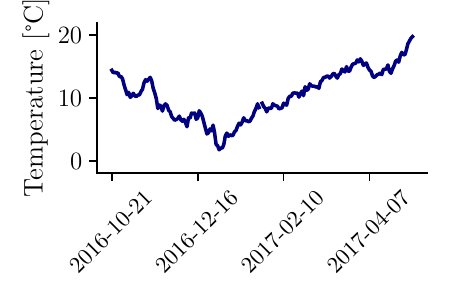}
    \end{subfigure}%
    \hfill
        \begin{subfigure}[b]{0.19\textwidth}
    \includegraphics[width=\textwidth]{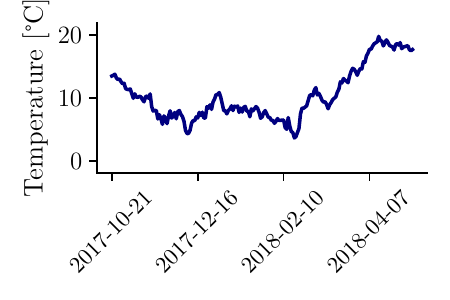}
    \end{subfigure}%
    \hfill
        \begin{subfigure}[b]{0.19\textwidth}
    \includegraphics[width=\textwidth]{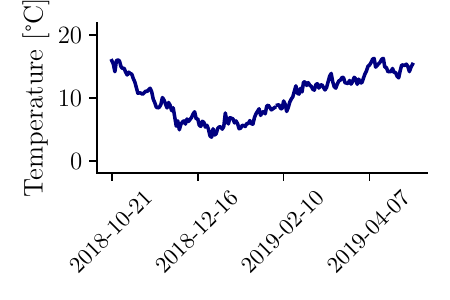}
    \end{subfigure}%
    \hfill
        \begin{subfigure}[b]{0.19\textwidth}
    \includegraphics[width=\textwidth]{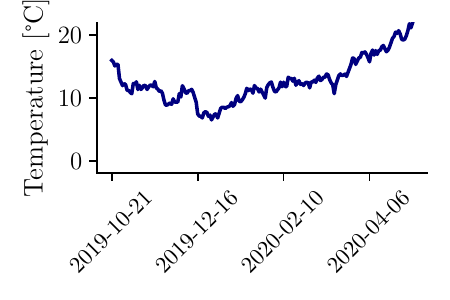}
    \end{subfigure}%
    \caption{Aggregated national data of temperature.}
    \label{fig:natTemp}

\end{figure}

\begin{figure}[t]
    \centering
    \begin{subfigure}[b]{0.19\textwidth}
    \includegraphics[width=\textwidth]{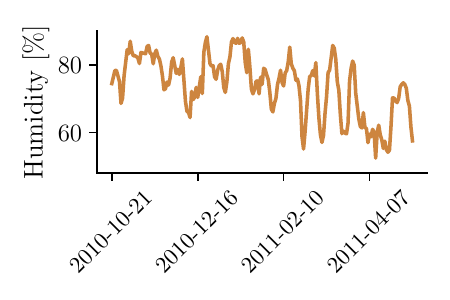}
    \end{subfigure}%
    \hfill
    \begin{subfigure}[b]{0.19\textwidth}
    \includegraphics[width=\textwidth]{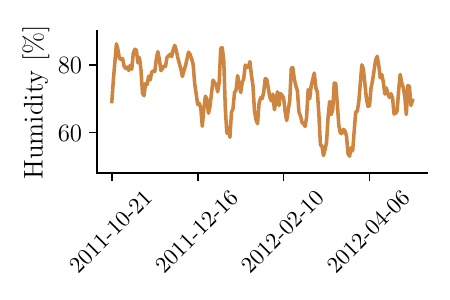}
    \end{subfigure}%
    \hfill
        \begin{subfigure}[b]{0.19\textwidth}
    \includegraphics[width=\textwidth]{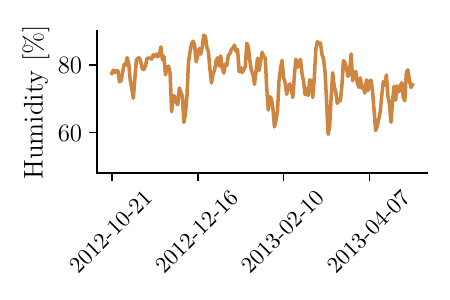}
    \end{subfigure}%
    \hfill
        \begin{subfigure}[b]{0.19\textwidth}
    \includegraphics[width=\textwidth]{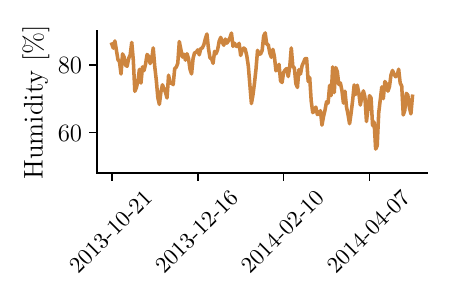}
    \end{subfigure}%
    \hfill
        \begin{subfigure}[b]{0.19\textwidth}
    \includegraphics[width=\textwidth]{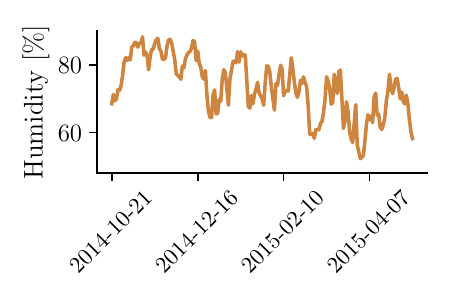}
    \end{subfigure}%
    \vspace{0.5cm} 
        \begin{subfigure}[b]{0.19\textwidth}
    \includegraphics[width=\textwidth]{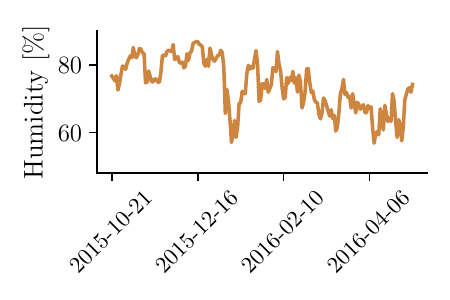}
    \end{subfigure}%
    \hfill
    \begin{subfigure}[b]{0.19\textwidth}
    \includegraphics[width=\textwidth]{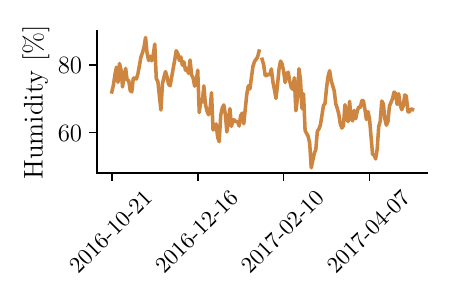}
    \end{subfigure}%
    \hfill
        \begin{subfigure}[b]{0.19\textwidth}
    \includegraphics[width=\textwidth]{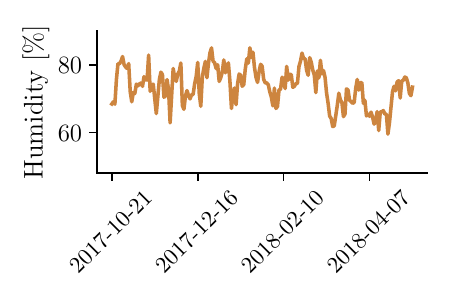}
    \end{subfigure}%
    \hfill
        \begin{subfigure}[b]{0.19\textwidth}
    \includegraphics[width=\textwidth]{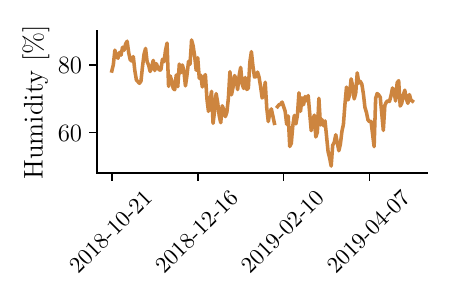}
    \end{subfigure}%
    \hfill
        \begin{subfigure}[b]{0.19\textwidth}
    \includegraphics[width=\textwidth]{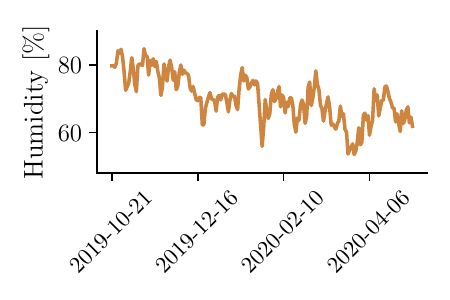}
    \end{subfigure}%
    \caption{Aggregated national data of relative humidity.}
            \label{fig:natUmid}

\end{figure}
The Influnet dataset provides extensive information on influenza epidemics, including seroepidemiological data that assess the population's immunity levels following the 2009 pandemic wave of the A/H1N1pdm09 strain, divided by age groups.
However, for the purposes of this study, we will disregard age-dependent factors and adopt the standard homogeneous \textit{SEIR} model in \eqref{eq:SEIRsys}, using only meteorological data as exogenous variables.
In particular, we took into account temperature and relative humidity as meteorological information, following the analysis of \cite{lowen2007influenza, mubayi2021analytical}, which studied the coupled interplay between relative humidity and temperature in the spreading of infectious diseases.
These data are available only for some cities in different administrative regions.
Therefore, in order to obtain the national data, they have been averaged after their extraction from a popular Italian meteorological website\footnote{\texttt{https://www.ilmeteo.it/portale/archivio-meteo}} as detailed below.

We dispose of meteorological data coming from 15 different cities belonging to different administrative Italian regions.
The remaining 5 regions (Abruzzo, Basilicata, Molise, Valle d'Aosta, Umbria), where meteorological data were not available for more than four of the ten considered years, have been paired with neighboring regions of similar latitude (cf. Figure \ref{fig:italianRegions}).
Then, the national averages for temperature and humidity were then calculated using a weighted mean, where each area's weight was determined by its relative population size.

\begin{figure}[t]
    \begin{subfigure}[b]{0.48\textwidth}
    \centering
    \includegraphics[width=\textwidth]{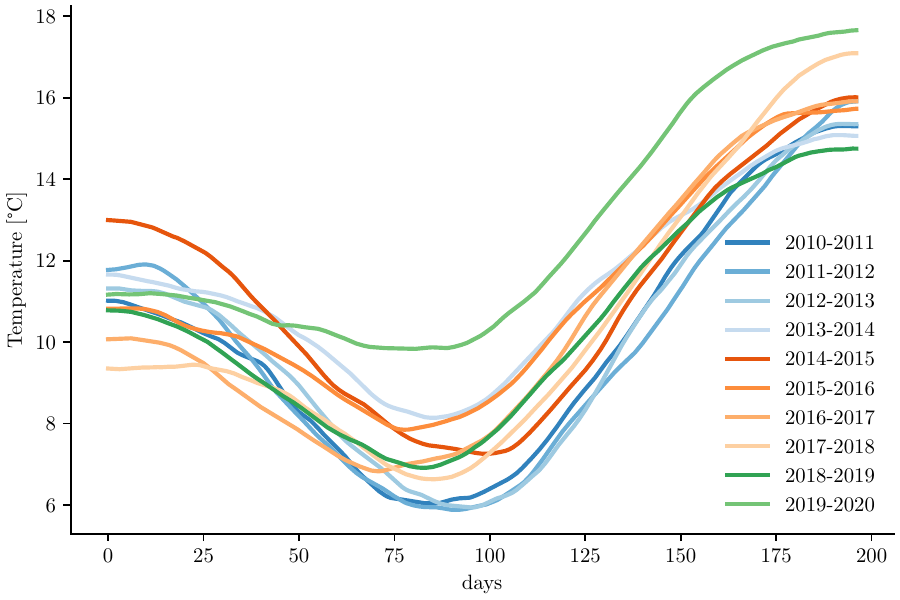}
    \end{subfigure}
    \hfill
    \begin{subfigure}[b]{0.48\textwidth}
    \centering
    \includegraphics[width=\textwidth]{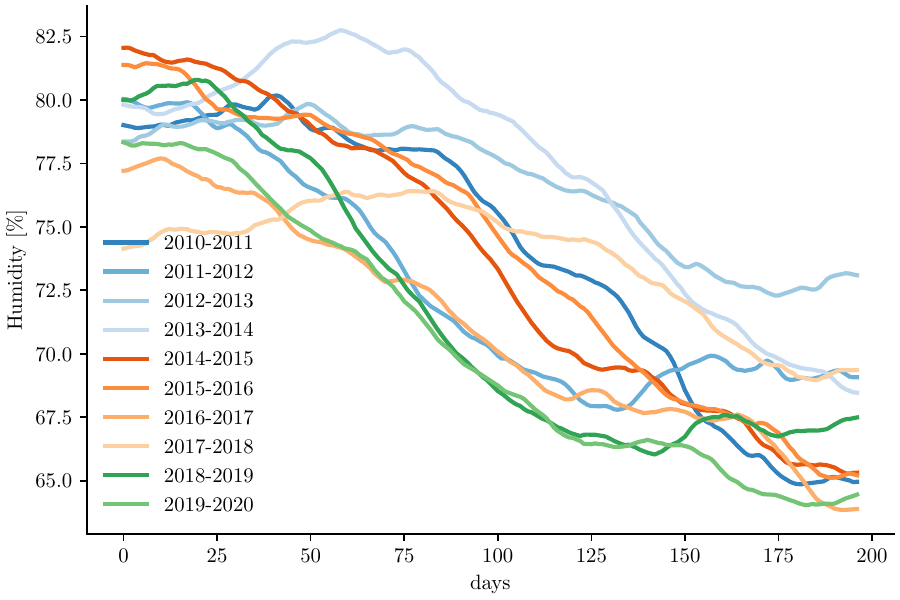}
    \end{subfigure}
    \caption{Input  time series of national temperature and relative humidity after applying the Savitzky-Golay filter.}
    \label{fig:temp_um_sg}
\end{figure}

Furthermore, the average data are quite noisy, as illustrated in Figures \ref{fig:natTemp} and \ref{fig:natUmid}.
In order to filter out higher frequencies, we apply to both temperature and relative humidity time series a Savitzky-Golay filter of order 2 \cite{krishnan2012selection}.
In this way we retrieve in each point a smoother least square approximation of these data which is more prone for training and testing the proposed approach (see Figure \ref{fig:temp_um_sg}).

Unlike other pandemic events over the past five centuries that occurred unexpectedly, influenza exhibits a certain cyclicality attributed to various genetic mechanisms responsible for triggering new outbreak waves.
The causative agent of influenza is not seen as a single, stable strain persisting over the years, but rather as a heterogeneous collection of viral evolutionary events that, while highly similar, are not identical \cite{morens2011pandemic}.
This justifies the need to estimate the parameter $\delta$ during training and testing stages.
Differences in transmissibility, mortality rates, and severity can be identified for previous influenza waves, largely due to genetic heterogeneity, which has been extensively studied \cite{biggerstaff2014estimates}.
Influenza circulating in Italy during the last decade is mainly constituted by three strains: influenza A/H1N1pdm09, influenza B and influenza A/H3N2.
The reproduction number for A/H1N1pdm09 is similar to that of seasonal influenza, even though this virus is highly transmissible, leading to a global pandemic in 2009.
Its spread has been largely boosted by lacking in pre-existing immunity in the population.
The A/H3N2 subtype often dominates during epidemic seasons, causing more frequent and intense outbreaks, in particular in older adults.
It has undergone significant antigenic drifts over the years, leading to more severe symptoms with respect to those of seasonal flu.
Instead, influenza B is generally less transmissible with respect to A wildtypes, affecting predominantly children and young adults.
However, it can lead to severe outcomes in older adults, especially in presence of comorbidities.

We aim at catching differences across waves of different seasons with the use of the latent parameter $\delta$, which will be confronted with the average transmissibility of the virus' lineages circulating across the years as studied by \cite{trentini2022characterizing}.
From this study inferring on effective reproduction numbers of the different strains in each season, estimated from the serological analyses of ILI cases in Lombardy during each year for determining strain-compositions, we derive a mean effective reproduction number that we compare with the reconstructed $\delta$ (see Figure \ref{fig:deltas_rec}).

We explicitly incorporate temperature and humidity into the model for the transmission rate, while other exogenous factors are implicitly included in factor $\delta$.

\subsubsection{Hyperparameters setup}
\label{subsec:realHSet}
\begin{table}[t]
\tiny
\centering
\begin{tabular}{|l|l|l|l|l|l|l|l|l|l|}
\hline
& Epochs & $\alpha_{\mathcal{A}}$ & $\alpha_{\mathcal{B}}$ & $\alpha_{\mathcal{D}}$ & $\alpha_{\mathcal{E},1}$ & $\alpha_{\mathcal{E},2}$ & $\alpha_{\mathcal{I}}$ & $\alpha_{\mathrm{reg}}$ & $\alpha_{\beta}$\\ \hline \hline
Training ($\mathcal{E}_1/\mathcal{E}_2$) Adam   &  500/1000 & $5 \cdot 10^{-3}$ & $10^{-7}$ & $10^{-7}$& $1.1$ & $1.1 \cdot 10^{2}$ & $10^{-6}$ & $10^{-7}$ & 0\\  \hline
Training BFGS  ($\mathcal{E}_1/\mathcal{E}_2$) &  5000/3000 & / & / & / & / & / & / & / & /\\  \hline
Estimation Adam &  8000 &  6.2 & $10^{-7}$& 0 & $8.0$ & 0& 0& 0& $5\cdot 10^{-1}$ \\  \hline          
Estimation BFGS &  1000 & /&/ &/ & /& /&/  & / &  / \\  \hline        
\end{tabular}
\caption{Table of the hyperparameters of each minimization stage (Test case 2).}
\label{tab:tabHyperParamReal}
\end{table}
Following the results of the sensitivity analysis in \textit{Test case 1}, we adopted the same 4-neurons two-layers neural network architecture.
We fix the learning rates at $\eta_{\mathrm{train}, \mathcal{E}_1} = 10^{-3}$, $\eta_{\mathrm{train}, \mathcal{E}_2} =10^{-7}$ and $\eta_{\mathrm{testg}} = 10^{-3}$.
Table \ref{tab:tabHyperParamReal} summarizes all the values of the other hyperparameters involved.
The observation window lasts 49 days, corresponding to 7 weeks, which is almost one-fourth of the total simulation time of 28 weeks or 196 days.
The choice of this width is also based on the sensitivity analysis performed on the synthetic test case.
Also the choice of the width is based on the sensitivity analysis performed in \textit{Test case 1}.
\subsubsection{Results}
\begin{figure}[t]
    \begin{subfigure}[b]{0.48\textwidth}
    \centering
    \includegraphics[width=\textwidth]{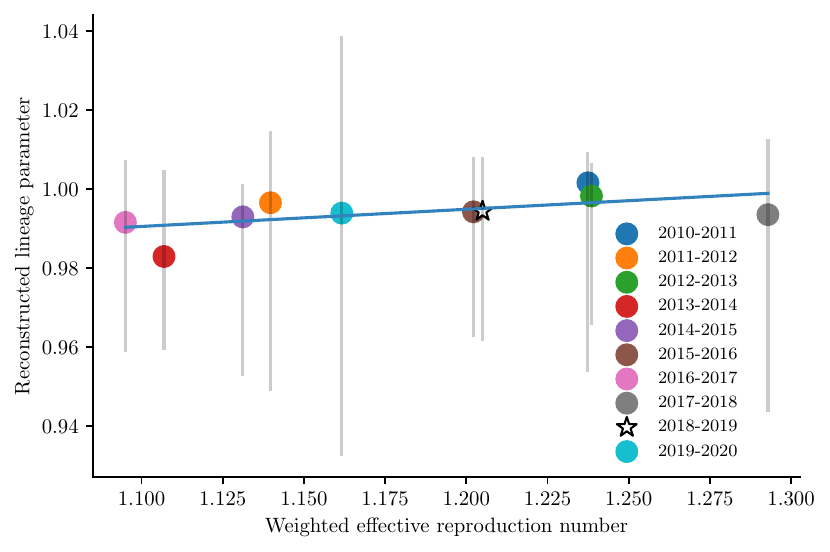}
    \caption{Case 1.}
    \end{subfigure}
    \hfill
    \begin{subfigure}[b]{0.48\textwidth}
    \centering
    \includegraphics[width=\textwidth]{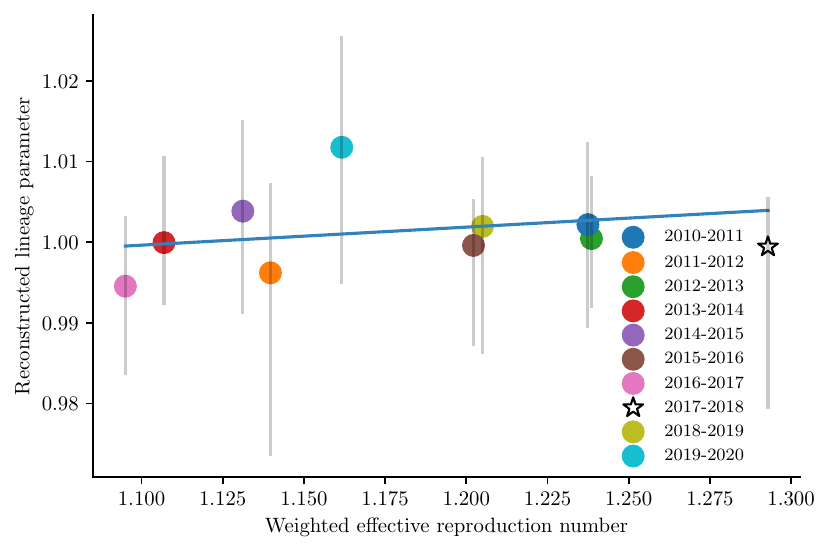}
    \caption{Case 2.}
    \end{subfigure}
    \vspace{0.5cm} 
    \begin{subfigure}[b]{\textwidth}
    \centering
    \includegraphics[width=0.48\textwidth]{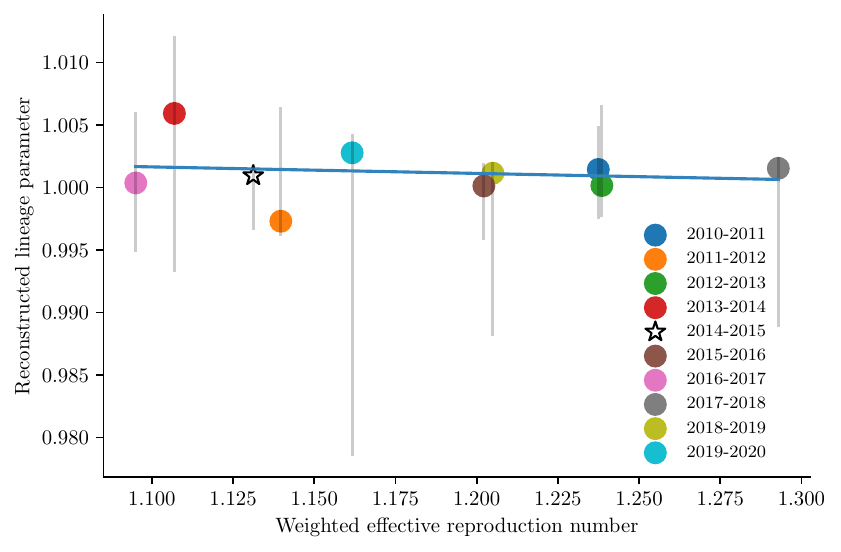}
    \caption{Case 3.}
    \end{subfigure}
    \caption{Reconstructed latent parameters ($\delta$) versus the reproductive effective number of influenza waves weighted by lineage composition retrieved starting from \cite{trentini2022characterizing}. Star-shaped point correspond to the testing sample. The blue line corresponds to the linear regression of training median values. Each gray bar corresponds to 0.1-0.9 percentile values of the respective reconstructed parameter (considering various random initializations).}
    \label{fig:deltas_rec}
\end{figure}
Among the 10 available influenza waves, we randomly select 9 samples for the training set and 1 for the testing set (leave-one-out approach); then, we run 20 different simulations with the same network topology described in Subsection \ref{subsec:realHSet}, with different initializations of weights and biases.
We present three representative cases of different scenarios by considering varying training and testing data:
\begin{itemize}
    \item Case 1: testing wave corresponds to 2018-2019;
    \item Case 2: testing wave corresponds to 2017-2018;
    \item Case 3: testing wave corresponds to 2014-2015.
\end{itemize}
In Figure \ref{fig:deltas_rec} we collect the mean retrieved values for the median latent parameter $\delta$ for each influenza wave in the three cases, together with 0.1-0.9 percentile bands.
Those quantities have been plotted against the weighted effective reproduction number for the different influenza waves, which depends on the dominant variants' composition as obtained in \cite{trentini2022characterizing}.
In all three cases, the linear correlation coefficients are, in absolute value, less than 0.5, indicating that the recovered unknown latent parameter cannot be linked to the weighted effective reproduction number through a linear transformation.
We conclude that the evolution of transmission rate cannot be solely attributed to the intrinsic reproducibility of the current strain; rather, other exogenous factors (whose effects are embedded in the latent parameters) significantly influence each single wave.
However, since the median values of $\delta$ for the same wave remain similar across the three different cases, training is robust when using the leave-one-out approach.
The clustering of $\delta$ values near 1 is also associated with a relatively high weight of the corresponding regularization term in the training and testing loss.

\begin{figure}[H]
    \centering
    \begin{subfigure}[b]{0.19\textwidth}
    \includegraphics[width=\textwidth]{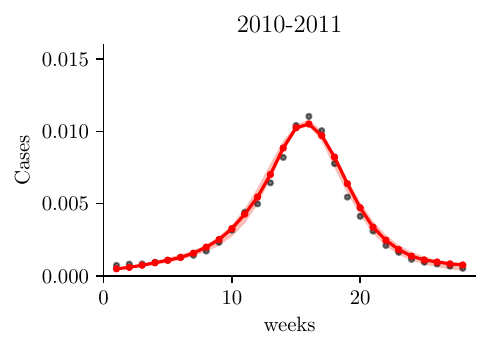}
    \end{subfigure}%
    \hfill
    \begin{subfigure}[b]{0.19\textwidth}
    \includegraphics[width=\textwidth]{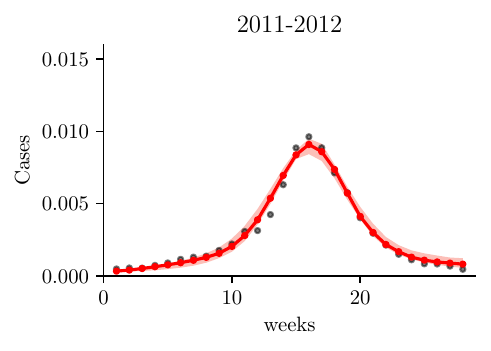}
    \end{subfigure}%
    \hfill
        \begin{subfigure}[b]{0.19\textwidth}
    \includegraphics[width=\textwidth]{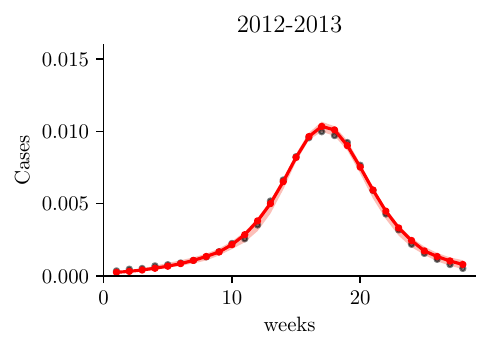}
    \end{subfigure}%
    \hfill
        \begin{subfigure}[b]{0.19\textwidth}
    \includegraphics[width=\textwidth]{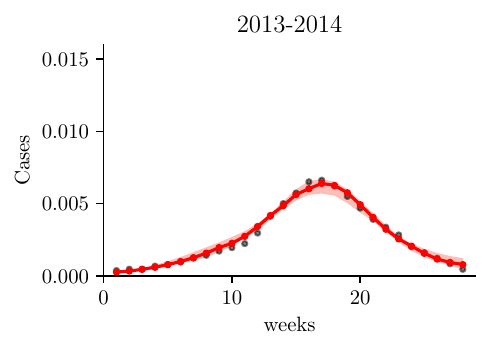}
    \end{subfigure}%
    \hfill
        \begin{subfigure}[b]{0.19\textwidth}
    \includegraphics[width=\textwidth]{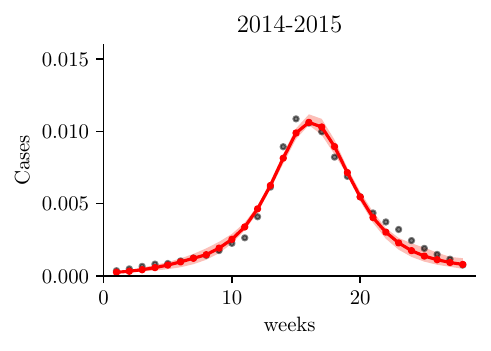}
    \end{subfigure}%
    \vspace{0.5cm} 
        \begin{subfigure}[b]{0.19\textwidth}
    \includegraphics[width=\textwidth]{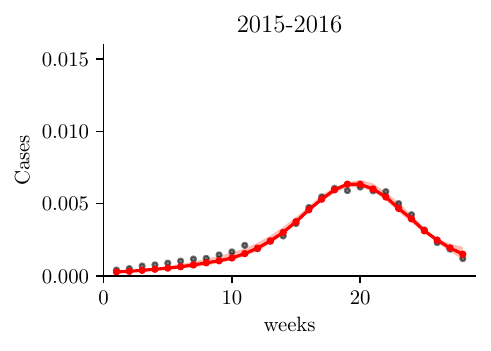}
    \end{subfigure}%
    \hfill
    \begin{subfigure}[b]{0.19\textwidth}
    \includegraphics[width=\textwidth]{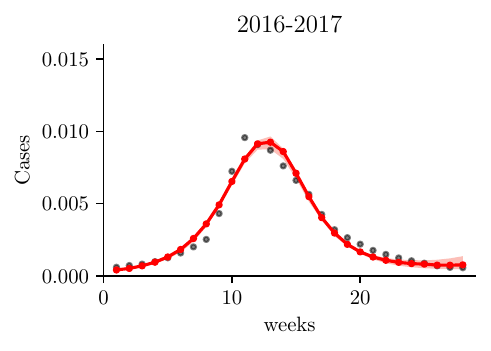}
    \end{subfigure}%
    \hfill
        \begin{subfigure}[b]{0.19\textwidth}
    \includegraphics[width=\textwidth]{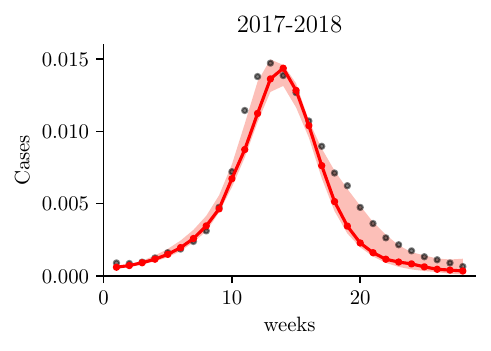}
    \end{subfigure}%
    \hfill
        \begin{subfigure}[b]{0.19\textwidth}
    \includegraphics[width=\textwidth]{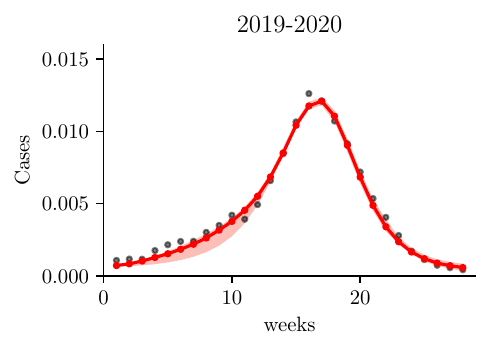}
    \end{subfigure}%
    \hfill
        \begin{subfigure}[b]{0.19\textwidth}
    \includegraphics[width=\textwidth]{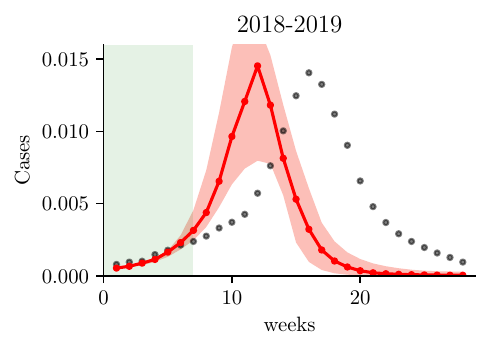}
    \end{subfigure}%

    \vspace{0.5cm} 

    \begin{subfigure}[b]{0.19\textwidth}
    \includegraphics[width=\textwidth]{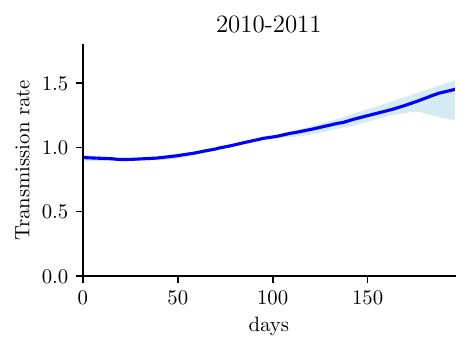}
    \end{subfigure}%
    \hfill
    \begin{subfigure}[b]{0.19\textwidth}
    \includegraphics[width=\textwidth]{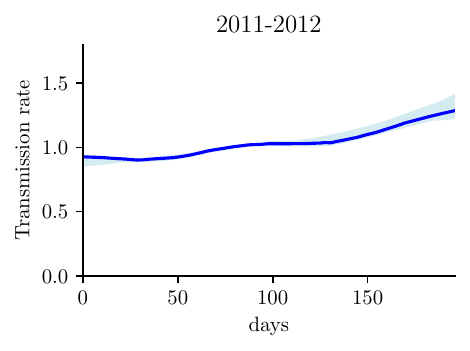}
    \end{subfigure}%
    \hfill
        \begin{subfigure}[b]{0.19\textwidth}
    \includegraphics[width=\textwidth]{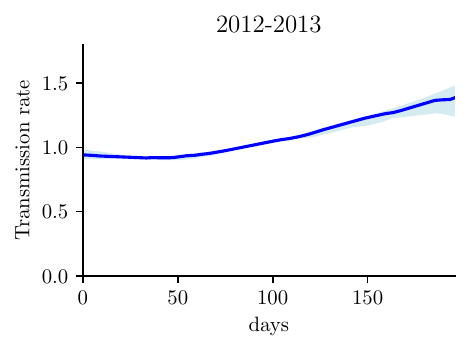}
    \end{subfigure}%
    \hfill
        \begin{subfigure}[b]{0.19\textwidth}
    \includegraphics[width=\textwidth]{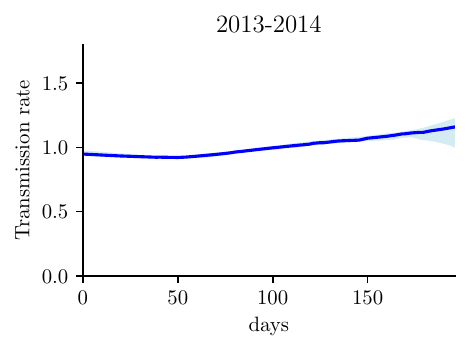}
    \end{subfigure}%
    \hfill
        \begin{subfigure}[b]{0.19\textwidth}
    \includegraphics[width=\textwidth]{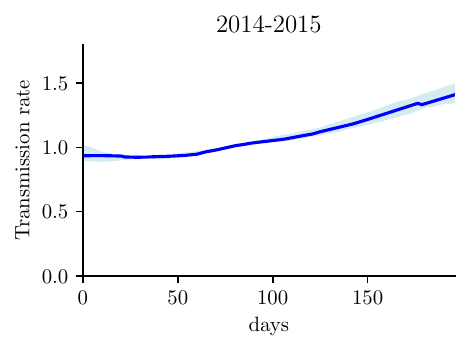}
    \end{subfigure}%
    \vspace{0.5cm} 
        \begin{subfigure}[b]{0.19\textwidth}
    \includegraphics[width=\textwidth]{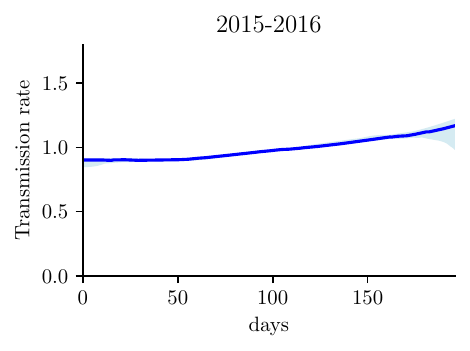}
    \end{subfigure}%
    \hfill
    \begin{subfigure}[b]{0.19\textwidth}
    \includegraphics[width=\textwidth]{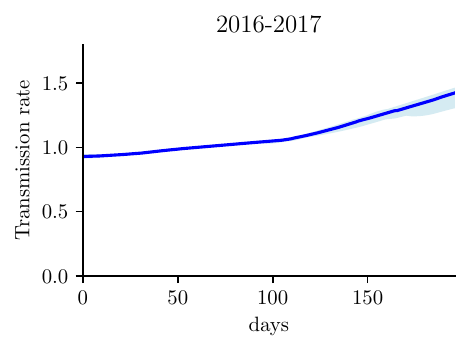}
    \end{subfigure}%
    \hfill
        \begin{subfigure}[b]{0.19\textwidth}
    \includegraphics[width=\textwidth]{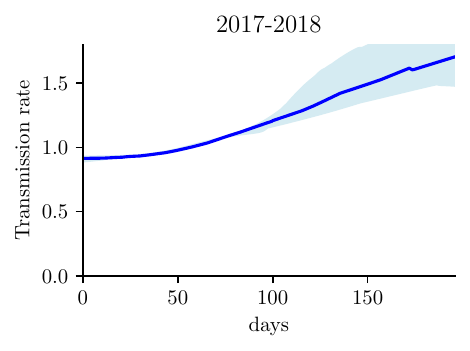}
    \end{subfigure}%
    \hfill
        \begin{subfigure}[b]{0.19\textwidth}
    \includegraphics[width=\textwidth]{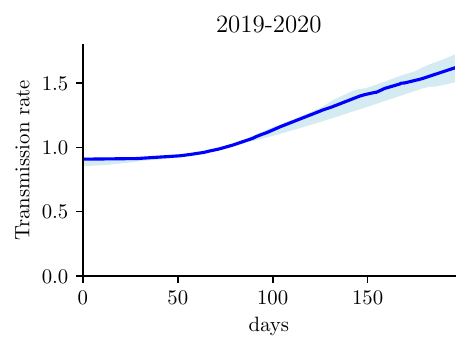}
    \end{subfigure}%
    \hfill
        \begin{subfigure}[b]{0.19\textwidth}
    \includegraphics[width=\textwidth]{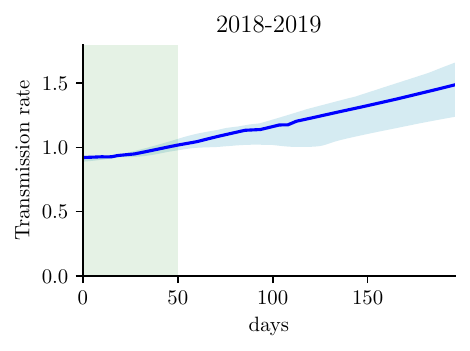}
    \end{subfigure}%

    \vspace{0.5cm} 

    \begin{subfigure}[b]{0.19\textwidth}
    \includegraphics[width=\textwidth]{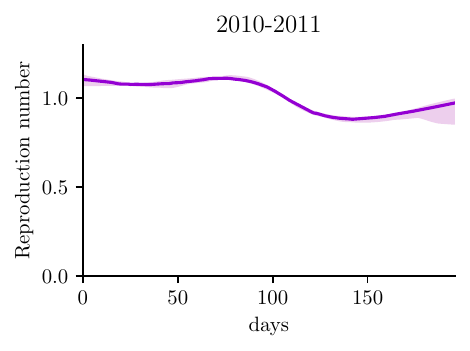}
    \end{subfigure}%
    \hfill
    \begin{subfigure}[b]{0.19\textwidth}
    \includegraphics[width=\textwidth]{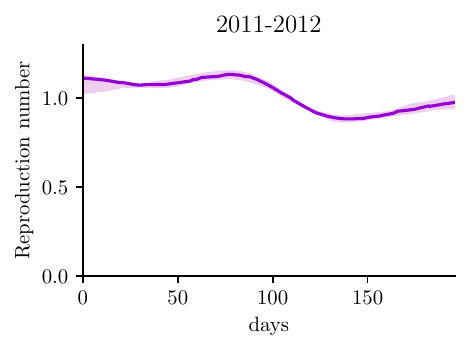}
    \end{subfigure}%
    \hfill
        \begin{subfigure}[b]{0.19\textwidth}
    \includegraphics[width=\textwidth]{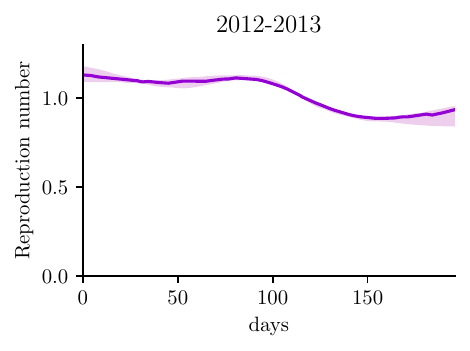}
    \end{subfigure}%
    \hfill
        \begin{subfigure}[b]{0.19\textwidth}
    \includegraphics[width=\textwidth]{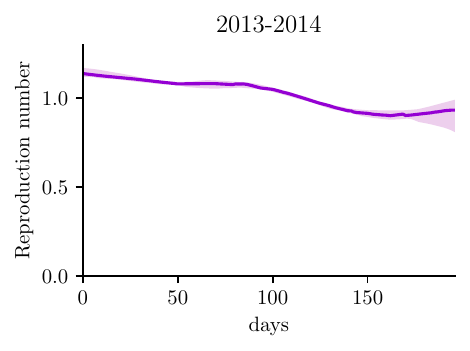}
    \end{subfigure}%
    \hfill
        \begin{subfigure}[b]{0.19\textwidth}
    \includegraphics[width=\textwidth]{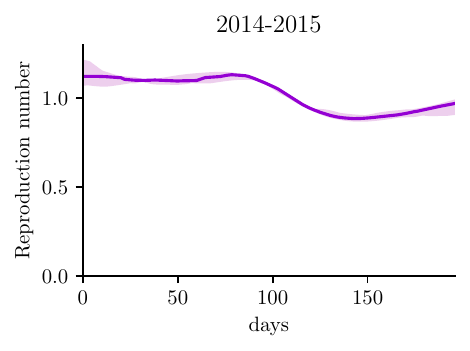}
    \end{subfigure}%
    \vspace{0.5cm} 
        \begin{subfigure}[b]{0.19\textwidth}
    \includegraphics[width=\textwidth]{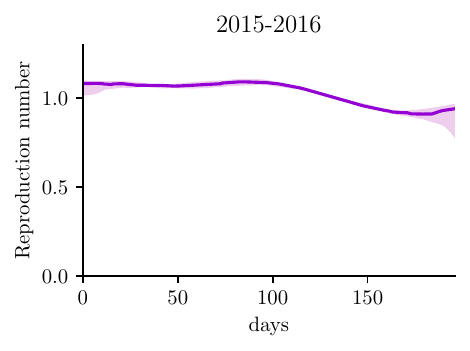}
    \end{subfigure}%
    \hfill
    \begin{subfigure}[b]{0.19\textwidth}
    \includegraphics[width=\textwidth]{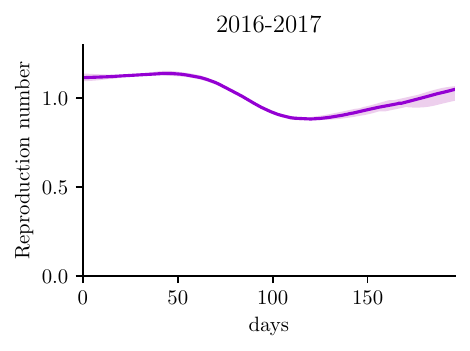}
    \end{subfigure}%
    \hfill
        \begin{subfigure}[b]{0.19\textwidth}
    \includegraphics[width=\textwidth]{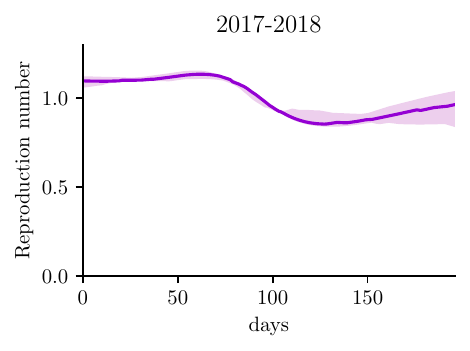}
    \end{subfigure}%
    \hfill
        \begin{subfigure}[b]{0.19\textwidth}
    \includegraphics[width=\textwidth]{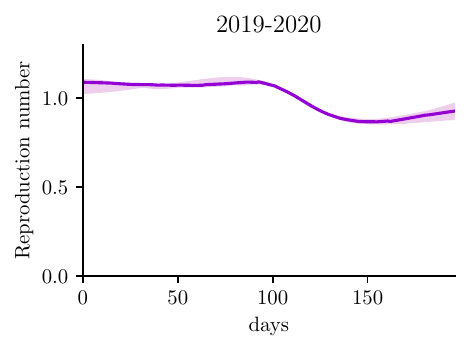}
    \end{subfigure}%
    \hfill
        \begin{subfigure}[b]{0.19\textwidth}
    \includegraphics[width=\textwidth]{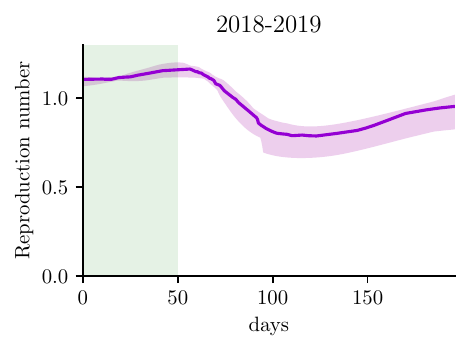}
    \end{subfigure}%
    \caption{\textbf{Case 1}. Reconstructions of cases, transmission rate, reproduction number exploiting influenza data. The training set is constituted by all epidemic waves from 2010-2011 to 2019-2020 except for 2018-2019, which belongs to the test set. The green window corresponds to the 49 days of observability in which we estimate the latent parameter for the testing sample.}
    \label{fig:infCase1}
\end{figure}

\begin{figure}[H]
    \centering
    \begin{subfigure}[b]{0.19\textwidth}
    \includegraphics[width=\textwidth]{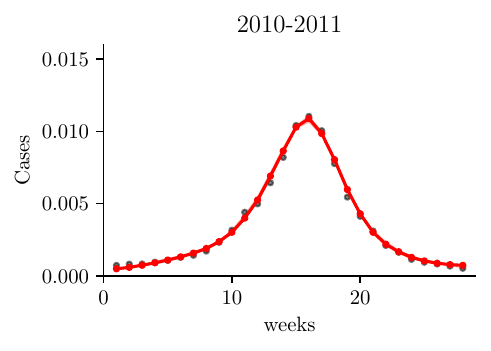}
    \end{subfigure}%
    \hfill
    \begin{subfigure}[b]{0.19\textwidth}
    \includegraphics[width=\textwidth]{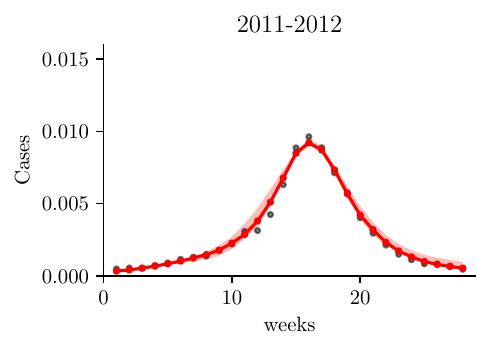}
    \end{subfigure}%
    \hfill
        \begin{subfigure}[b]{0.19\textwidth}
    \includegraphics[width=\textwidth]{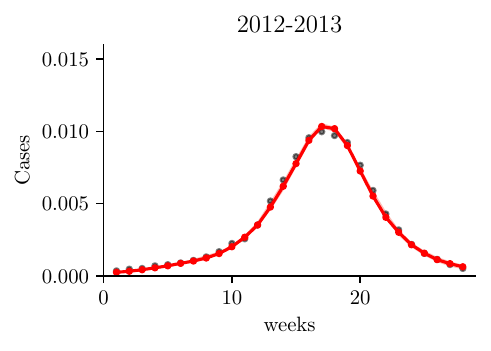}
    \end{subfigure}%
    \hfill
        \begin{subfigure}[b]{0.19\textwidth}
    \includegraphics[width=\textwidth]{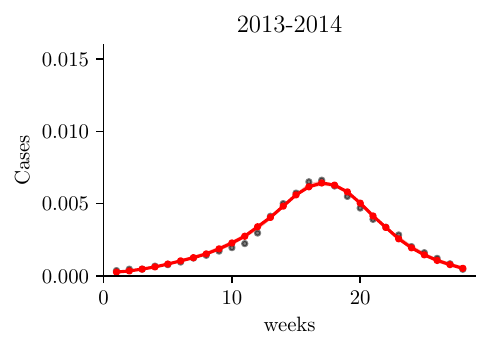}
    \end{subfigure}%
    \hfill
        \begin{subfigure}[b]{0.19\textwidth}
    \includegraphics[width=\textwidth]{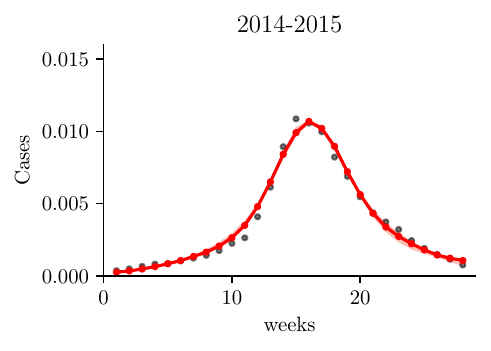}
    \end{subfigure}%
    \vspace{0.5cm} 
        \begin{subfigure}[b]{0.19\textwidth}
    \includegraphics[width=\textwidth]{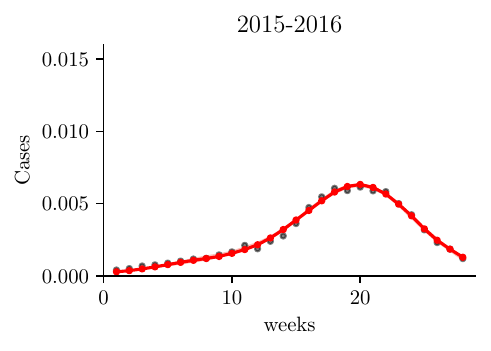}
    \end{subfigure}%
    \hfill
    \begin{subfigure}[b]{0.19\textwidth}
    \includegraphics[width=\textwidth]{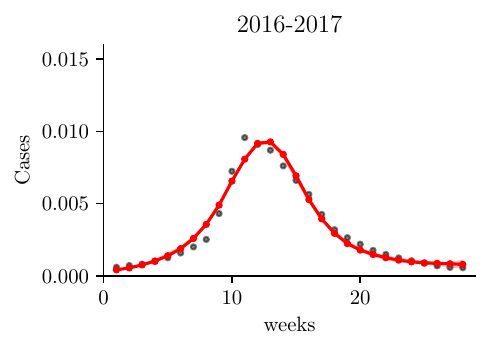}
    \end{subfigure}%
    \hfill
        \begin{subfigure}[b]{0.19\textwidth}
    \includegraphics[width=\textwidth]{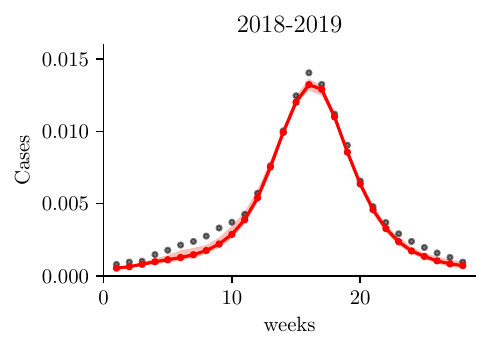}
    \end{subfigure}%
    \hfill
        \begin{subfigure}[b]{0.19\textwidth}
    \includegraphics[width=\textwidth]{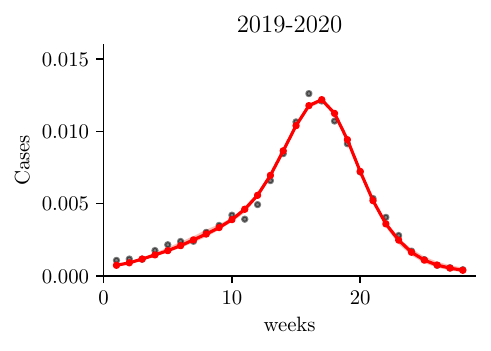}
    \end{subfigure}%
    \hfill
        \begin{subfigure}[b]{0.19\textwidth}
    \includegraphics[width=\textwidth]{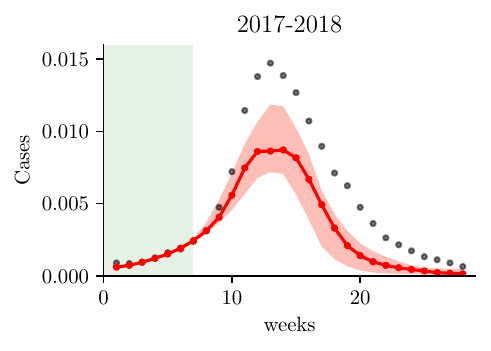}
    \end{subfigure}%

    \vspace{0.5cm} 

    \begin{subfigure}[b]{0.19\textwidth}
    \includegraphics[width=\textwidth]{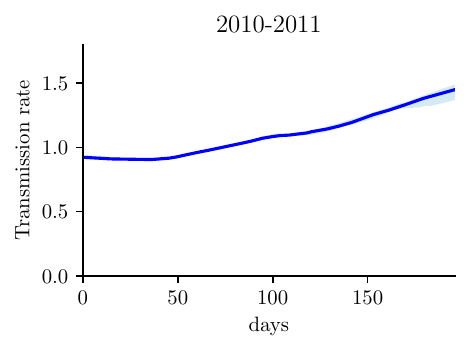}
    \end{subfigure}%
    \hfill
    \begin{subfigure}[b]{0.19\textwidth}
    \includegraphics[width=\textwidth]{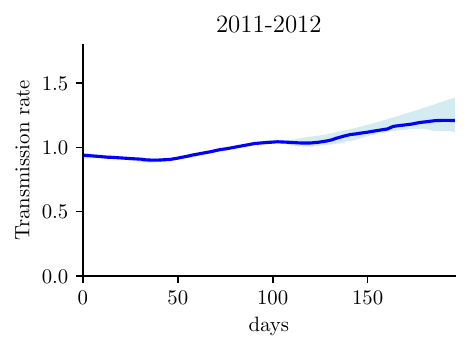}
    \end{subfigure}%
    \hfill
        \begin{subfigure}[b]{0.19\textwidth}
    \includegraphics[width=\textwidth]{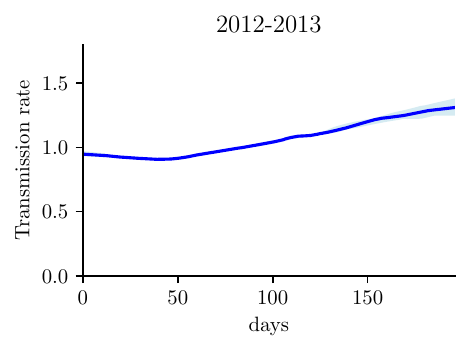}
    \end{subfigure}%
    \hfill
        \begin{subfigure}[b]{0.19\textwidth}
    \includegraphics[width=\textwidth]{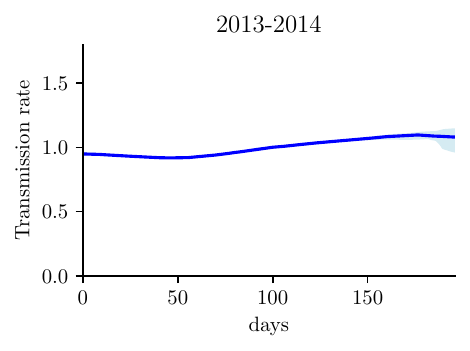}
    \end{subfigure}%
    \hfill
        \begin{subfigure}[b]{0.19\textwidth}
    \includegraphics[width=\textwidth]{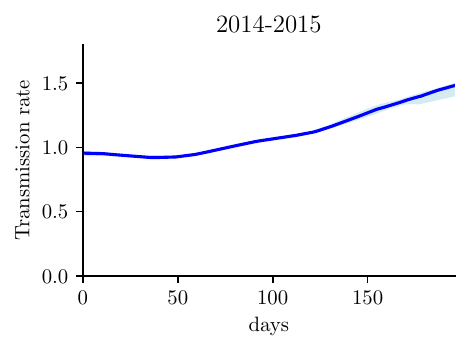}
    \end{subfigure}%
    \vspace{0.5cm} 
        \begin{subfigure}[b]{0.19\textwidth}
    \includegraphics[width=\textwidth]{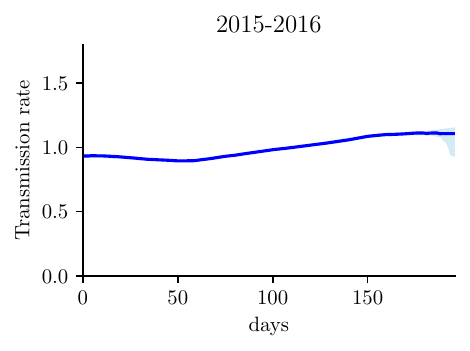}
    \end{subfigure}%
    \hfill
    \begin{subfigure}[b]{0.19\textwidth}
    \includegraphics[width=\textwidth]{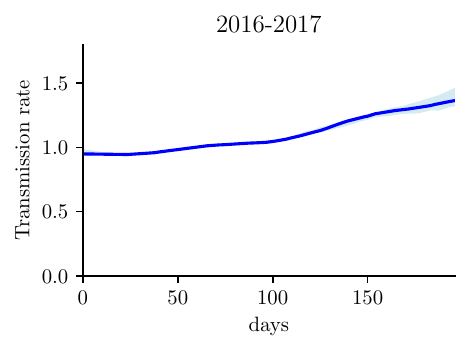}
    \end{subfigure}%
    \hfill
        \begin{subfigure}[b]{0.19\textwidth}
    \includegraphics[width=\textwidth]{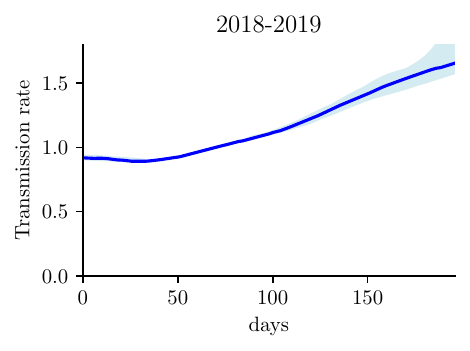}
    \end{subfigure}%
    \hfill
        \begin{subfigure}[b]{0.19\textwidth}
    \includegraphics[width=\textwidth]{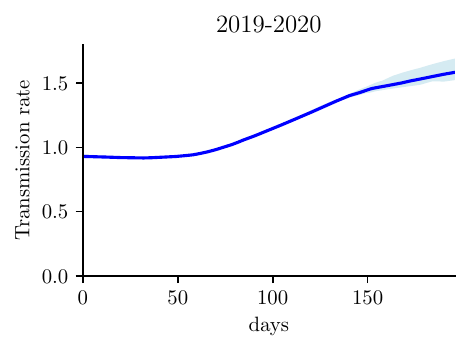}
    \end{subfigure}%
    \hfill
        \begin{subfigure}[b]{0.19\textwidth}
    \includegraphics[width=\textwidth]{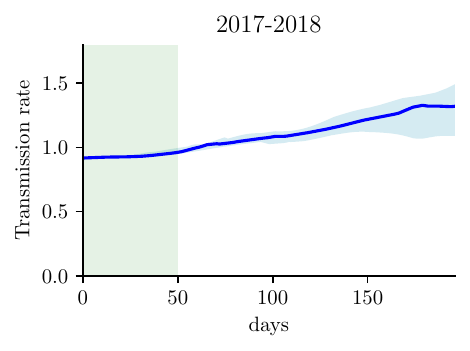}
    \end{subfigure}%

    \vspace{0.5cm} 

    \begin{subfigure}[b]{0.19\textwidth}
    \includegraphics[width=\textwidth]{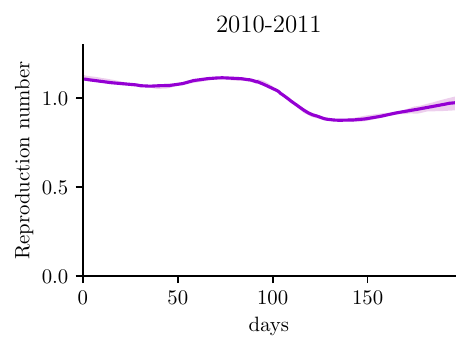}
    \end{subfigure}%
    \hfill
    \begin{subfigure}[b]{0.19\textwidth}
    \includegraphics[width=\textwidth]{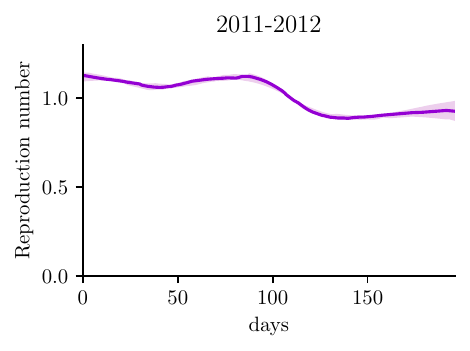}
    \end{subfigure}%
    \hfill
        \begin{subfigure}[b]{0.19\textwidth}
    \includegraphics[width=\textwidth]{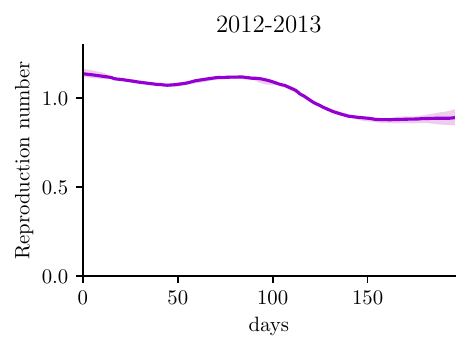}
    \end{subfigure}%
    \hfill
        \begin{subfigure}[b]{0.19\textwidth}
    \includegraphics[width=\textwidth]{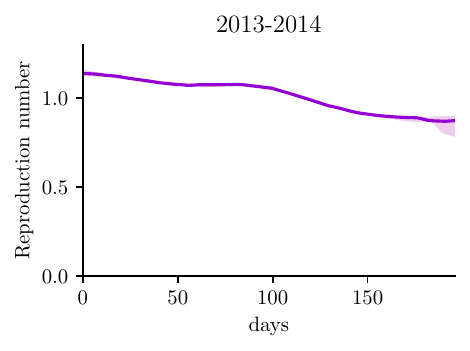}
    \end{subfigure}%
    \hfill
        \begin{subfigure}[b]{0.19\textwidth}
    \includegraphics[width=\textwidth]{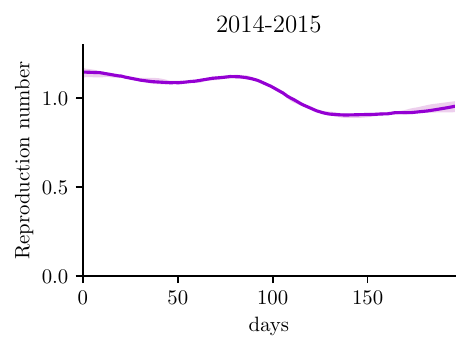}
    \end{subfigure}%
    \vspace{0.5cm} 
        \begin{subfigure}[b]{0.19\textwidth}
    \includegraphics[width=\textwidth]{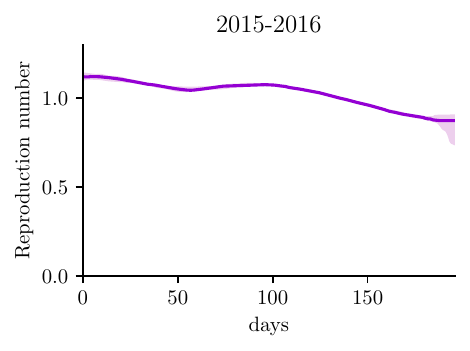}
    \end{subfigure}%
    \hfill
    \begin{subfigure}[b]{0.19\textwidth}
    \includegraphics[width=\textwidth]{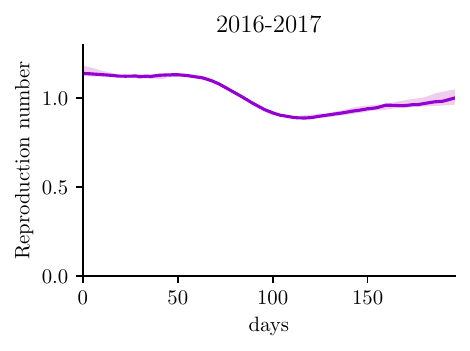}
    \end{subfigure}%
    \hfill
        \begin{subfigure}[b]{0.19\textwidth}
    \includegraphics[width=\textwidth]{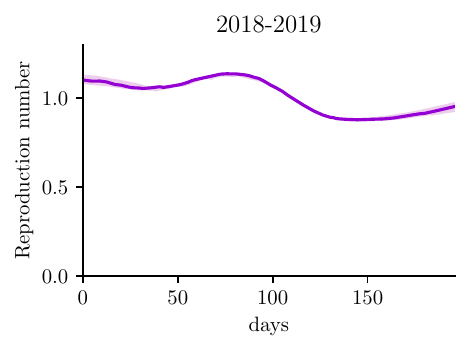}
    \end{subfigure}%
    \hfill
        \begin{subfigure}[b]{0.19\textwidth}
    \includegraphics[width=\textwidth]{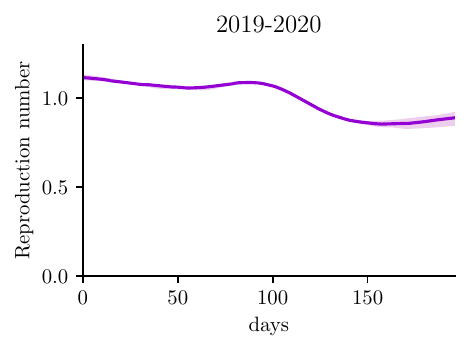}
    \end{subfigure}%
    \hfill
        \begin{subfigure}[b]{0.19\textwidth}
    \includegraphics[width=\textwidth]{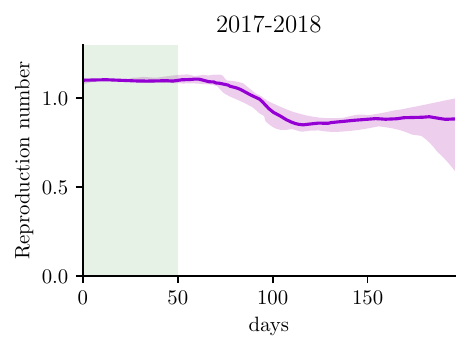}
    \end{subfigure}%
    \caption{\textbf{Case 2}. Reconstructions of cases, transmission rate, reproduction number exploiting influenza data. The training set is constituted by all epidemic waves from 2010-2011 to 2019-2020 except for 2017-2018, which belongs to the test set. The green window corresponds to the 49 days of observability in which we estimate the latent parameter for the testing sample.}
    \label{fig:infCase2}

\end{figure}

\begin{figure}[H]
    \centering
    \begin{subfigure}[b]{0.19\textwidth}
    \includegraphics[width=\textwidth]{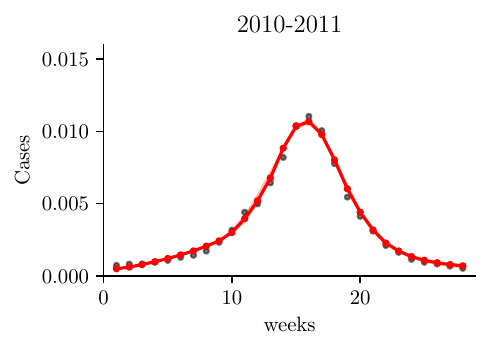}
    \end{subfigure}%
    \hfill
    \begin{subfigure}[b]{0.19\textwidth}
    \includegraphics[width=\textwidth]{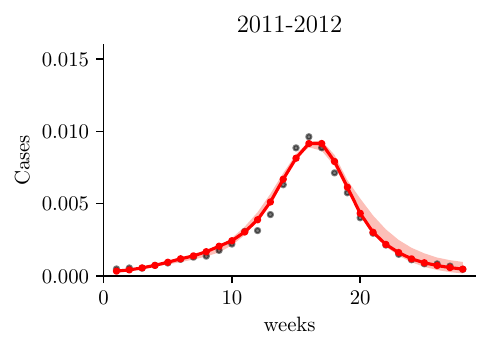}
    \end{subfigure}%
    \hfill
        \begin{subfigure}[b]{0.19\textwidth}
    \includegraphics[width=\textwidth]{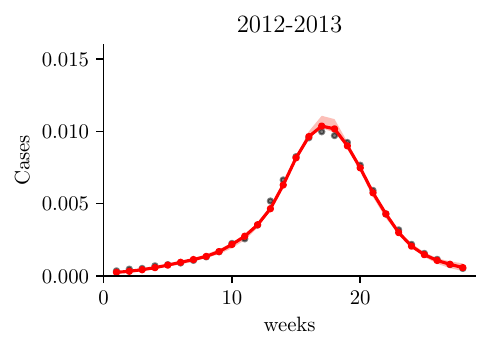}
    \end{subfigure}%
    \hfill
        \begin{subfigure}[b]{0.19\textwidth}
    \includegraphics[width=\textwidth]{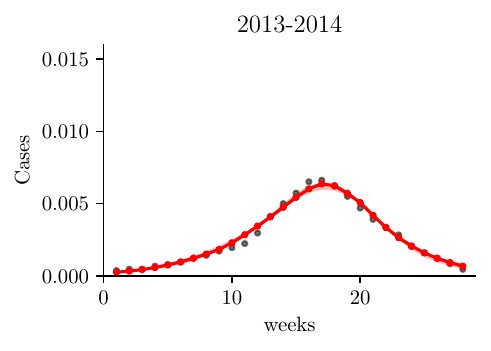}
    \end{subfigure}%
    \hfill
        \begin{subfigure}[b]{0.19\textwidth}
    \includegraphics[width=\textwidth]{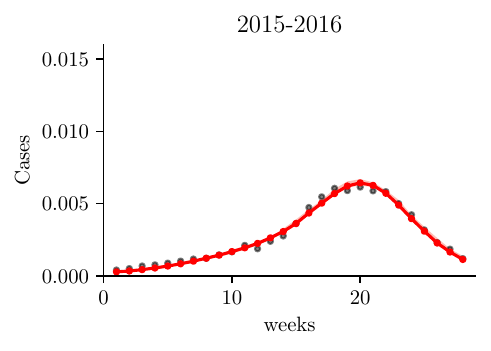}
    \end{subfigure}%
    \vspace{0.5cm} 
        \begin{subfigure}[b]{0.19\textwidth}
    \includegraphics[width=\textwidth]{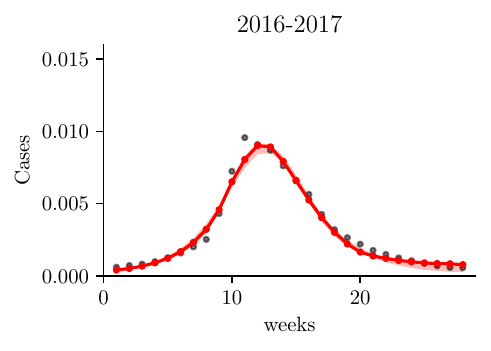}
    \end{subfigure}%
    \hfill
    \begin{subfigure}[b]{0.19\textwidth}
    \includegraphics[width=\textwidth]{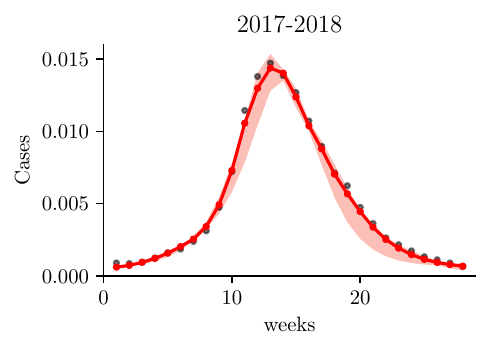}
    \end{subfigure}%
    \hfill
        \begin{subfigure}[b]{0.19\textwidth}
    \includegraphics[width=\textwidth]{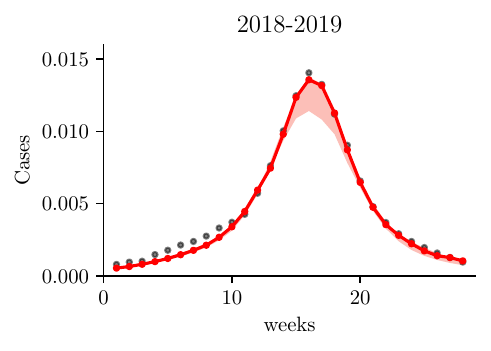}
    \end{subfigure}%
    \hfill
        \begin{subfigure}[b]{0.19\textwidth}
    \includegraphics[width=\textwidth]{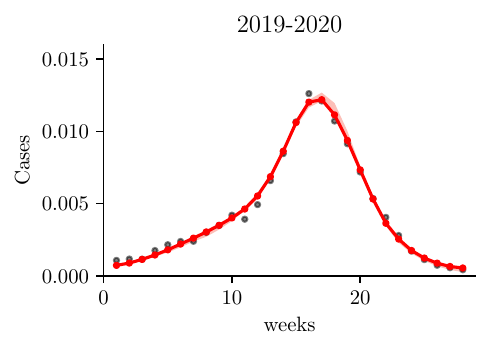}
    \end{subfigure}%
    \hfill
        \begin{subfigure}[b]{0.19\textwidth}
    \includegraphics[width=\textwidth]{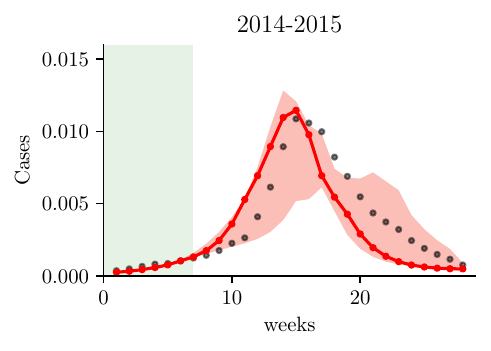}
    \end{subfigure}%

    \vspace{0.5cm} 

    \begin{subfigure}[b]{0.19\textwidth}
    \includegraphics[width=\textwidth]{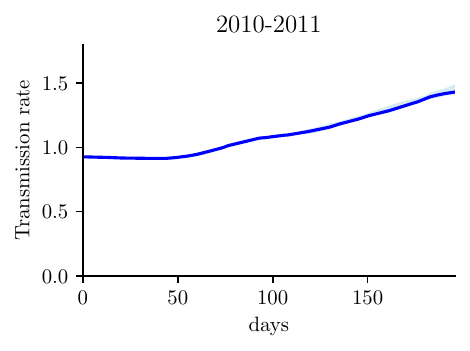}
    \end{subfigure}%
    \hfill
    \begin{subfigure}[b]{0.19\textwidth}
    \includegraphics[width=\textwidth]{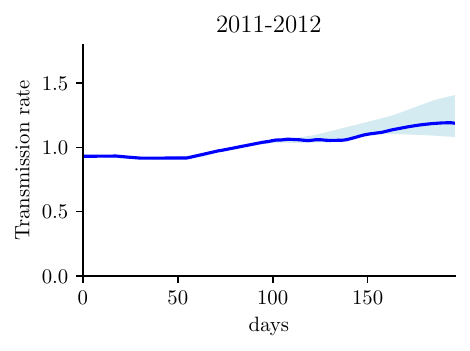}
    \end{subfigure}%
    \hfill
        \begin{subfigure}[b]{0.19\textwidth}
    \includegraphics[width=\textwidth]{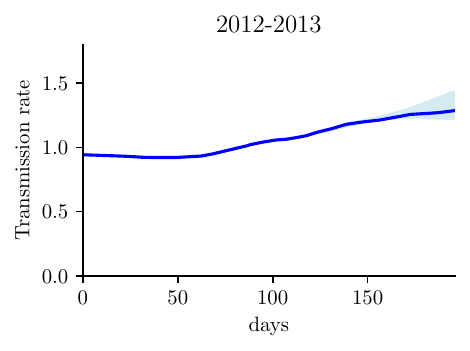}
    \end{subfigure}%
    \hfill
        \begin{subfigure}[b]{0.19\textwidth}
    \includegraphics[width=\textwidth]{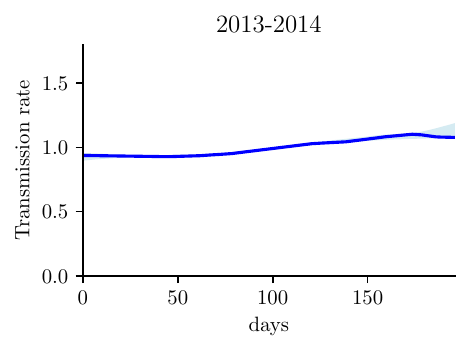}
    \end{subfigure}%
    \hfill
        \begin{subfigure}[b]{0.19\textwidth}
    \includegraphics[width=\textwidth]{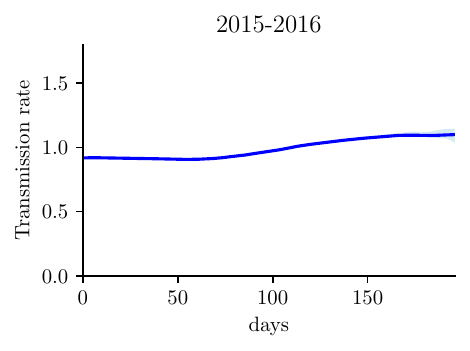}
    \end{subfigure}%
    \vspace{0.5cm} 
        \begin{subfigure}[b]{0.19\textwidth}
    \includegraphics[width=\textwidth]{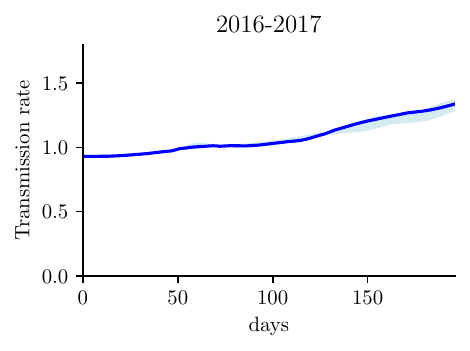}
    \end{subfigure}%
    \hfill
    \begin{subfigure}[b]{0.19\textwidth}
    \includegraphics[width=\textwidth]{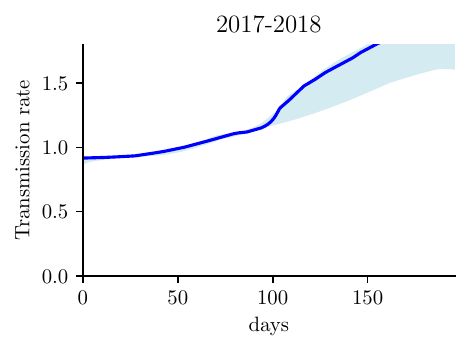}
    \end{subfigure}%
    \hfill
        \begin{subfigure}[b]{0.19\textwidth}
    \includegraphics[width=\textwidth]{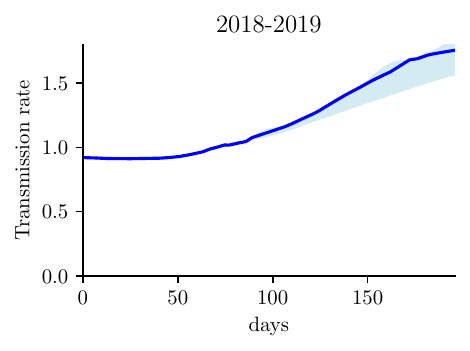}
    \end{subfigure}%
    \hfill
        \begin{subfigure}[b]{0.19\textwidth}
    \includegraphics[width=\textwidth]{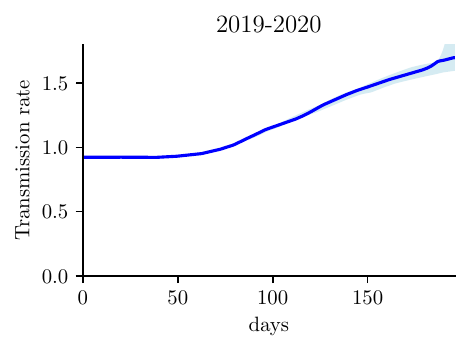}
    \end{subfigure}%
    \hfill
        \begin{subfigure}[b]{0.19\textwidth}
    \includegraphics[width=\textwidth]{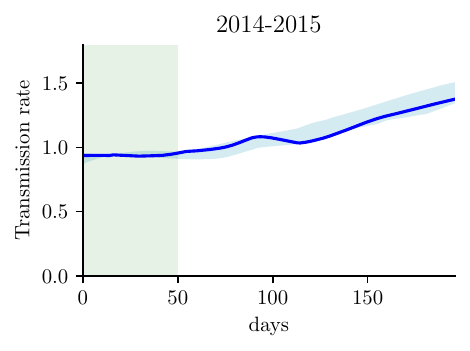}
    \end{subfigure}%

    \vspace{0.5cm} 

    \begin{subfigure}[b]{0.19\textwidth}
    \includegraphics[width=\textwidth]{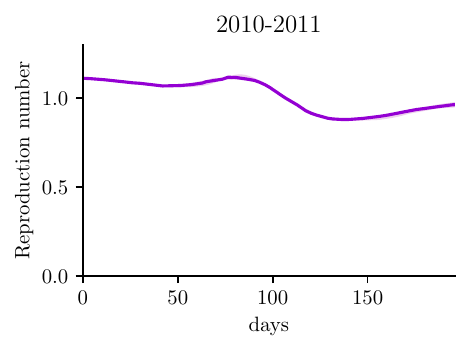}
    \end{subfigure}%
    \hfill
    \begin{subfigure}[b]{0.19\textwidth}
    \includegraphics[width=\textwidth]{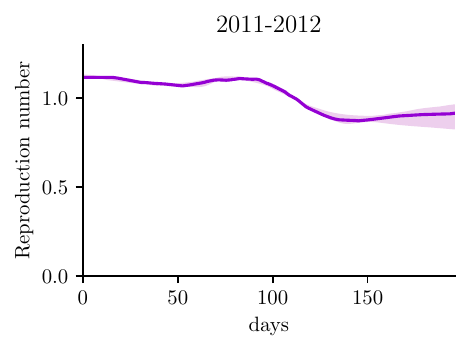}
    \end{subfigure}%
    \hfill
        \begin{subfigure}[b]{0.19\textwidth}
    \includegraphics[width=\textwidth]{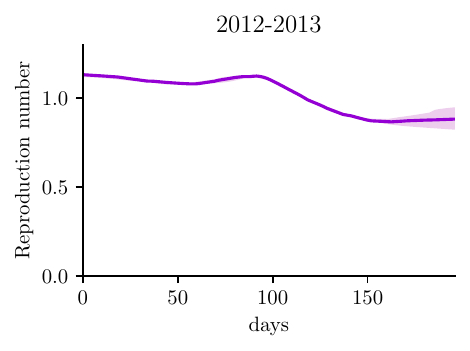}
    \end{subfigure}%
    \hfill
        \begin{subfigure}[b]{0.19\textwidth}
    \includegraphics[width=\textwidth]{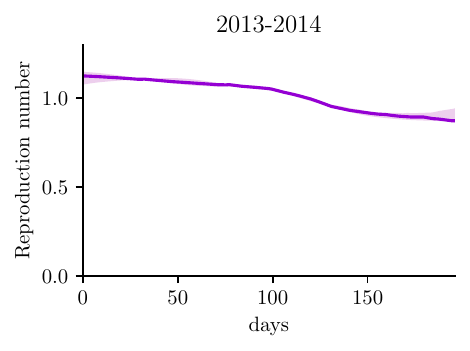}
    \end{subfigure}%
    \hfill
        \begin{subfigure}[b]{0.19\textwidth}
    \includegraphics[width=\textwidth]{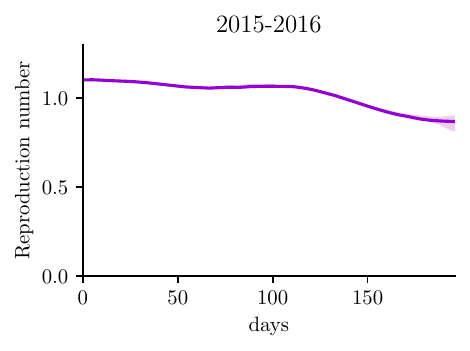}
    \end{subfigure}%
    \vspace{0.5cm} 
        \begin{subfigure}[b]{0.19\textwidth}
    \includegraphics[width=\textwidth]{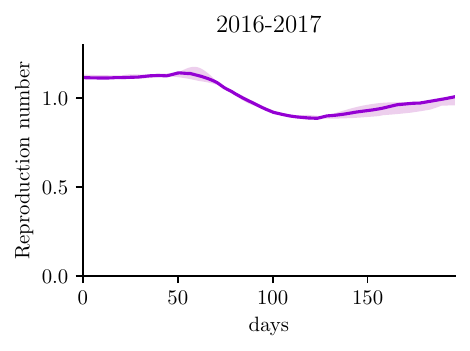}
    \end{subfigure}%
    \hfill
    \begin{subfigure}[b]{0.19\textwidth}
    \includegraphics[width=\textwidth]{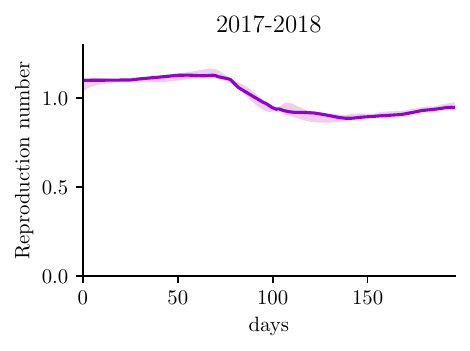}
    \end{subfigure}%
    \hfill
        \begin{subfigure}[b]{0.19\textwidth}
    \includegraphics[width=\textwidth]{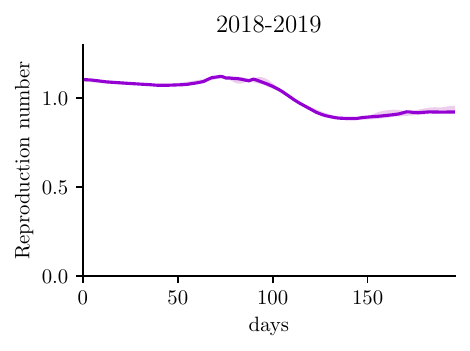}
    \end{subfigure}%
    \hfill
        \begin{subfigure}[b]{0.19\textwidth}
    \includegraphics[width=\textwidth]{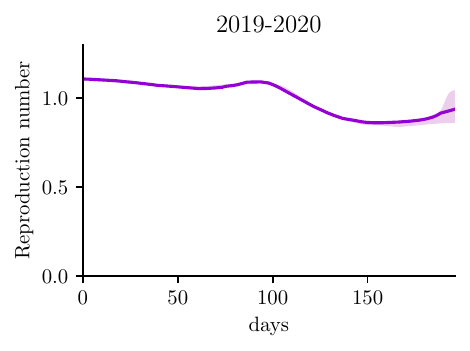}
    \end{subfigure}%
    \hfill
        \begin{subfigure}[b]{0.19\textwidth}
    \includegraphics[width=\textwidth]{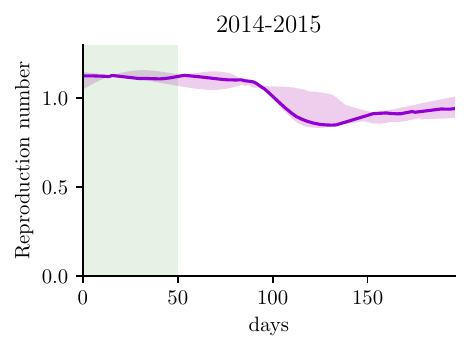}
    \end{subfigure}%
    \caption{\textbf{Case 3}. Reconstructions of cases, transmission rate, reproduction number exploiting influenza data. The training set is constituted by all epidemic waves from 2010-2011 to 2019-2020 except for 2014-2015, which belongs to the test set. The green window corresponds to the 49 days of observability in which we estimate the latent parameter for the testing sample.}

    \label{fig:infCase3}

\end{figure}

\begin{figure}[t]
 \centering
    \begin{subfigure}[b]{0.48\textwidth}
    \includegraphics[width=\textwidth]{real_case_1/cases_2018-2019_testg.pdf}
    \caption{}
    \end{subfigure}%
    \hfill
        \begin{subfigure}[b]{0.48\textwidth}
    \includegraphics[width=\textwidth]{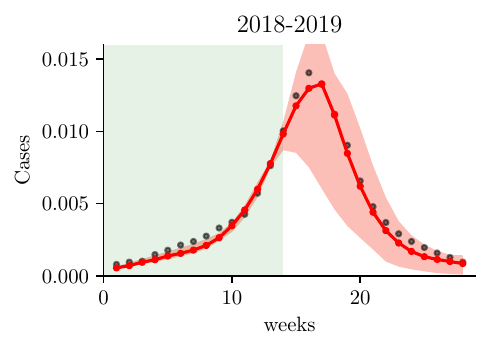}
    \caption{}
    \end{subfigure}%
    \caption{\textbf{Case 1}. Reconstruction of new cases of the wave corresponding to 2018-2019 (test set) with two width of the observation window (green area), respectively (a) $T_{\mathrm{obs}} = 49$ days and (b) $T_{\mathrm{obs}} = 98$ days.}
    \label{fig:confWindTC1}
\end{figure}

Figures \ref{fig:infCase1}-\ref{fig:infCase3} represent the new cases, transmission rate and reproducing number for both training and testing samples in the three cases respectively.
The number of new cases obtained on the training samples is always caught in terms of absolute values and, furthermore, the time at which the peak happens coincide with the ground truth.
In Case 1 the absolute value of the peak for the testing wave is retrieved, even though the peak is reached almost three weeks in advance.
In this case increasing the time-frame of observability (green region in the figure) for the estimation phase would help in making predictions more adherent with the amount of cases actually counted in 2018-2019 (see Figure \ref{fig:confWindTC1}).
Besides, in Case 2 the time at which the peak occurs is accurately predicted, although the median amount of cases is underestimated with respect of the ground truth. 
Indeed, the attained values at the peak for this epidemic wave are higher with respect to the other training waves.
Peak values this high have not been learnt by the model and, hence, they seem to be hardly predictable.
Finally, the median trajectory of new cases in Case 3 successfully catches the behavior of real data for the epidemic wave of 2014-2015.
The difficulties of the proposed architecture during the estimation and, consequently, during the testing phase can be ascribed to the limited amount of data coming from these epidemic waves, combined with high variability and similarities of input data as one can deduce from Figure \ref{fig:temp_um_sg}.
The latter impact on the prior used for determining the new $\delta$ parameter during the estimation phase.

From the transmission rate reconstructions, for which we do not dispose of a ground truth to compare our results with, we can infer some qualitative dependencies on the input variables: in accordance with literature \cite{lowen2007influenza, mubayi2021analytical} we observe an inverse growth of the transmission rate with respect to relative humidity (see, \textit{e.g.} \cite{mecenas2020effects} where it emerges that drier environments enhance transmission spread, similarly to air pollution \cite{cohen2017estimates}).
On the other hand, the transmission rate on temperature is not obvious in our case, since temperature's time series correspond to the same seasonal periods in different waves, and therefore the input functions can hardly be distinguished.
It is worth remarking that in our results peaks always happen during the colder months of wintertime.
Finally, the reconstructed transmission rate continues to grow in the long-term horizon, even the outbreak is declining.
This growing trend in the transmission rate was also observed in \textit{Test case 1} (cf. Figure \ref{fig:cases_unc_s}).
Note that no explicit relationship between the behavior (whether increasing or decreasing) of the transmission rate and the behavior of the infected population is imposed in the loss function.
However, we could explicitly enforce a long-term decline in the transmission rate within the loss function when the epidemic wave reaches its tail to prevent any unexpected upward trends.

The estimated reproduction number $\mathcal{R}_t$ always lies in the confidence interval estimated by \cite{trentini2022characterizing} for each influenza epidemic.
After the amount of new cases has reached its peak, the reproduction number always registers a sudden drop under the bifurcation value $\mathcal{R}_t = 1$, indicating that the epidemic is proceeding to the equilibrium depleting infectious.
Larger uncertainty bands characterize long-time behavior for both transmission rate and, consequently, reproduction number.

\begin{figure}[H]
    \centering
    \begin{subfigure}[b]{0.44\textwidth}
    \includegraphics[width=\textwidth]{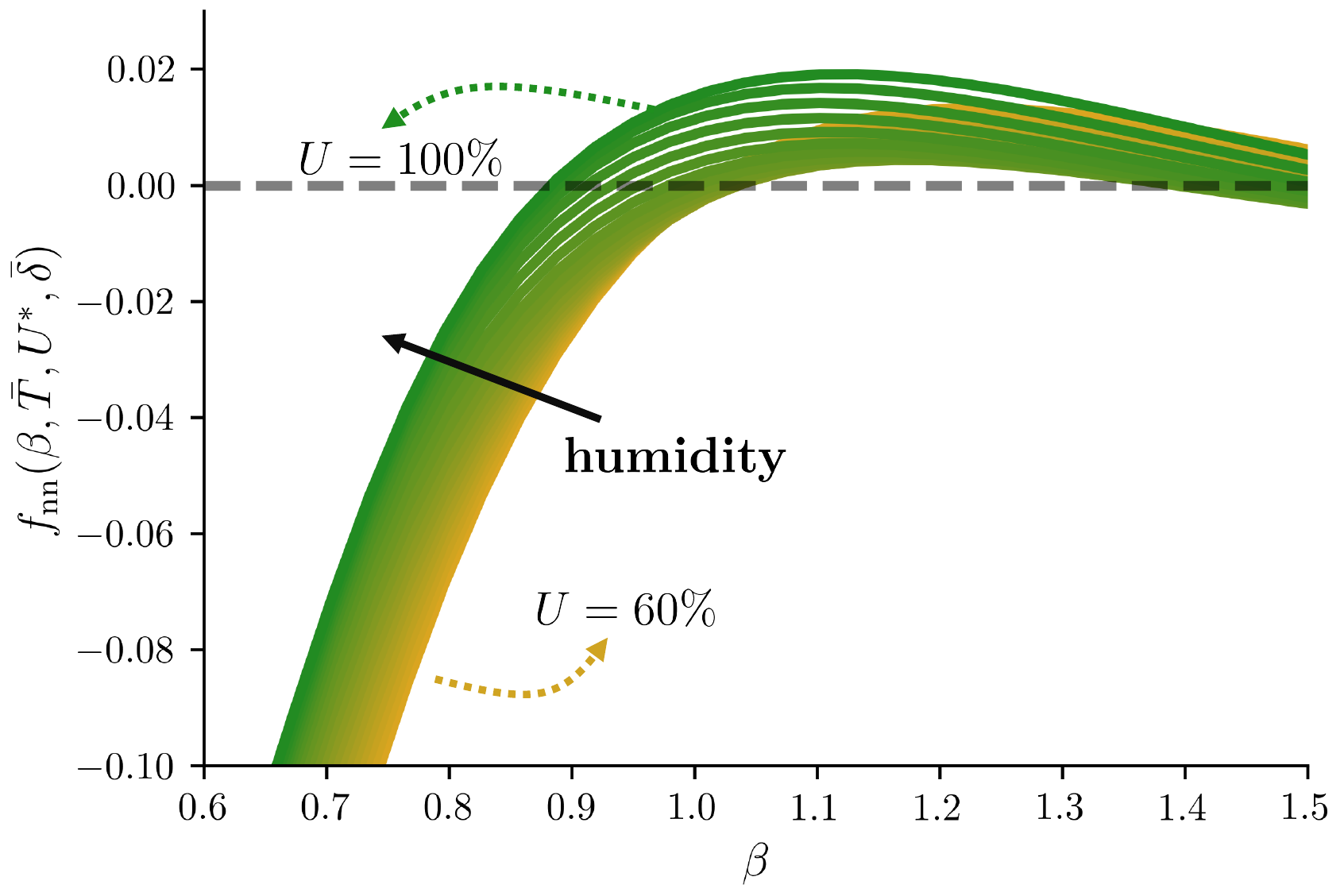}
    \caption{$\bar{T} = 12^{\circ}$, $\bar{\delta} = 1$.}
    \end{subfigure}%
    \hfill
    \begin{subfigure}[b]{0.44\textwidth}
    \includegraphics[width=\textwidth]{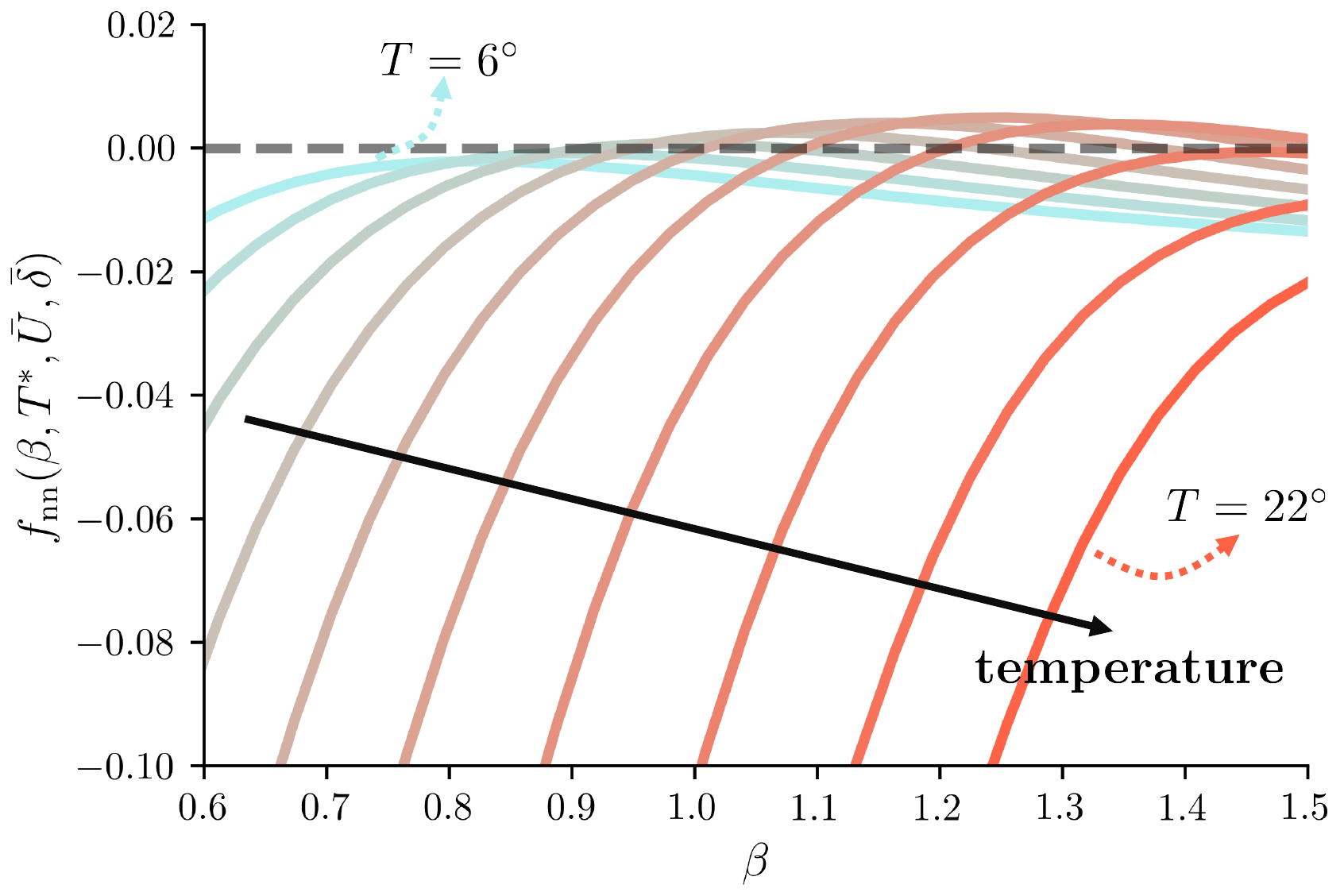}
    \caption{$\bar{U} = 72.5\%$, $\bar{\delta} = 1$.}
    \end{subfigure}%
    \vspace{0.5cm} 
    \begin{subfigure}[b]{0.44\textwidth}
    \includegraphics[width=\textwidth]{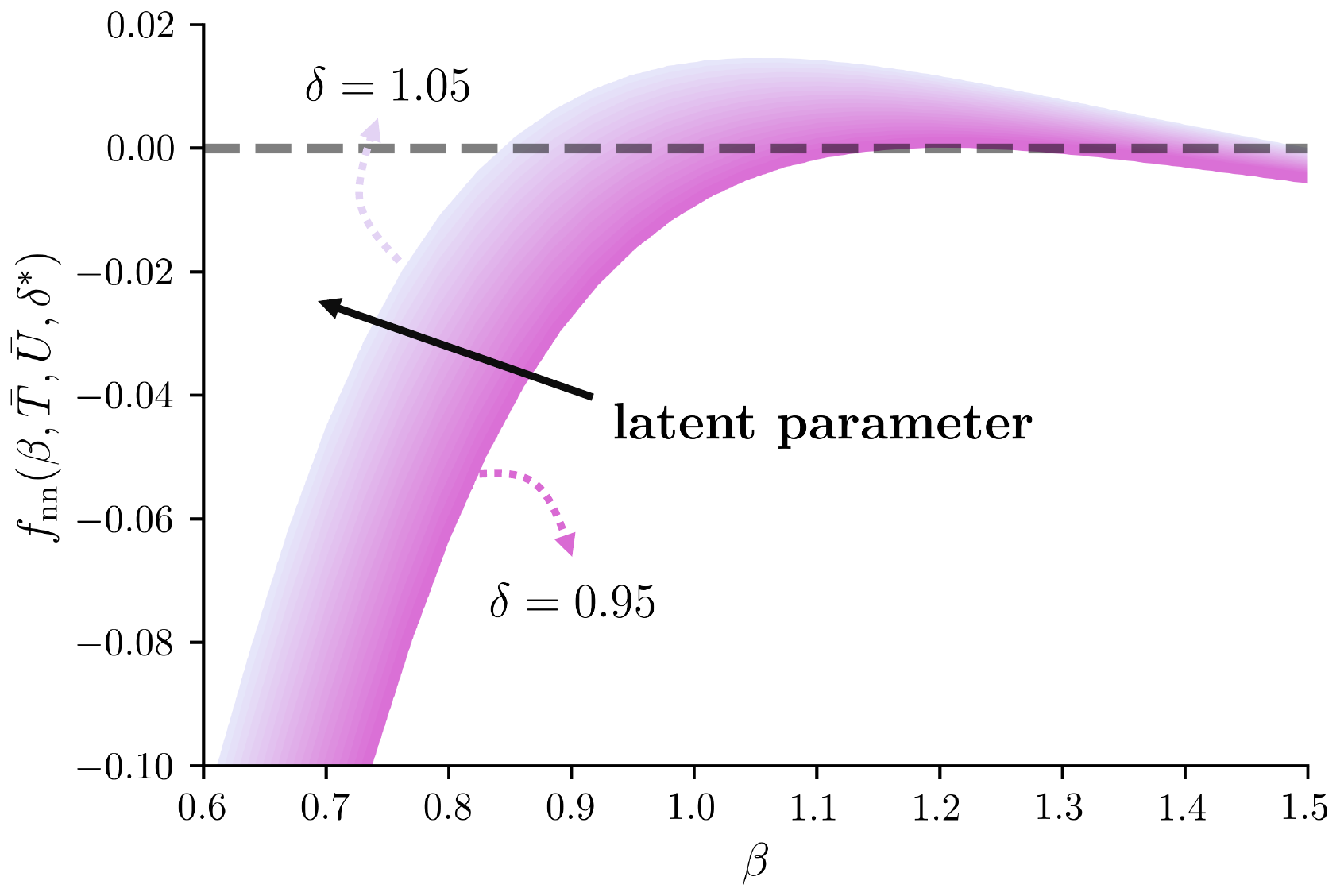}
    \caption{$\bar{T} = 12^{\circ}$, $\bar{U} = 72.5 \%$.}
    \end{subfigure}%
    \caption{Analysis of the reconstructed model for the transmission rate: plot of the right hand side of the transmission rate ($f_{nn}$) in terms of $\beta$ where two input exogenous parameters are cyclically kept fixed, while the third varies.  Each curve corresponds to different values of this latter varying parameter: relative humidity in (a), temperature in (b) and the reconstructed latent parameter in (c).}
    \label{fig:forw_anal_cparam}
\end{figure}

To gain more insight about the interpratability of our learned model, we analyze the phase-plane of the reconstructed transmission rate model.
Specifically, we focus on the median model of Case 3, and we analyse the value of the surrogate right-hand side ($f_{nn}$, cf. \eqref{eq:sys1}) at different values of the transmission rate.
In this case, we keep two of the three input parameters ($T$, $U$, $\delta$) cyclically fixed, while the third varies generating a different curve in the phase plane $(\beta, \, f_{nn})$, cf. \eqref{eq:anal_synt_db}.
In this way we explore how the unstable equilibrium point behaves in relation to different constant values of this latter parameter (see Figure \ref{fig:forw_anal_cparam}).
Hence, as humidity decreases, the unstable equilibrium shifts to higher values, and a similar effect is observed when temperature increases (cf Subsection \ref{subsec:real_dg}).
This behavior aligns with the literature (cf. \cite{lowen2007influenza, mubayi2021analytical}), as the range of values where the transmission rate has negative derivative expands with rising temperatures and decreasing humidity.
In contrast, the latent parameter, which differentiates between waves, shifts the equilibrium point further as its value decreases.

\section{Conclusions}
\label{sec:concLD}
In this work, we propose an innovative scientific machine learning method to infer the unknown differential evolution governing the transmission rate during epidemics, depending on exogenous parameters, which are known to influence transmission mechanisms.
By learning a neural dynamical model of the transmission rate, we aim to address the critical issue of extrapolating this parameter's values in order to make reliable forecasts beyond the observation interval.
Our framework incorporates a data assimilation approach to estimate, online, a latent parameter that distinguishes different waves of the same disease.
This latent parameter embodies the averaged effects of external factors that are not considered explicitly in the chosen set of exogenous factors, but still influence the transmission rate.

We present an extensive overview of numerical results for two distinct test cases.
In a synthetic scenario, we assess performance in terms of test error and the ability to recover the prescribed hidden dynamics.
The results prove that the proposed architecture achieves good accuracy, even when handling noisy data.
Additionally, we investigate the recovered model for transmission rates influenced by temperature and relative humidity, using real data from influenza waves in Italy between 2010 and 2020.
Despite data scarcity, the leave-one-out results indicate accurate peak predictions.
Furthermore, the results show that this tool is effective especially for short-term forecasts and, in some cases, successfully captures the entire yearly dynamics of new cases.
We explore the interpretability of the reconstructed model and confirm that it aligns with literature findings on the influence of increasing temperature and relative humidity on transmission rates.
However, as with many other computational frameworks, we remark that long-term predictions are less reliable and can sometimes be inaccurate.
Notably, it is impossible to obtain direct observations of the transmission rate’s evolution, and the epidemic data used for calibration are generally less affected by monitoring uncertainties near the infection peak compared to later stages of the outbreak.
Moreover, in our context this limitation arises from the few data available for training and the modest set of exogenous variables considered.
Extending this work to include additional time series is straightforward and will be the focus of future studies.

Possible future developments of this work include expanding the model to incorporate a broader set of key exogenous variables, such as non-pharmaceutical interventions, vaccinations, and immunity profiles.
Additionally, the data assimilation approach can be enhanced by using real-time moving windows to better estimate the latent parameter as more data becomes available.
Finally, the initial conditions of the epidemic, which were artificially fixed in both cases, could be learned online using the same approach.
This latter extension necessitates further investigation from a numerical standpoint.

We emphasize that this approach is general and can be extended to other models aimed at discovering the hidden dynamics of key parameters, even beyond the field of epidemiology.

\section*{Acknowledgments}
SP, FR, MV and GZ are members of INdAM-GNCS.
The present reasearch is part of the activities of "Dipartimento di Eccellenza 2023-2027", MUR, Italy.
FR and SP have received support from the project PRIN2022, MUR, Italy, 2023-2025, P2022N5ZNP “SIDDMs: shape-informed data-driven models for parametrized PDEs, with application to computational cardiology”, founded by the European Union (Next Generation EU, Mission 4 Component 2).

\section*{Data Availability}
The computational framework is available on GitHub together with the numerical results of this work: \url{https://github.com/giovanniziarelli/inferringTRdynamicsML/}.
\pagebreak

\printbibliography

@article{parolini2021suihter,
  title={SUIHTER: A new mathematical model for COVID-19. Application to the analysis of the second epidemic outbreak in Italy},
  author={Parolini, Nicola and Ded{\`e}, Luca and Antonietti, Paola F and Ardenghi, Giovanni and Manzoni, Andrea and Miglio, Edie and Pugliese, Andrea and Verani, Marco and Quarteroni, Alfio},
  journal={Proceedings of the Royal Society A},
  volume={477},
  number={2253},
  pages={20210027},
  year={2021},
  publisher={The Royal Society}
}

@book{martcheva2015introduction,
  title={An introduction to mathematical epidemiology},
  author={Martcheva, Maia},
  volume={61},
  year={2015},
  publisher={Springer}
}

@article{regazzoni2019machine,
  title={Machine learning for fast and reliable solution of time-dependent differential equations},
  author={Regazzoni, Francesco and Ded{\`e}, Luca and Quarteroni, Alfio},
  journal={Journal of Computational physics},
  volume={397},
  pages={108852},
  year={2019},
  publisher={Elsevier}
}

@article{parolini2022modelling,
  title={Modelling the COVID-19 epidemic and the vaccination campaign in Italy by the SUIHTER model},
  author={Parolini, Nicola and Ded{\`e}, Luca and Ardenghi, Giovanni and Quarteroni, Alfio},
  journal={Infectious Disease Modelling},
  volume={7},
  number={2},
  pages={45--63},
  year={2022},
  publisher={Elsevier}
}

@article{bertuzzo2020geography,
  title={The geography of COVID-19 spread in Italy and implications for the relaxation of confinement measures},
  author={Bertuzzo, Enrico and Mari, Lorenzo and Pasetto, Damiano and Miccoli, Stefano and Casagrandi, Renato and Gatto, Marino and Rinaldo, Andrea},
  journal={Nature communications},
  volume={11},
  number={1},
  pages={42--64},
  year={2020},
  publisher={Nature Publishing Group UK London}
}

@article{ziarelli2023optimized,
  title={Optimized numerical solutions of SIRDVW multiage model controlling SARS-CoV-2 vaccine roll out: An application to the Italian scenario},
  author={Ziarelli, Giovanni and Ded{\`e}, Luca and Parolini, Nicola and Verani, Marco and Quarteroni, Alfio},
  journal={Infectious Disease Modelling},
  volume={8},
  number={3},
  pages={672--703},
  year={2023},
  publisher={Elsevier}
}

@article{lemaitre2022optimal,
  title={Optimal control of the spatial allocation of COVID-19 vaccines: Italy as a case study},
  author={Lemaitre, Joseph C and Pasetto, Damiano and Zanon, Mario and Bertuzzo, Enrico and Mari, Lorenzo and Miccoli, Stefano and Casagrandi, Renato and Gatto, Marino and Rinaldo, Andrea},
  journal={PLoS computational biology},
  volume={18},
  number={7},
  pages={e1010237},
  year={2022},
  publisher={Public Library of Science San Francisco, CA USA}
}

@article{millevoi2023physics,
  title={A Physics-Informed Neural Network approach for compartmental epidemiological models},
  author={Millevoi, Caterina and Pasetto, Damiano and Ferronato, Massimiliano},
  journal={PLOS Computational Biology},
  volume={20},
  number={9},
  pages={e1012387},
  year={2024},
  publisher={Public Library of Science San Francisco, CA USA}
}

@article{he2023combining,
  title={Combining the dynamic model and deep neural networks to identify the intensity of interventions during COVID-19 pandemic},
  author={He, Mengqi and Tang, Sanyi and Xiao, Yanni},
  journal={PLOS Computational Biology},
  volume={19},
  number={10},
  pages={e1011535},
  year={2023},
  publisher={Public Library of Science San Francisco, CA USA}
}

@article{marziano2021effect,
  title={The effect of COVID-19 vaccination in Italy and perspectives for living with the virus},
  author={Marziano, Valentina and Guzzetta, Giorgio and Mammone, Alessia and Riccardo, Flavia and Poletti, Piero and Trentini, Filippo and Manica, Mattia and Siddu, Andrea and Bella, Antonino and Stefanelli, Paola and others},
  journal={Nature communications},
  volume={12},
  number={1},
  pages={7272},
  year={2021},
  publisher={Nature Publishing Group UK London}
}

@article{sherratt2023predictive,
  title={Predictive performance of multi-model ensemble forecasts of COVID-19 across European nations},
  author={Sherratt, Katharine and Gruson, Hugo and Johnson, Helen and Niehus, Rene and Prasse, Bastian and Sandmann, Frank and Deuschel, Jannik and Wolffram, Daniel and Abbott, Sam and Ullrich, Alexander and others},
  journal={Elife},
  volume={12},
  pages={e81916},
  year={2023},
  publisher={eLife Sciences Publications Limited}
}

@article{lowen2014roles,
  title={Roles of humidity and temperature in shaping influenza seasonality},
  author={Lowen, Anice C and Steel, John},
  journal={Journal of virology},
  volume={88},
  number={14},
  pages={7692--7695},
  year={2014},
  publisher={American Society for Microbiology}
}

@article{regazzoni2021combining,
  title={Combining data assimilation and machine learning to build data-driven models for unknown long time dynamics—applications in cardiovascular modeling},
  author={Regazzoni, Francesco and Chapelle, Dominique and Moireau, Philippe},
  journal={International Journal for Numerical Methods in Biomedical Engineering},
  volume={37},
  number={7},
  pages={e3471},
  year={2021},
  publisher={Wiley Online Library}
}

@article{johnsen2022seasonal,
  title={Seasonal variation in the transmission rate of covid-19 in a temperate climate can be implemented in epidemic population models by using daily average temperature as a proxy for seasonal changes in transmission rate},
  author={Johnsen, Morten Guldborg and Christiansen, Lasse Engbo and Gr{\ae}sb{\o}ll, Kaare},
  journal={Microbial Risk Analysis},
  volume={22},
  pages={100235},
  year={2022},
  publisher={Elsevier}
}

@article{pasetto2018near,
  title={Near real-time forecasting for cholera decision making in Haiti after Hurricane Matthew},
  author={Pasetto, Damiano and Finger, Flavio and Camacho, Anton and Grandesso, Francesco and Cohuet, Sandra and Lemaitre, Joseph C and Azman, Andrew S and Luquero, Francisco J and Bertuzzo, Enrico and Rinaldo, Andrea},
  journal={PLoS computational biology},
  volume={14},
  number={5},
  pages={e1006127},
  year={2018},
  publisher={Public Library of Science San Francisco, CA USA}
}

@article{souto2022assessing,
  title={Assessing the best time interval between doses in a two-dose vaccination regimen to reduce the number of deaths in an ongoing epidemic of SARS-CoV-2},
  author={Souto Ferreira, Leonardo and Canton, Otavio and Da Silva, Rafael Lopes Paix{\~a}o and Poloni, Silas and Sudbrack, V{\'\i}tor and Borges, Marcelo Eduardo and Franco, Caroline and Marquitti, Flavia Maria Darcie and de Moraes, Jos{\'e} C{\'a}ssio and Veras, Maria Am{\'e}lia de Sousa Mascena and others},
  journal={PLoS computational biology},
  volume={18},
  number={3},
  pages={e1009978},
  year={2022},
  publisher={Public Library of Science San Francisco, CA USA}
}

@article{giordano2021modeling,
  title={Modeling vaccination rollouts, SARS-CoV-2 variants and the requirement for non-pharmaceutical interventions in Italy},
  author={Giordano, Giulia and Colaneri, Marta and Di Filippo, Alessandro and Blanchini, Franco and Bolzern, Paolo and De Nicolao, Giuseppe and Sacchi, Paolo and Colaneri, Patrizio and Bruno, Raffaele},
  journal={Nature medicine},
  volume={27},
  number={6},
  pages={993--998},
  year={2021},
  publisher={Nature Publishing Group US New York}
}

@article{richard2021age,
  title={Age-structured non-pharmaceutical interventions for optimal control of COVID-19 epidemic},
  author={Richard, Quentin and Alizon, Samuel and Choisy, Marc and Sofonea, Mircea T and Djidjou-Demasse, Rams{\`e}s},
  journal={PLoS computational biology},
  volume={17},
  number={3},
  pages={e1008776},
  year={2021},
  publisher={Public Library of Science San Francisco, CA USA}
}

@article{hinch2021openabm,
  title={OpenABM-Covid19—An agent-based model for non-pharmaceutical interventions against COVID-19 including contact tracing},
  author={Hinch, Robert and Probert, William JM and Nurtay, Anel and Kendall, Michelle and Wymant, Chris and Hall, Matthew and Lythgoe, Katrina and Bulas Cruz, Ana and Zhao, Lele and Stewart, Andrea and others},
  journal={PLoS computational biology},
  volume={17},
  number={7},
  pages={e1009146},
  year={2021},
  publisher={Public Library of Science San Francisco, CA USA}
}

@article{hazelbag2020calibration,
  title={Calibration of individual-based models to epidemiological data: A systematic review},
  author={Hazelbag, C Marijn and Dushoff, Jonathan and Dominic, Emanuel M and Mthombothi, Zinhle E and Delva, Wim},
  journal={PLoS computational biology},
  volume={16},
  number={5},
  pages={e1007893},
  year={2020},
  publisher={Public Library of Science San Francisco, CA USA}
}

@article{gleeson2022calibrating,
  title={Calibrating COVID-19 susceptible-exposed-infected-removed models with time-varying effective contact rates},
  author={Gleeson, James P and Brendan Murphy, Thomas and O’Brien, Joseph D and Friel, Nial and Bargary, Norma and O'Sullivan, David JP},
  journal={Philosophical Transactions of the Royal Society A},
  volume={380},
  number={2214},
  pages={20210120},
  year={2022},
  publisher={The Royal Society}
}

@article{sainte2006modeling,
  title={Modeling and estimation of the cardiac electromechanical activity},
  author={Sainte-Marie, Jacques and Chapelle, Dominique and Cimrman, Robert and Sorine, Michel},
  journal={Computers \& structures},
  volume={84},
  number={28},
  pages={1743--1759},
  year={2006},
  publisher={Elsevier}
}

@article{regazzoni2020biophysically,
  title={Biophysically detailed mathematical models of multiscale cardiac active mechanics},
  author={Regazzoni, Francesco and Ded{\`e}, Luca and Quarteroni, Alfio},
  journal={PLoS computational biology},
  volume={16},
  number={10},
  pages={e1008294},
  year={2020},
  publisher={Public Library of Science San Francisco, CA USA}
}

@article{saiteja2022critical,
  title={Critical review on optimal regenerative braking control system architecture, calibration parameters and development challenges for EVs},
  author={Saiteja, Pemmareddy and Ashok, Bragadeshwaran and Wagh, Atharva Sanjay and Farrag, Mohamed Emad},
  journal={International Journal of Energy Research},
  volume={46},
  number={14},
  pages={20146--20179},
  year={2022},
  publisher={Wiley Online Library}
}

@article{sanso2009statistical,
  title={Statistical calibration of climate system properties},
  author={Sans{\'o}, Bruno and Forest, Chris},
  journal={Journal of the Royal Statistical Society Series C: Applied Statistics},
  volume={58},
  number={4},
  pages={485--503},
  year={2009},
  publisher={Oxford University Press}
}

@article{dukic2012tracking,
  title={Tracking epidemics with Google flu trends data and a state-space SEIR model},
  author={Dukic, Vanja and Lopes, Hedibert F and Polson, Nicholas G},
  journal={Journal of the American Statistical Association},
  volume={107},
  number={500},
  pages={1410--1426},
  year={2012},
  publisher={Taylor \& Francis}
}

@article{lekone2006statistical,
  title={Statistical inference in a stochastic epidemic SEIR model with control intervention: Ebola as a case study},
  author={Lekone, Phenyo E and Finkenst{\"a}dt, B{\"a}rbel F},
  journal={Biometrics},
  volume={62},
  number={4},
  pages={1170--1177},
  year={2006},
  publisher={Oxford University Press}
}

@article{buonomo2011simple,
  title={A simple analysis of vaccination strategies for rubella},
  author={Buonomo, Bruno},
  journal={Mathematical Biosciences \& Engineering},
  volume={8},
  number={3},
  pages={677--687},
  year={2011},
  publisher={Mathematical Biosciences \& Engineering}
}

@article{kumar2021wavelet,
  title={A wavelet based numerical scheme for fractional order SEIR epidemic of measles by using Genocchi polynomials},
  author={Kumar, Sunil and Kumar, Ranbir and Osman, Mohamed S and Samet, Bessem},
  journal={Numerical methods for partial differential equations},
  volume={37},
  number={2},
  pages={1250--1268},
  year={2021},
  publisher={Wiley Online Library}
}

@article{regazzoni2024learning,
  title={Learning the intrinsic dynamics of spatio-temporal processes through Latent Dynamics Networks},
  author={Regazzoni, Francesco and Pagani, Stefano and Salvador, Matteo and Ded{\`e}, Luca and Quarteroni, Alfio},
  journal={Nature Communications},
  volume={15},
  number={1},
  pages={1834},
  year={2024},
  publisher={Nature Publishing Group UK London}
}

@article{hoertel2020stochastic,
  title={A stochastic agent-based model of the SARS-CoV-2 epidemic in France},
  author={Hoertel, Nicolas and Blachier, Martin and Blanco, Carlos and Olfson, Mark and Massetti, Marc and Rico, Marina S{\'a}nchez and Limosin, Fr{\'e}d{\'e}ric and Leleu, Henri},
  journal={Nature medicine},
  volume={26},
  number={9},
  pages={1417--1421},
  year={2020},
  publisher={Nature Publishing Group US New York}
}

@article{mecenas2020effects,
  title={Effects of temperature and humidity on the spread of COVID-19: A systematic review},
  author={Mecenas, Paulo and Bastos, Renata Travassos da Rosa Moreira and Vallinoto, Antonio C R and Normando, David},
  journal={PLoS one},
  volume={15},
  number={9},
  pages={e0238339},
  year={2020},
  publisher={Public Library of Science San Francisco, CA USA}
}

@article{cohen2017estimates,
  title={Estimates and 25-year trends of the global burden of disease attributable to ambient air pollution: an analysis of data from the Global Burden of Diseases Study 2015},
  author={Cohen, Aaron J and Brauer, Michael and Burnett, Richard and Anderson, H Ross and Frostad, Joseph and Estep, Kara and Balakrishnan, Kalpana and Brunekreef, Bert and Dandona, Lalit and Dandona, Rakhi and others},
  journal={The Lancet},
  volume={389},
  number={10082},
  pages={1907--1918},
  year={2017},
  publisher={Elsevier}
}

@article{chinazzi2020effect,
  title={The effect of travel restrictions on the spread of the 2019 novel coronavirus (COVID-19) outbreak},
  author={Chinazzi, Matteo and Davis, Jessica T and Ajelli, Marco and Gioannini, Corrado and Litvinova, Maria and Merler, Stefano and Pastore y Piontti, Ana and Mu, Kunpeng and Rossi, Luca and Sun, Kaiyuan and others},
  journal={Science},
  volume={368},
  number={6489},
  pages={395--400},
  year={2020},
  publisher={American Association for the Advancement of Science}
}

@article{kerr2021covasim,
  title={Covasim: an agent-based model of COVID-19 dynamics and interventions},
  author={Kerr, Cliff C and Stuart, Robyn M and Mistry, Dina and Abeysuriya, Romesh G and Rosenfeld, Katherine and Hart, Gregory R and N{\'u}{\~n}ez, Rafael C and Cohen, Jamie A and Selvaraj, Prashanth and Hagedorn, Brittany and others},
  journal={PLOS Computational Biology},
  volume={17},
  number={7},
  pages={e1009149},
  year={2021},
  publisher={Public Library of Science San Francisco, CA USA}
}

@article{lima2021impact,
  title={Impact of mobility restriction in COVID-19 superspreading events using agent-based model},
  author={Lima, Larissa L and Atman, Allbens PF},
  journal={Plos one},
  volume={16},
  number={3},
  pages={e0248708},
  year={2021},
  publisher={Public Library of Science San Francisco, CA USA}
}

@article{lasser2022assessing,
  title={Assessing the impact of SARS-CoV-2 prevention measures in Austrian schools using agent-based simulations and cluster tracing data},
  author={Lasser, Jana and Sorger, Johannes and Richter, Lukas and Thurner, Stefan and Schmid, Daniela and Klimek, Peter},
  journal={Nature communications},
  volume={13},
  number={1},
  pages={554},
  year={2022},
  publisher={Nature Publishing Group UK London}
}

@article{olumoyin2021data,
  title={Data-driven deep-learning algorithm for asymptomatic COVID-19 model with varying mitigation measures and transmission rate},
  author={Olumoyin, Kayode D and Khaliq, Abdulqayyum M and Furati, Khaled M},
  journal={Epidemiologia},
  volume={2},
  number={4},
  pages={471--489},
  year={2021},
  publisher={MDPI}
}

@article{shaier2022data,
  title={Data-Driven Approaches for Predicting Spread of Infectious Diseases Through DINNs: Disease Informed Neural Networks},
  author={Shaier, Sagi and Raissi, Maziar and Seshaiyer, Padmanabhan},
  journal={Letters in Biomathematics},
  volume={9},
  number={1},
  pages={71--105},
  year={2022}
}

@article{bertaglia2022asymptotic,
  title={Asymptotic-Preserving Neural Networks for multiscale hyperbolic models of epidemic spread},
  author={Bertaglia, Giulia and Lu, Chuan and Pareschi, Lorenzo and Zhu, Xueyu},
  journal={Mathematical Models and Methods in Applied Sciences},
  volume={32},
  number={10},
  pages={1949--1985},
  year={2022},
  publisher={World Scientific}
}

@article{virlogeux2015estimating,
  title={Estimating the distribution of the incubation periods of human avian influenza A (H7N9) virus infections},
  author={Virlogeux, Victor and Li, Ming and Tsang, Tim K and Feng, Luzhao and Fang, Vicky J and Jiang, Hui and Wu, Peng and Zheng, Jiandong and Lau, Eric HY and Cao, Yu and others},
  journal={American Journal of Epidemiology},
  volume={182},
  number={8},
  pages={723--729},
  year={2015},
  publisher={Oxford University Press}
}

@article{lowen2007influenza,
  title={Influenza virus transmission is dependent on relative humidity and temperature},
  author={Lowen, Anice C and Mubareka, Samira and Steel, John and Palese, Peter},
  journal={PLoS pathogens},
  volume={3},
  number={10},
  pages={e151},
  year={2007},
  publisher={Public Library of Science San Francisco, USA}
}

@article{mubayi2021analytical,
  title={Analytical estimation of data-motivated time-dependent disease transmission rate: An application to ebola and selected public health problems},
  author={Mubayi, Anuj and Pandey, Abhishek and Brasic, Christine and Mubayi, Anamika and Ghosh, Parijat and Ghosh, Aditi},
  journal={Tropical Medicine and Infectious Disease},
  volume={6},
  number={3},
  pages={141},
  year={2021},
  publisher={MDPI}
}

@article{morens2011pandemic,
  title={Pandemic influenza: certain uncertainties},
  author={Morens, David M and Taubenberger, Jeffery K},
  journal={Reviews in medical virology},
  volume={21},
  number={5},
  pages={262--284},
  year={2011},
  publisher={Wiley Online Library}
}

@article{world2003influenza,
  title={Influenza fact sheet: Overview},
  author={World Health Organization and others},
  journal={Weekly Epidemiological Record= Relev{\'e} {\'e}pid{\'e}miologique hebdomadaire},
  volume={78},
  number={11},
  pages={77--80},
  year={2003}
}

@article{white2010reporting,
  title={Reporting errors in infectious disease outbreaks, with an application to Pandemic Influenza A/H1N1},
  author={White, Laura F and Pagano, Marcello},
  journal={Epidemiologic Perspectives \& Innovations},
  volume={7},
  pages={1--12},
  year={2010},
  publisher={Springer}
}

@article{krishnan2012selection,
  title={On the selection of optimum Savitzky-Golay filters},
  author={Krishnan, Sunder R and Seelamantula, Chandra S},
  journal={IEEE transactions on signal processing},
  volume={61},
  number={2},
  pages={380--391},
  year={2012},
  publisher={IEEE}
}

@article{trentini2022characterizing,
  title={Characterizing the transmission patterns of seasonal influenza in Italy: lessons from the last decade},
  author={Trentini, Filippo and Pariani, Elena and Bella, Antonino and Diurno, Giulio and Crottogini, Lucia and Rizzo, Caterina and Merler, Stefano and Ajelli, Marco},
  journal={BMC public health},
  volume={22},
  number={1},
  pages={19},
  year={2022},
  publisher={Springer}
}

@article{biggerstaff2014estimates,
  title={Estimates of the reproduction number for seasonal, pandemic, and zoonotic influenza: a systematic review of the literature},
  author={Biggerstaff, Matthew and Cauchemez, Simon and Reed, Carrie and Gambhir, Manoj and Finelli, Lyn},
  journal={BMC infectious diseases},
  volume={14},
  number={1},
  pages={1--20},
  year={2014},
  publisher={Springer}
}

@book{boyce2012elementary,
  title={Elementary differential equations},
  author={Boyce, William E and DiPrima, Richard C},
  volume={6},
  year={2012},
  publisher={Wiley New York}
}

@article{jing2021covid,
  title={COVID-19 modelling by time-varying transmission rate associated with mobility trend of driving via Apple Maps},
  author={Jing, Min and Ng, Kok Yew and Mac Namee, Brian and Biglarbeigi, Pardis and Brisk, Rob and Bond, Raymond and Finlay, Dewar and McLaughlin, James},
  journal={Journal of Biomedical Informatics},
  volume={122},
  pages={103905},
  year={2021},
  publisher={Elsevier}
}

@article{chowell2016characterizing,
  title={Characterizing the reproduction number of epidemics with early subexponential growth dynamics},
  author={Chowell, Gerardo and Viboud, C{\'e}cile and Simonsen, Lone and Moghadas, Seyed M},
  journal={Journal of The Royal Society Interface},
  volume={13},
  number={123},
  pages={20160659},
  year={2016},
  publisher={The Royal Society}
}

@article{wallinga2007generation,
  title={How generation intervals shape the relationship between growth rates and reproductive numbers},
  author={Wallinga, Jacco and Lipsitch, Marc},
  journal={Proceedings of the Royal Society B: Biological Sciences},
  volume={274},
  number={1609},
  pages={599--604},
  year={2007},
  publisher={The Royal Society London}
}

@article{lessler2009incubation,
  title={Incubation periods of acute respiratory viral infections: a systematic review},
  author={Lessler, Justin and Reich, Nicholas G and Brookmeyer, Ron and Perl, Trish M and Nelson, Kenrad E and Cummings, Derek AT},
  journal={The Lancet infectious diseases},
  volume={9},
  number={5},
  pages={291--300},
  year={2009},
  publisher={Elsevier}
}

@article{fiandrino2024collaborative,
  title={Collaborative forecasting of influenza-like illness in Italy: the Influcast experience},
  author={Fiandrino, Stefania and Bizzotto, Andrea and Guzzetta, Giorgio and Merler, Stefano and Baldo, Federico and Valdano, Eugenio and Mateo-Urdiales, Alberto and Bella, Antonino and Celino, Francesco and Zino, Lorenzo and others},
  journal={medRxiv},
  pages={2024--09},
  year={2024},
  publisher={Cold Spring Harbor Laboratory Press}
}

\newpage
\appendix
\section{Transmission rate: derivation of the synthetic model}
\label{app:der_dl}
In \cite{johnsen2022seasonal} the authors assume that the scaling factor for the transmission rate $\psi(T(t))$ evolves following a logistic temperature dependency for modelling the dependency of temperature of COVID19 epidemic spread:
\begin{equation}
    \psi(T(t)) = \dfrac{a}{1 + b e^{c T(t)}} + d.
\end{equation}
The four parameters have been estimated through 500 maximum likelihood parameter estimators for establishing informed prior intervals for the approximate Bayesian approximation algorithm.
Hereafter, we show how the transmission rate equation \eqref{eq:modelTRsynth} has been derived.
Thus, the latter model assumes that the transmission factor depends on the instantaneous value of temperature, whilst we assume that the whole history of temperature up to a given time influences the transmission factor at the prescribed time, \textit{i.e.} we assume that the transmission rate depends on temperature as:
\begin{equation}
    \beta(t) = \dfrac{a}{be^{-\int_{t_0}^t \left (\frac{T(\zeta)}{T_m} - 1 \right ) \diff \zeta} + c } +d.
        \label{eq:requestedModelTRSynt}
\end{equation}
Solving equation \eqref{eq:modelTRsynth}, which is a Bernoulli nonlinear differential ODE with non-constant coefficients \cite{boyce2012elementary}, we obtain the following solution
\begin{equation}
    \beta(t) = \dfrac{\beta_r \frac{\beta_0}{\beta_r - \beta_0}}{\frac{\beta_0}{\beta_r - \beta_0} - e^{-\int_{t_0}^t \left (\frac{T(\zeta)}{T_m} - 1 \right ) \diff \zeta}},
    \label{eq:derivedModelTRSynt}
\end{equation}
where $\beta_r$ is a reference value for the transmission rate, $\beta_0$ the initial value, $T_m$ a mean temperature value. 
By solving the nonlinear Bernoulli equation \eqref{eq:modelTRsynth}, we find that it corresponds to \eqref{eq:derivedModelTRSynt}.
This solution is in the requested form \eqref{eq:requestedModelTRSynt}.
\pagebreak

\section{Additional results on Test case 2}
\begin{figure}[t]
    \centering
    \includegraphics[width=0.9\textwidth]{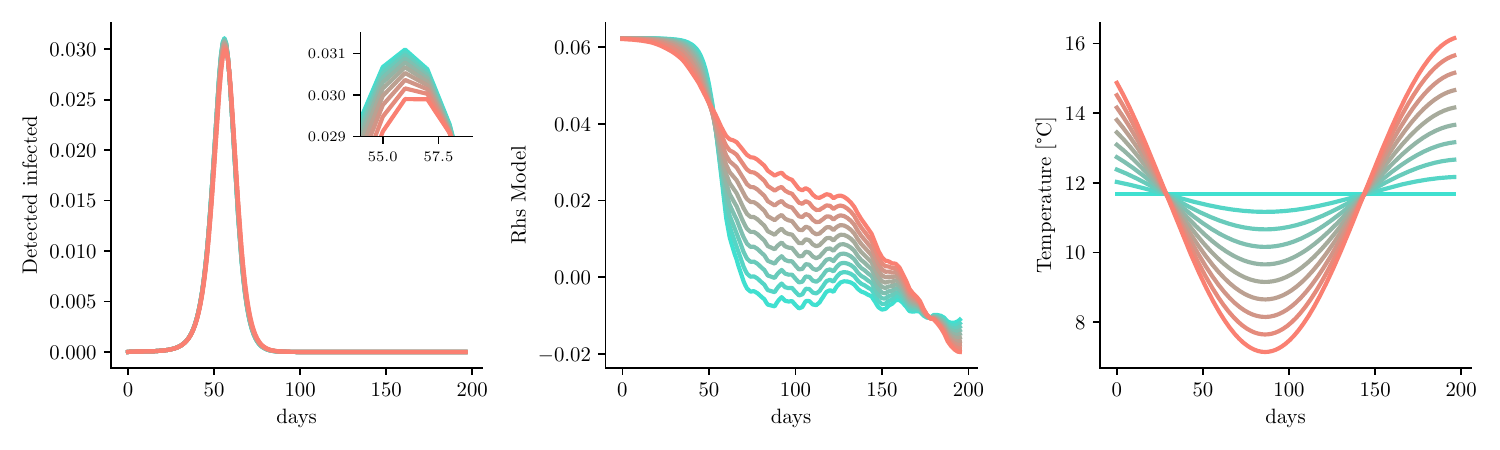}
    \caption{Total infected (left), learnt neural network model (center) corresponding to forward evaluations of harmonic temperature strategies varying amplitude (right). We consider as relative humidity time series the one corresponding to influenza wave of 2014/2015.}
    \label{fig:temps_eval_amp}
\end{figure}

\begin{figure}[H]
    \centering
    \includegraphics[width=0.9\textwidth]{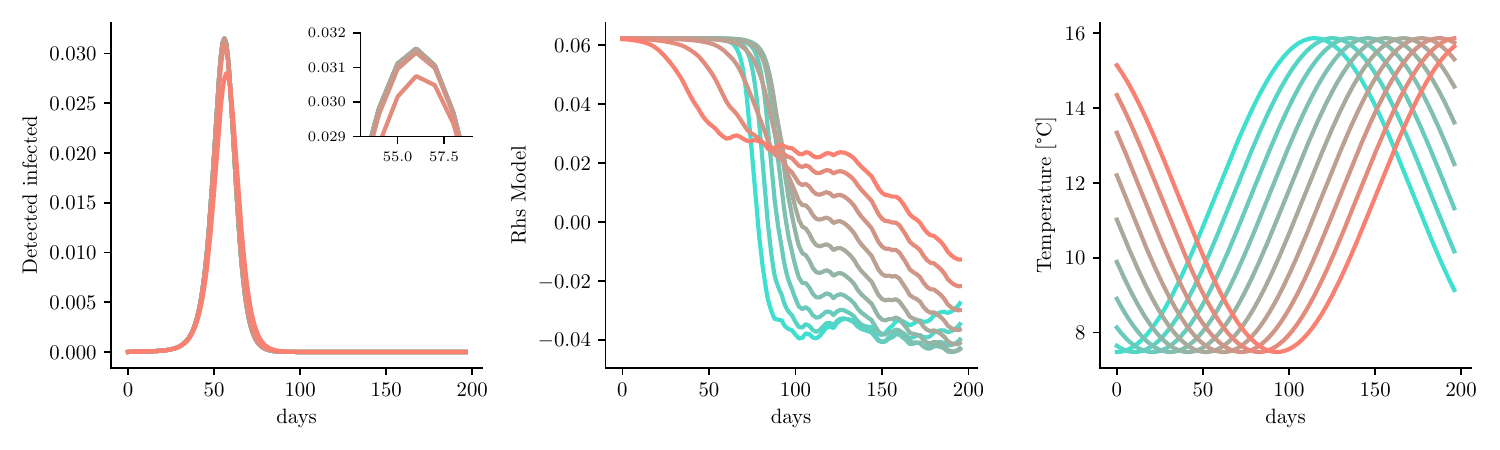}
    \caption{Total infected (left), learnt neural network model (center) corresponding to forward evaluations of harmonic temperature strategies varying phase (right). We consider as relative humidity time series the one corresponding to influenza wave of 2014/2015.}
    \label{fig:temps_eval_phase}
\end{figure}

\begin{figure}[H]
    \centering
    \includegraphics[width=0.9\textwidth]{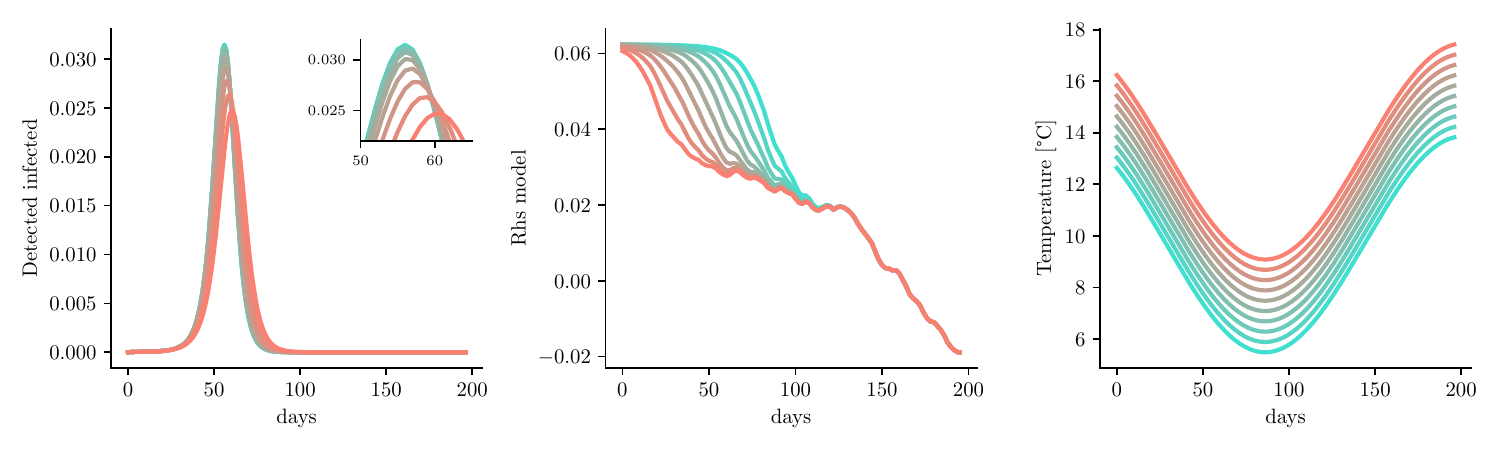}
    \caption{Total infected (left), learnt neural network model (center) corresponding to forward evaluations of harmonic temperature strategies varying mean value (right). We consider as relative humidity time series the one corresponding to influenza wave of 2014/2015.}
    \label{fig:temps_eval_mean}
\end{figure}

\begin{figure}[H]
    \centering
    \includegraphics[width=0.9\textwidth]{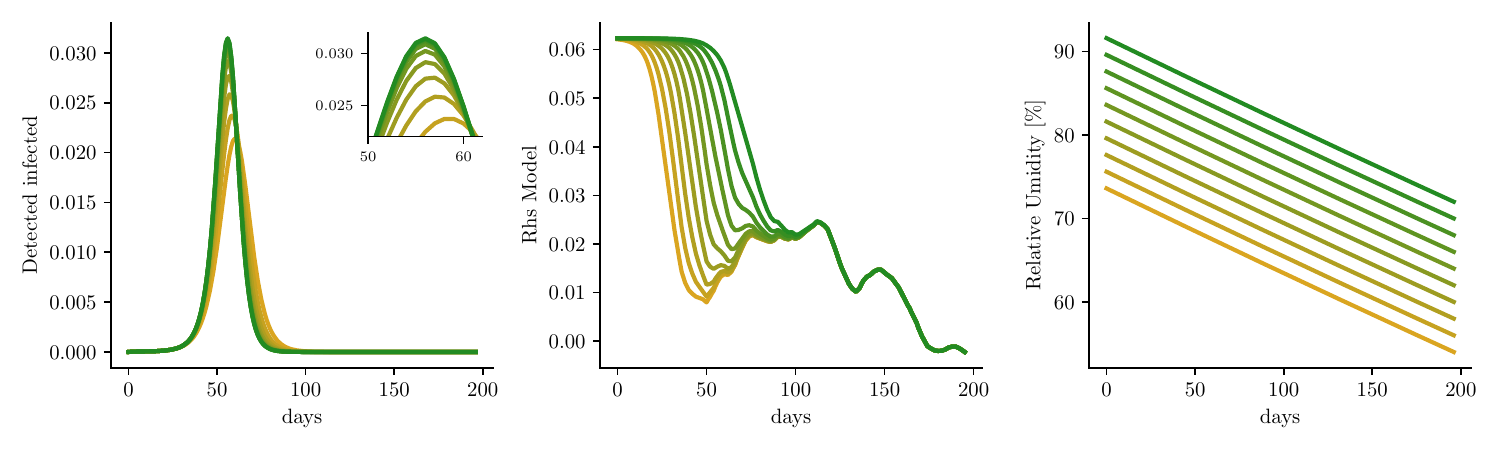}
    \caption{Total infected (left), learnt neural network model (center) corresponding to forward evaluations of linear humidity strategies (right). We consider as relative temperature time series the one corresponding to influenza wave of 2014/2015.}
    \label{fig:umids_eval_mean}
\end{figure}

In this Appendix, we present numerical results to analyze the reconstructed model using different input time series. Specifically, Figures \ref{fig:temps_eval_amp}-\ref{fig:umids_eval_mean} focus on the trained model from Case 3 (Section \ref{subsec:realInflu}) and test it forwardly, varying one parameterized input (either temperature or humidity) while keeping the other fixed at the original Savitzky-Golay filtered time series.
The latent parameter $\delta$ is held constant, as determined from testing Case 3 with both the actual temperature and humidity time series.

In order to have a parametrized family of input functions, we fit a sinusoidal least squares model with the temperature data referring to 2014-2015 wave.
Then, we consider different input functions belonging to this family varying the amplitude of the sinusoidal signal (Figure \ref{fig:temps_eval_amp}, right), its phase (Figure \ref{fig:temps_eval_phase}, right), or the mean value (Figure \ref{fig:temps_eval_mean}, right).
Each figure illustrates the behavior of detected infectious individuals for each considered temperature input (left), the neural network's trend (center), and the corresponding parameterized input trajectory (right).

We note that the oscillating temperature pattern positively affects the infectious peaks, which decrease as amplitude increases (cf. Figure \ref{fig:temps_eval_amp}, left). However, if the same data are shifted forward, \textit{i.e.} assuming new cases occur later in the winter season, higher peak values are observed  (cf. Figure \ref{fig:temps_eval_phase}, left).
At last, reducing the mean temperature value influences the model by increasing and advancing the peak (cf. Figure \ref{fig:temps_eval_mean}, left).

Additionally, we assess the impact of relative humidity by fitting a linear least squares model to real data from 2014-2015, while maintaining the real temperature time series.
Then, we consider different relative humidity signals belonging to this parametrized family as in Figure \ref{fig:umids_eval_mean} (right), and considering as temperature signal the one referring to 2014-2015 wave.
Consistently with findings from \cite{lowen2007influenza, mubayi2021analytical}, drier conditions tend to reduce peak values and delay the epidemic’s progression of the surrogate model.

\pagebreak
\end{document}